\documentclass[sigconf]{acmart} 

\usepackage{amsmath}
\usepackage{multirow}
\usepackage{float}
\usepackage{tikz}
\definecolor{OliveGreen}{rgb}{0.22, 0.6, 0.21}
\definecolor{LightBlue}{rgb}{0.267, 0.447, 0.769}
\definecolor{DarkWhite}{rgb}{0.75, 0.75, 0.75}

\usepackage{placeins}
\usepackage{afterpage}

\usepackage{enumitem} 
\setlist{noitemsep,nolistsep} 

\newenvironment{myitemize}{
\begin{itemize}[noitemsep,leftmargin=5.75pt,itemindent=0pt,labelsep=4pt,topsep=0pt]
  }
  {\end{itemize}
}

\AtBeginDocument{%
  \providecommand\BibTeX{{%
    \normalfont B\kern-0.5em{\scshape i\kern-0.25em b}\kern-0.8em\TeX}}}

\copyrightyear{2023}
\acmYear{2023}
\setcopyright{acmlicensed}\acmConference[SA Conference Papers '23]{SIGGRAPH Asia 2023 Conference Papers}{December 12--15, 2023}{Sydney, NSW, Australia}
\acmBooktitle{SIGGRAPH Asia 2023 Conference Papers (SA Conference Papers '23), December 12--15, 2023, Sydney, NSW, Australia}
\acmPrice{15.00}
\acmDOI{10.1145/3610548.3618203}
\acmISBN{979-8-4007-0315-7/23/12}

\begin{document}

\title{Efficient Graphics Representation with Differentiable Indirection}


\author{Sayantan Datta}
\email{sayantan.datta@mail.mcgill.ca}
\affiliation{%
    \institution{McGill Univeristy}
    \city{Montreal}
    \country{Canada}
}
\affiliation{
  \institution{Meta Reality Labs}
 \city{Redmond}
 \country{USA}
}


\author{Carl Marshall}
\email{csmarshall@meta.com}
\author{Zhao Dong}
\email{zhaodong@meta.com}
\author{Zhengqin Li}
\email{zhl@meta.com}
\affiliation{%
\institution{Meta Reality Labs}
\city{Redmond}
\country{USA}
}


\author{Derek Nowrouzezahrai}
\email{derek@cim.mcgill.ca}
\affiliation{%
  \institution{McGill University}
  \city{Montreal}
  \country{Canada}
}


\newcommand{\zl}[1]{{\color{red}#1}}
\newcommand{\cm}[1]{{\color{blue}[cm: #1]}}
\newcommand{\zd}[1]{{\color{green}[zd: #1]}}

\renewcommand{\shortauthors}{Datta et al.}
\newcommand{\din}{{\textit{DIn}}}
\newcommand{\mrhe}{{\textit{MRHE}}}
\newcommand{\mlp}{{\textit{MLP}}}
\newcommand{\mlps}{{\textit{MLPs}}}
\newcommand{\sdf}{{\textit{SDF}}}
\newcommand{\sdfs}{{\textit{SDFs}}}
\newcommand{\sh}{{\textit{SH}}}
\newcommand{\flop}{{FLOP}}
\newcommand{\flops}{{FLOPs}}

\begin{abstract}
   We introduce \textit{differentiable indirection} -- a novel learned primitive that employs differentiable multi-scale lookup tables as an effective substitute for traditional compute and data operations across the graphics pipeline. We demonstrate its flexibility on a number of graphics tasks, i.e., geometric and image representation, texture mapping, shading, and radiance field representation. In all cases, \textit{differentiable indirection} seamlessly integrates into existing architectures, trains rapidly, and yields both versatile and efficient results.
\end{abstract}



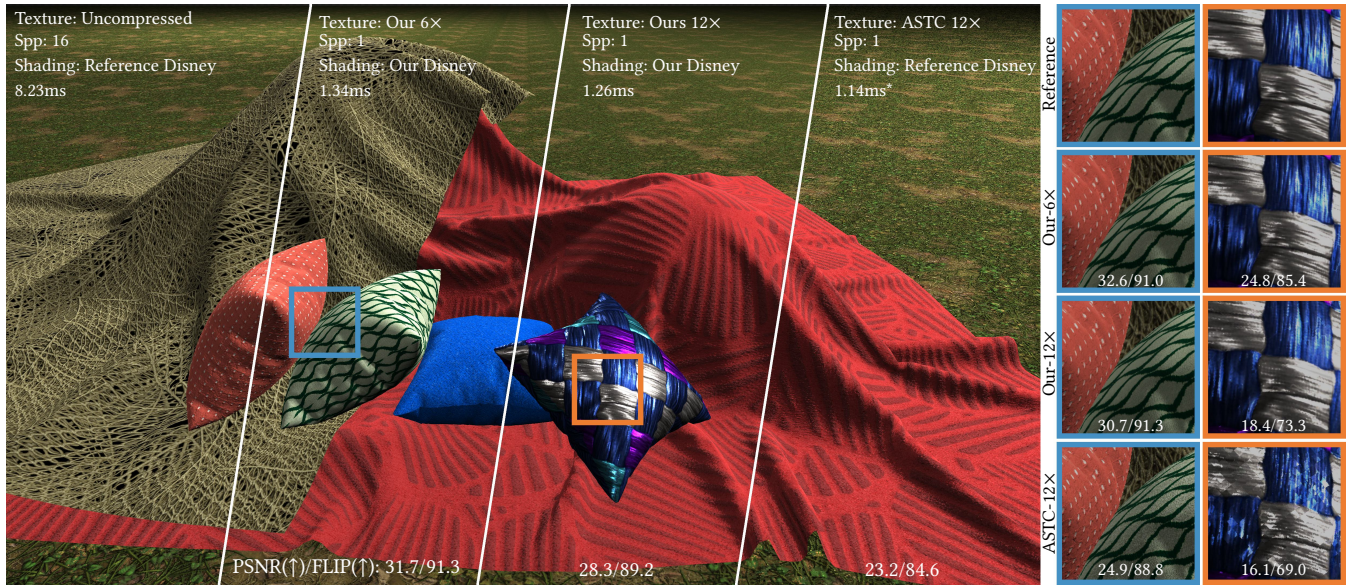
\begin{teaserfigure}
\begin{tikzpicture}
    \node[anchor=south west,inner sep=0] at (0,0){
    \includegraphics[width=\textwidth,trim={0cm 0cm 0cm 0cm},clip]{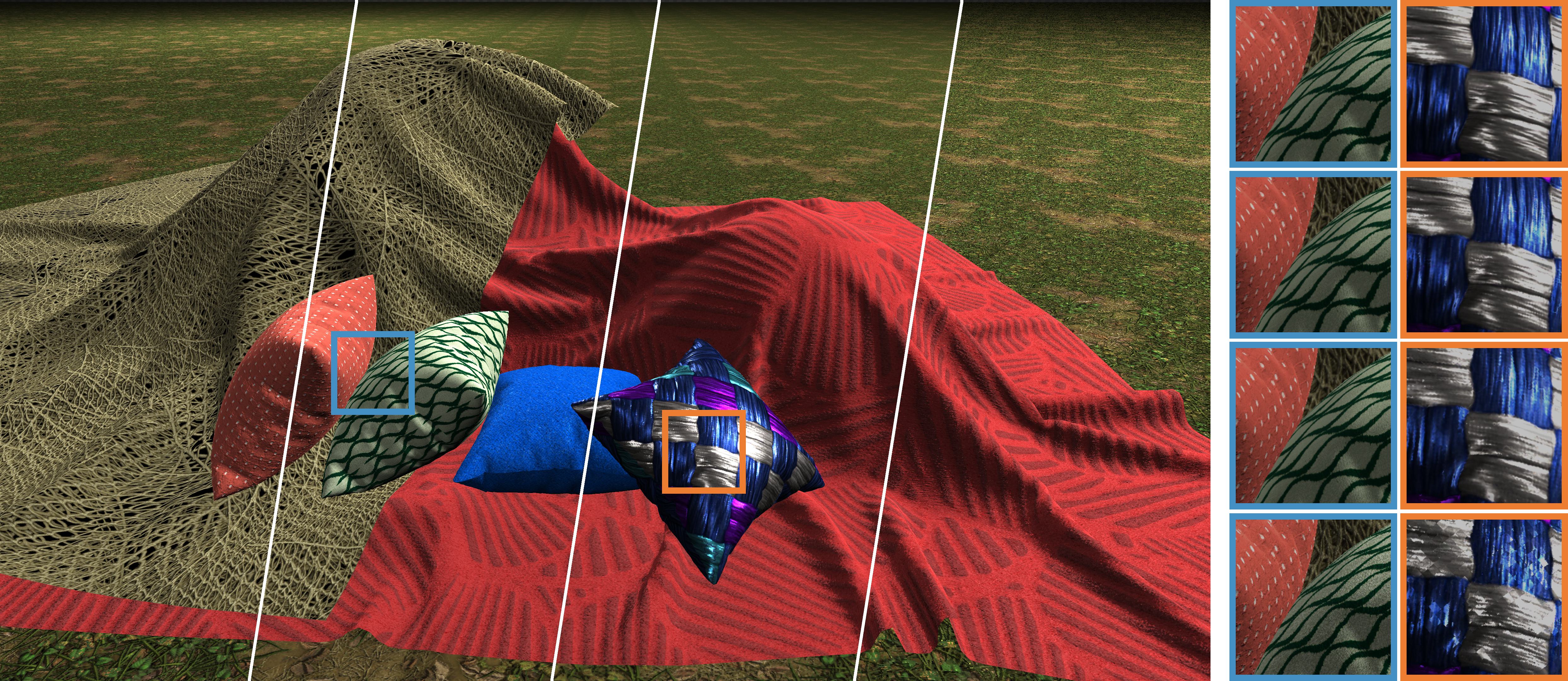}}; ;
        \node[anchor=south west] at (0.0,7.3) {\textcolor{white}{\footnotesize Texture: Uncompressed}};
    \node[anchor=south west] at (0.0,7.0) {\textcolor{white}{\footnotesize Spp: 16}};
    \node[anchor=south west] at (0.0,6.67) {\textcolor{white}{\footnotesize Shading: Reference Disney}};
    \node[anchor=south west] at (0.0,6.4) {\textcolor{white}{\footnotesize 8.23ms}};
    \node[anchor=south west] at (4.05,7.3) {\textcolor{white}{\footnotesize Texture: Our 6$\times$}};
    \node[anchor=south west] at (4.05,7.0) {\textcolor{white}{\footnotesize Spp: 1}};
    \node[anchor=south west] at (4.05,6.67) {\textcolor{white}{\footnotesize Shading: Our Disney}};
    \node[anchor=south west] at (4.05,6.4) {\textcolor{white}{\footnotesize 1.34ms}};
    \node[anchor=south west] at (7.55,7.3) {\textcolor{white}{\footnotesize Texture: Ours 12$\times$}};
    \node[anchor=south west] at (7.55,7.0) {\textcolor{white}{\footnotesize Spp: 1}};
    \node[anchor=south west] at (7.55,6.67) {\textcolor{white}{\footnotesize Shading: Our Disney}};
    \node[anchor=south west] at (7.55,6.4) {\textcolor{white}{\footnotesize 1.26ms}};
    \node[anchor=south west] at (10.9,7.3) {\textcolor{white}{\footnotesize Texture: ASTC 12$\times$}};
    \node[anchor=south west] at (10.9,7.0) {\textcolor{white}{\footnotesize Spp: 1}};
    \node[anchor=south west] at (10.9,6.67) {\textcolor{white}{\footnotesize Shading: Reference Disney}};
    \node[anchor=south west] at (10.9,6.4) {\textcolor{white}{\footnotesize 1.14ms*}};
    \node[anchor=south west] at (2.9,0.0) {\textcolor{white}{\small PSNR($\uparrow$)/FLIP($\uparrow$): 31.7/91.3}};
    \node[anchor=south west] at (7.5,0.0) {\textcolor{white}{\small 28.3/89.2}};
    \node[anchor=south west] at (11.3,0.0) {\textcolor{white}{\small 23.2/84.6}};
     \node[anchor=south west, rotate=90] at (14.056,6.2) {\textcolor{black}{\footnotesize Reference}};
     \node[anchor=south west, rotate=90] at (14.056,4.4) {\textcolor{black}{\footnotesize Our-6$\times$}};
     \node[anchor=south west, rotate=90] at (14.056,2.4) {\textcolor{black}{\footnotesize Our-12$\times$}};
     \node[anchor=south west, rotate=90] at (14.056,0.3) {\textcolor{black}{\footnotesize ASTC-12$\times$}};
    {
    \newcommand{\wLeft}{14.4}
    \newcommand{\wRight}{16.3}
    \newcommand{\hOffset}{3.85}
    \newcommand{\pSkip}{1.92}
    \node[anchor=south west] at (\wLeft, \hOffset - 0*\pSkip) {\textcolor{white}{\footnotesize 32.6/91.0} };
    \node[anchor=south west] at (\wRight, \hOffset - 0*\pSkip) {\textcolor{white}{\footnotesize 24.8/85.4} };
    \node[anchor=south west] at (\wLeft, \hOffset - 1*\pSkip) {\textcolor{white}{\footnotesize 30.7/91.3} };
    \node[anchor=south west] at (\wRight, \hOffset - 1*\pSkip) {\textcolor{white}{\footnotesize 18.4/73.3} };
    \node[anchor=south west] at (\wLeft, \hOffset - 2*\pSkip) {\textcolor{white}{\footnotesize 24.9/88.8} };
    \node[anchor=south west] at (\wRight, \hOffset - 2*\pSkip) {\textcolor{white}{\footnotesize 16.1/69.0} };
    }
\end{tikzpicture}
\caption{Figure shows the use of \textit{differentiable indirection} for texture compression/sampling and parametric shading at 4K screen resolution. Our technique relies on a few linearly interpolated indirect memory lookups and applies to a wide range of tasks in the graphics pipeline including distance and radiance field compression.}
\label{fig:teaser}
\end{teaserfigure}

\maketitle

\section{Introduction}

Neural primitives are the fundamental building block of neural networks and used for a variety of purposes in graphics applications, such as appearance capture \cite{iron22}, shading \cite{renderML22}, radiance caching \cite{nrc21}, view-synthesis \cite{NeRF20}, and shadows \cite{neuralShadow22}. Having \textit{efficient} neural primitives is vital due to their impact on latency, power, and training speed. Achieving high runtime performance with neural primitives is essential to the adoption of neural networks in real-time and low-power applications, such as AR/VR. 

We introduce a simple primitive with excellent runtime characteristics, featuring \textit{low compute \flops}, \textit{minimal memory reads per query}, and \textit{a compact parameter size}. Many neural networks rely on \textit{multi-layer perceptrons} (\mlp) due to their appeal as universal function approximators; however, \mlp~ layers are often the most computationally expensive component of a network and scale quadratically (both in \flops~and bytes transferred) with quality due to large matrix operations \cite{deep15}. Conversely, combining memory grids with fixed function non-linearities such as \textit{Spherical Harmonics (SH)} \cite{fridovich2022plenoxels} or \textit{ReLUs} \cite{ReluField22} reduces compute and memory transfer but incurs a large parameter cost. Our novel primitive --  \textit{differentiable indirection} -- strikes a balance across these criteria and \textcolor{black}{useful for a variety of data compression and compute representation tasks}. It is compatible with any differentiable logic, such as \mlps~ or fixed function approaches, but significantly reduces or even eliminates reliance on \mlps. Notably, all of our examples are \mlp-free, thereby eliminating the need for specialized hardware \cite{nvCoopMat} acceleration in real-time applications. \textit{Differentiable indirection} draws its expressive power solely from memory indirections and linear interpolation. This approach aligns well with the emerging computing paradigm of \textit{compute in memory} \cite{cimMac21,cimSram22}, which departs from traditional \textit{von Neumann} model that \mlps~ are modelled on. We apply \textit{differentiable indirection} to various tasks in the (neural) graphics pipeline, showcasing its potential as an efficient and flexible primitive for improving runtime efficiency. 
\section{Related work}
Neural primitives serve as the fundamental building blocks for modern neural techniques. We provide an overview of existing neural primitives and explore their applications in graphics.

\mlp~architectures provide a compact implicit representation that seamlessly scales up to higher dimensional inputs, such as signed distance field \cite{deepSdf}, neural radiance field \cite{NeRF20}, and neural BRDF \cite{bi2020neural,boss2021neural,boss2021nerd,zhang2021nerfactor}. They trade parameter size for compute, memory bandwidth, and a relatively longer training time as each training example affects all network weights. Even small MLPs (2 layer deep, 64 unit wide) are computationally and memory-intensive, requiring thousands of \flops~and bytes transferred per query. In contrast, our differentiable indirection relies on memory indirections and interpolations, resulting in reduced computational demands.

Grid-based representations explicitly store trainable parameters on a regular grid \cite{localSDF20, ReluField22} or a tree \cite{plenoxels21, ngLod} and then retrieve them at run-time using the input coordinate as key. The stored features are processed further using a non-linearity, such as \textit{ReLU} \cite{ReluField22} or \sh~ \cite{fridovich2022plenoxels}. While suitable for fast, localized updates, explicit representations tend to have large parameter sizes and has difficulty scaling up to higher dimensional inputs. 

Recent works combine \mlp~and grid representations to balance between memory and computational cost, achieving complex neural shading \cite{zeltner2023realtime,neumip}, neural material texture compression \cite{ntc2023}, and efficient and high-quality neural radiance field rendering \cite{dvgo22, chen2022tensorf, variableNeuralField, mul22instant}. Particularly, \textit{instant-NGP} combines \textit{multi-resolution hash encoding}  \cite{mul22instant} with pyramid of latent features, demonstrating to be effective for a wide-range of reconstruction and compression applications. Similarly, we show the effectiveness of our primitive in the broader context of neural rendering. 
However, our technique is also effective as a standalone unit without an \mlp. This provides unparalleled efficiency advantages, making it particularly suitable for low-power applications like neural shading on mobile devices.

\section{Overview \label{sec:ovrview}}

\textit{Differentiable indirection} (\din) is a flexible and powerful tool for representing compute/data problems in both the modern and neural graphics pipelines. \din~is similar to a pointer indirection - we query a memory location that contains a pointer to a secondary location containing the final output. However, we also make pointer indirection differentiable, hence \textit{differentiable indirection}. Our algorithm learns the pointer values stored as an array using gradient descent. \din~is a flexible and simple-to-integrate representation, as demonstrated in our applications to geometric and image representation, texture mapping, shading, and radiance field compression.

\begin{figure}[t!]
 \begin{tikzpicture}
    \node[anchor=south west,inner sep=0] at (0.0,0.0) {
    \includegraphics[width=23cm, trim={0.0cm 14.1cm 0.0cm 0.0cm}, clip]{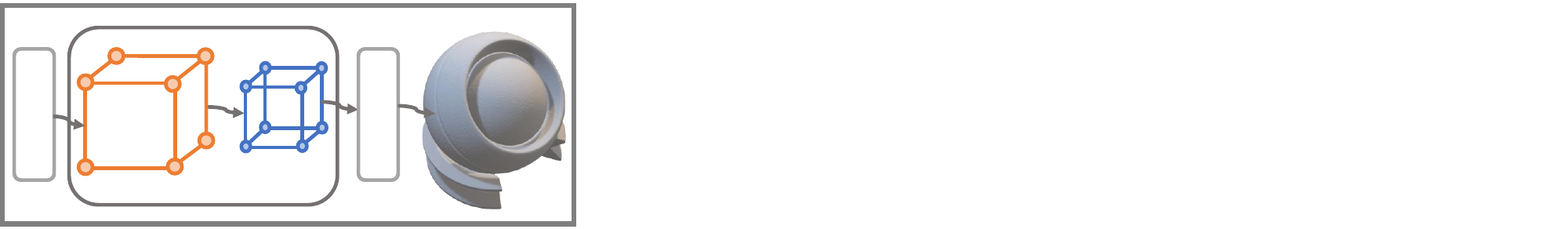}};
    \node[anchor=south west, rotate=90] at (0.79, 0.74) { \textcolor{black}{\small Spatial coords} };
    \node[anchor=south west] at (1.7, 2.5) { \textcolor{black}{\small Primary} };
    \node[anchor=south west] at (1.3, 1.3) { \textcolor{black}{\small 3D-Array} };
    \node[anchor=south west] at (3.6, 2.5) { \textcolor{black}{\small Cascaded} };
    \node[anchor=south west] at (3.9, 1.5) { \textcolor{black}{\small 3D} };
    \node[anchor=south west] at (1.1, 0.4) { \textcolor{black}{\small SDF network, total size: 24MB} };
     \node[anchor=south west, rotate=90] at (5.77, 0.65) { \textcolor{black}{\small Marching cubes} };
\end{tikzpicture}
\caption{\label{fig:sdfArch} \sdf~representation using \din. Primary and cascaded resolutions are $200^3\times 3$ and $64^3\times 1$ respectively.}
\end{figure}
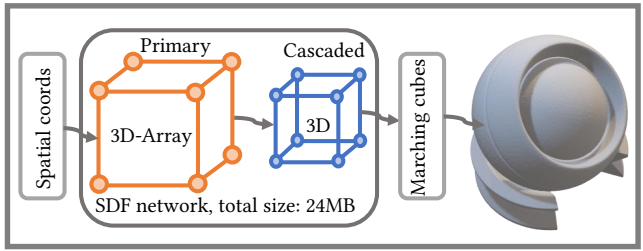

Our technique in its simplest form requires two arrays -- a \textbf{primary} array and a \textbf{cascaded} array. The article uses the same terminology throughout. Lookup into the primary array returns a pointer into the cascaded array. The corresponding location in the cascaded array contains the output. Figure \ref{fig:sdfArch} shows a visual representation of \textit{differentiable indirection} where the primary and the cascaded arrays are highlighted in orange and blue respectively. 

\paragraph{Differentiable Arrays} We introduce fully differentiable arrays - the main building block of our technique. A key requirement of our technique is that the arrays are not only differentiable w.r.t.\ to the content of each cell but also w.r.t.\ its indices or the uv-coordinates. The latter allows the gradients to backpropagate though the cascaded array to learn the pointer values stored in the primary array. The differentiability is achieved by linearly interpolating the array cells. For an input coordinate $\mathbf{x} \in [0,1)^d$, the coordinate is scaled by the array resolution ($N$) and rounding down $\lfloor{\mathbf{x}\cdot\mathbf{N}}\rfloor$ and up $\lceil{\mathbf{x}\cdot\mathbf{N}}\rceil$ - forming a $d$-dimensional voxel encapsulating $\mathbf{x}$. The vertices of the voxel is interpolated according to the distance of the query point from the corners to produce the output ($o$) given by: 
\begin{equation}
    \mathbf{o} = \textstyle\sum\limits_{i=0}^{2^d-1}\alpha_i \mathcal{F}(\mathbf{c}[i]),
\end{equation}
where $\alpha_i$, and $\mathbf{c}[i]$, are the interpolation weights, and the array cell content at the voxel-vertex $i$ respectively. We pass the cell contents through an additional non-linearity $\mathcal{F}$ with some specific characteristics as discussed in the next section. The gradients w.r.t.\ the input coordinate $\mathbf{x}$ and cell content $\mathbf{c}[i]$ is given by
\begin{equation}
\frac{d\mathbf{o}}{d\mathbf{x}} = \sum\limits_{i=0}^{2^d-1} \frac{d\alpha_i}{d\mathbf{x}}\mathcal{F}(\mathbf{c}[i]),\,\,\text{and}\,\, \frac{d\mathbf{o}}{d\mathbf{c}[i]} = \alpha_i \frac{d\mathcal{F}}{d\mathbf{c}[i]}
\end{equation}
respectively. The gradients are plugged into \textit{autodiff} framework such as \textsc{PyTorch} \cite{citePytorch}, enabling backpropagation though arrays. Prior techniques - \textit{Multi Resolution Hash Encoding}, abbreviation \mrhe~ \cite{mul22instant} and \textit{ReLU Fields} \cite{ReluField22} only compute gradient w.r.t.\ cell contents $\mathbf{c}[i]$.

\paragraph{Non-linearity} We apply a periodic non-linearity $\mathcal{F}$ to the primary array and an optional periodic/aperiodic non-linearity to the cascaded array. The purpose of the periodic non-linearity is to bound and continuously \textit{wrap-around} the array content for use as an input in the next layer. We test two periodic functions - a sinusoid given by $(1 + \sin(nx))/2$\ and a non-negative triangle wave with a period 2 with peak output 1 at input 1. We use the triangle wave for all cases except \textit{Disney-BRDF} where we use the sinusoid. Note that the non-linearity $\mathcal{F}$ is applied before interpolation, thus can be removed during inference and baked directly into the array cells for improved efficiency. The output is also bounded to $[0, 1)$ when using a periodic non-linearity, lending the opportunity to quantize the array values to 8/16-bit for inference without significant impact on quality; a property we utilize heavily for all tasks.


\paragraph{Initialization} \textit{Differentiable indirection} is thus a cascade of multi-dimensional differentiable arrays. We initialize the primary array using a linear ramp resembling a standard uv-map in 2D or a similar analog in higher dimension. Initially the primary array is an identity map between input and output. Gradient descent simultaneously distorts the identity map in the primary and updates the values stored in the cascaded. Linear interpolation puts an implicit constraint on the values the primary array may accommodate. If we imagine the uv-map as a fabric, gradient descent is only allowed to locally wrinkle the fabric. The effect is illustrated in figure \ref{fig:tex2d2d}. The choice of initialization for the cascaded array is application specific.  

\section{Applications}

We demonstrate the applications of \textit{differentiable indirection} across various stages of the graphics pipeline -- starting with geometric representations, followed by examples in deferred shading such as image and texture compression, and parametric shading. Finally, to compress implicit representation such as neural radiance field.     

\subsection{Compact geometric representation}

Our technique is easily adapted to implicit geometry representation task using a \textit{Signed Distance Function} (\sdf) representation. \sdfs~volumetrically encode the zero-level set of a spatial 3D shape function. \din~readily applies to \sdf~representation, compressing the volumetric information in the 3D arrays. Figure \ref{fig:sdfArch} shows the primary array queried by a spatial 3D-coordinate that points to the cascaded array containing the signed distances from the zero-level set; we train one network per \sdf~and use marching cubes~\cite{mcubes} for illustrative surface reconstruction.

\paragraph{Implementation details} We generate training samples pairs -- a position and its corresponding signed distance using an \sdf~dataset generator \cite{sdfGen22}. We preferentially sample points closer to the surface zero-level set \cite{ngLod}; one billion samples from near the surface and 20 million  uniformly distributed over the volumetric domain. We use 100 million near-surface samples to compute test time error statistics. While \mrhe~ proposes \textit{MAPE} as the loss function, this results in many ``floater'' artifacts near the surface; we believe this is due to \textit{MAPE} over-emphasizing on-surface sample importance at a cost of distorting the distance field slightly off-surface. We solve this problem by quantizing signed distances to $\pm1$ a.k.a.\ \textit{Truncated SDF} and applying an \textit{MAE} loss. In a grid based technique like ours, only having two discrete values allows the transition-boundary (representing a surface) to adapt more freely. 
Figure \ref{fig:sdfArch} illustrates a 3D shape generated using our training methodology. We include more results in section \ref{sec:implSdf} and training details in supplemental section \ref{sec:sdf}. 

\begin{figure}[t!]
 \begin{tikzpicture}
    \node[anchor=south west,inner sep=0] at (0.0,0.0) {
    \includegraphics[width=8.5cm, trim={0.0cm 6.0cm 0.0cm 0.0cm}, clip]{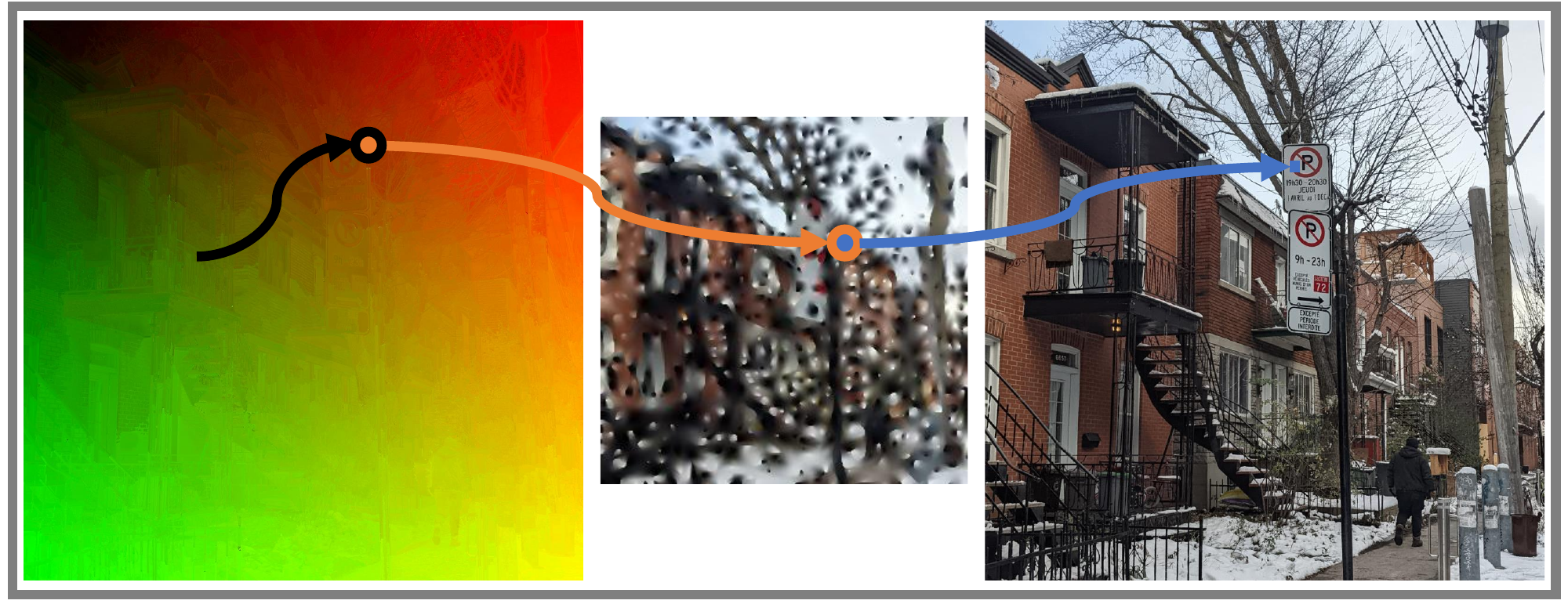}};
     \node[anchor=south west] at (0.75, 1.45) { \textcolor{black}{\small UV-coords} };
     \node[anchor=south west] at (0.75, 0.75) { \textcolor{black}{\small Primary array} };
     \node[anchor=south west] at (0.85, 0.35) { \textcolor{black}{\small ($976^2\times2$)} };
     \node[anchor=south west] at (3.23, 2.55) { \textcolor{black}{\small Cascaded array} };
     \node[anchor=south west] at (3.5, 0.2) { \textcolor{black}{\small ($244^2\times3$)} };
\end{tikzpicture}
\caption{\label{fig:tex2d2d} Visualizing learned primary (2D) and cascaded (2D) arrays, compressing a 2k image by 6$\times$. A 2D-primary and 4D-cascaded produce better results in practice.}
\end{figure}
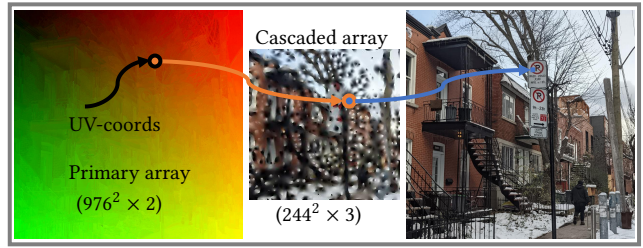

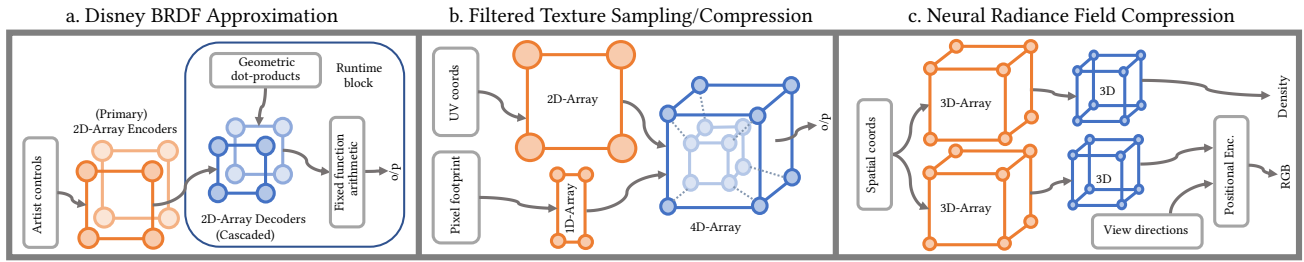
\begin{figure*}[t!]
\begin{tikzpicture}
    \node[anchor=south west,inner sep=0] at (0,0){
    \includegraphics[width=17.4cm,trim={0cm 13.0cm 0.05cm 0.05cm},clip]{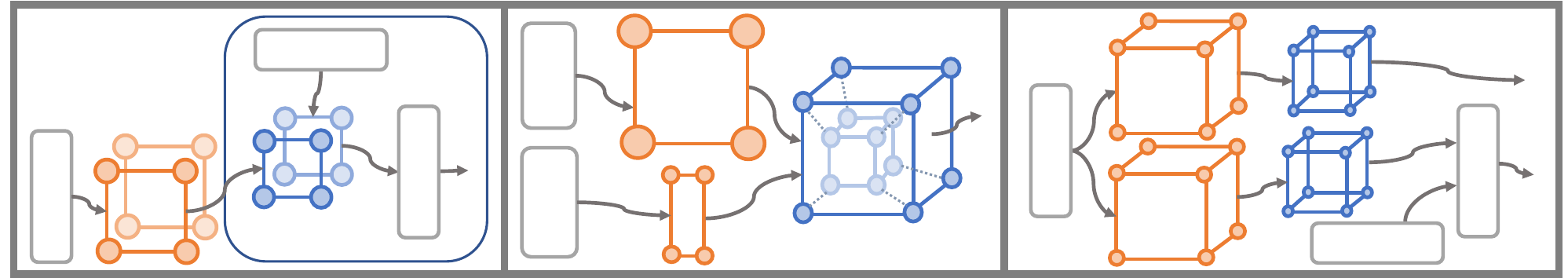}}; ;
    \node[anchor=south west] at (0.8, 3.0) { \textcolor{black}{\small a. Disney BRDF Approximation} };
    \node[anchor=south west] at (5.9, 3.0) { \textcolor{black}{\small b. Filtered Texture Sampling/Compression} };
    \node[anchor=south west] at (12.0, 3.0) { \textcolor{black}{\small c. Neural Radiance Field Compression} };

    \node[anchor=south west, rotate=90] at (0.7, 0.25) { \textcolor{black}{\tiny Artist controls} };
    \node[anchor=south west] at (1.2, 1.75) { \textcolor{black}{\tiny (Primary)} };
    \node[anchor=south west] at (0.9, 1.55) { \textcolor{black}{\tiny 2D-Array Encoders} };
    \node[anchor=south west] at (3.0, 2.26) { \textcolor{black}{\tiny dot-products} };
    \node[anchor=south west] at (3.1, 2.48) { \textcolor{black}{\tiny Geometric} };
    \node[anchor=south west] at (2.6, 0.35) { \textcolor{black}{\tiny 2D-Array Decoders} };
    \node[anchor=south west] at (2.8, 0.15) { \textcolor{black}{\tiny (Cascaded)} };
    \node[anchor=south west, rotate=90] at (4.72, 0.55) { \textcolor{black}{\tiny Fixed function} };
    \node[anchor=south west, rotate=90] at (4.9, 0.75) { \textcolor{black}{\tiny arithmetic} };
    \node[anchor=south west, rotate=90] at (5.48, 0.95) { \textcolor{black}{\tiny o/p} };
    \node[anchor=south west] at (4.4, 2.4) { \textcolor{black}{\tiny Runtime} };
    \node[anchor=south west] at (4.5, 2.2) { \textcolor{black}{\tiny block} };

    \node[anchor=south west, rotate=90] at (6.25, 1.75) { \textcolor{black}{\tiny UV coords} };
    \node[anchor=south west, rotate=90] at (6.3, 0.25) { \textcolor{black}{\tiny Pixel footprint} };
    \node[anchor=south west] at (7.2, 1.95) { \textcolor{black}{\tiny 2D-Array} };
    \node[anchor=south west, rotate=90] at (7.82, 0.25) { \textcolor{black}{\tiny 1D-Array }};
    \node[anchor=south west] at (9.1, 0.25) { \textcolor{black}{\tiny 4D-Array }};
    \node[anchor=south west, rotate=90] at (11.2, 1.625) { \textcolor{black}{\tiny o/p }};

    \node[anchor=south west, rotate=90] at (11.83, 0.8) { \textcolor{black}{\tiny Spatial coords} };
    \node[anchor=south west] at (12.43, 1.9) { \textcolor{black}{\tiny 3D-Array} };
    \node[anchor=south west] at (12.43, 0.5) { \textcolor{black}{\tiny 3D-Array} };
    \node[anchor=south west] at (14.55, 2.1) { \textcolor{black}{\tiny 3D} };
    \node[anchor=south west] at (14.5, 0.93) { \textcolor{black}{\tiny 3D} };
    \node[anchor=south west] at (14.6, 0.25) { \textcolor{black}{\tiny View directions} };
    \node[anchor=south west, rotate=90] at (16.55, 0.5) { \textcolor{black}{\tiny Positional Enc.} };
    \node[anchor=south west, rotate=90] at (17.3, 1.8) { \textcolor{black}{\tiny Density} };
    \node[anchor=south west, rotate=90] at (17.3, 0.85) { \textcolor{black}{\tiny RGB} };
\end{tikzpicture}
\caption{\label{fig:arch} 
From left to right, parametric shading, compact texture sampling, and radiance field based on \textit{differentiable indirection.}}
\end{figure*}

\subsection{Compact texture and image representations}
We apply \textit{differentiable indirection} to real-time texture and natural image (de-)compression, and filtered texture sampling. 
\subsubsection{Compact image representation\label{sec:texComp}} In this task we query the primary array with 2D uv-coordinates (figure \ref{fig:tex2d2d}). The cascaded array encodes the corresponding color value at the uv location. Figure \ref{fig:tex2d2d} visualizes the learned arrays' contents -- note the primary array favors high frequency details as a pseudo-distorted uv-map. We however recommend alternative network configurations than those in figure \ref{fig:tex2d2d} which produce improved results, as described next.

\paragraph{Implementation details} We use a 2D 4-channel array as primary and a 4D 3-channels array containing RGB values as the cascaded. For \textit{PBR} textures, we extend the number of channels in the cascaded array to include various shading parameters. During training, the network maps the queried uv-coordinates to the corresponding color/parameter value. A higher dimensional (> 2D) cascaded array results in better compression quality. For a mixture of textures containing natural images and several \textit{PBR} materials from \textit{Adobe Substance}, we noticed a 15\% improvement in PSNR moving from 2D to 3D and 3\% improvement from 3D to 4D. We generate the training uv-coordinates using stratified random sampling where each strata corresponds to a texel in the base texture. Target color values are obtained with bi-linear sampling of the reference image at the queried uv-coordinates. Details in supplemental section \ref{sec:compactImg}.

\subsubsection{Neural Texture Sampler\label{sec:texureSampler}} For shading applications, we also need to account for texture filtering \cite{texFilter} based on the projected pixel footprint onto a geometry surface. With minor modifications to the previous compression-only network, we can additionally treat filtered texture sampling. Figure \ref{fig:arch}(b) shows our texture sampling/filtering network configuration, requiring two inputs -- a uv-coordinate, and a pixel footprint magnitude, both of which are readily available in modern interactive renderers.

\begin{figure}[b!]
 \begin{tikzpicture}
    \node[anchor=south west,inner sep=0] at (0.0,0.0) {
    \includegraphics[width=8cm, trim={0cm 0cm 0cm 0.0cm},clip]{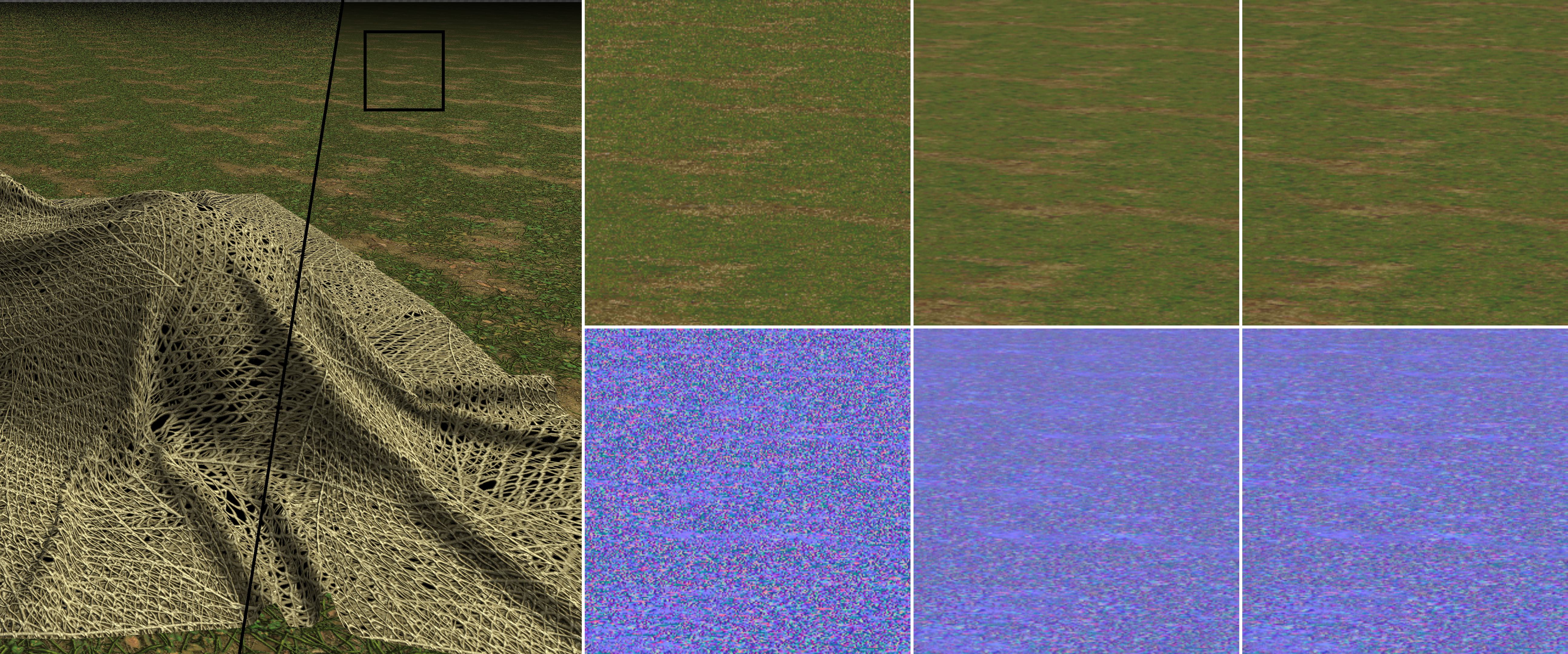}};
    \node[anchor=south west, rotate=0] at (0.3, 3.3) { \textcolor{black}{\small without} };
    \node[anchor=south west, rotate=0] at (2.0, 3.3) { \textcolor{black}{\small with} };
    \node[anchor=south west, rotate=0] at (3.3, 3.3) { \textcolor{black}{\small without} };
    \node[anchor=south west, rotate=0] at (5.2, 3.3) { \textcolor{black}{\small with} };
    \node[anchor=south west, rotate=0] at (6.8, 3.3) { \textcolor{black}{\small Ref} };
    {
     \newcommand{\wOffset}{3.2}
    \newcommand{\pSkip}{1.8}
    \newcommand{\hUp}{2.95}
    \newcommand{\hDn}{1.3}
    \node[anchor=south west] at (\wOffset + 0*\pSkip, \hUp) {\textcolor{black}{\footnotesize 26.7/89.9} };
    \node[anchor=south west] at (\wOffset + 0*\pSkip-0.3, \hUp-0.3) {\textcolor{black}{\footnotesize (Psnr($\uparrow$)/Flip($\uparrow$))} };
     
    \node[anchor=south west] at (\wOffset + 0*\pSkip, \hDn) {\textcolor{black}{\footnotesize 15.7/72.6} };
    \node[anchor=south west] at (\wOffset + 1*\pSkip, \hUp) {\textcolor{black}{\footnotesize 32.6/92.2} };
    \node[anchor=south west] at (\wOffset + 1*\pSkip, \hDn) {\textcolor{black}{\footnotesize 27.8/90.6} };
    }
\end{tikzpicture}
\caption{\label{fig:filterOnOff} 
Comparing texture compression with and without pixel footprint. Without the footprint information, the output at grazing angle is noisy as shown in the cutouts.}
\end{figure}
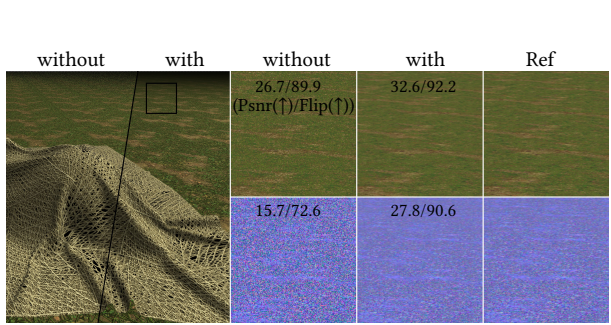

\paragraph{Implementation details} We use two primary arrays - a 2D array with 3 channels for input uv-coordinates, and a 1D array with 1 channel for pixel footprint. The output of the two primary arrays is concatenated and used as the input for the cascaded 4-D array. The network approximates a \textit{trilinear texture sampler} similar to those available on modern GPUs but unlike GPUs, we do not store an explicit mip-chain. We soft-emulate a GPU texture sampler to generate our target data. We train our network on random uv-samples and pixel-footprint values with corresponding target generated with the emulated trilinear sampler. Details in supplemental \ref{sec:texSamp}.

The cost of evaluating the new network only increases marginally compared to section \ref{sec:texComp} while providing higher quality per sample for shading tasks, especially for pixels with large footprint at grazing angles. The effect is illustrated in Figure \ref{fig:filterOnOff}. 

\subsection{Efficient parametric shading models}
So far, we have seen the applications of our technique in data representation. This section introduces parametric shading as compute approximation task. We use our technique on two different \textit{BRDFs} - a simple \textit{isotropic GGX} and a more complex \textit{Disney BRDF}. 

\subsubsection{Isotropic GGX approximation\label{sec:isotropicGGxApprox}}
\begin{figure}[b!]
 \begin{tikzpicture}
    \node[anchor=south west,inner sep=0] at (0.0,0.0) {
    \includegraphics[width=11.5cm, trim={0.3cm 11.1cm 10.75cm 0.2cm}, clip]{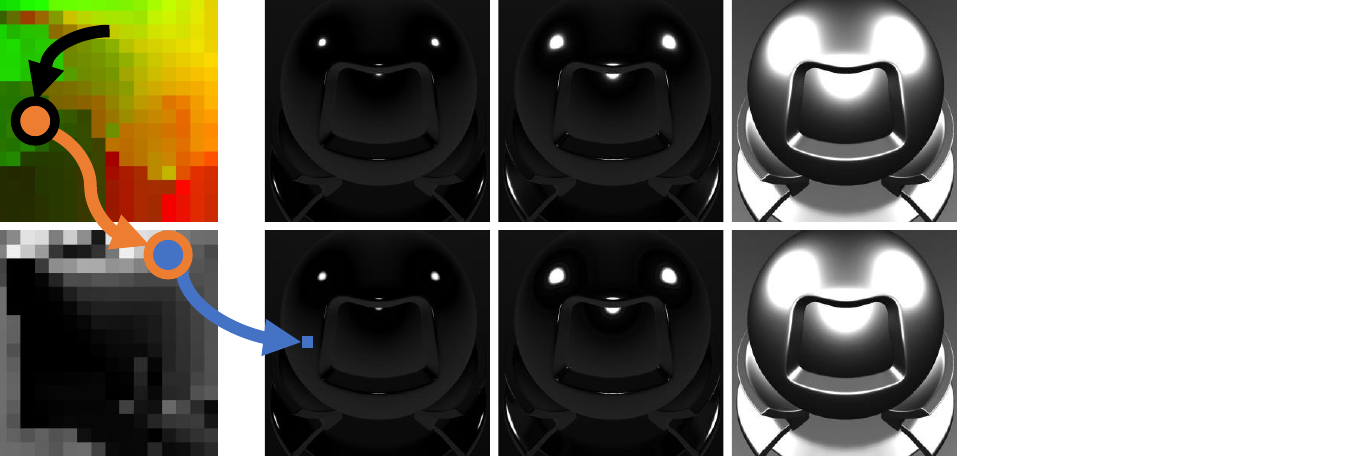}};
      \node[anchor=south west] at (-0.075, 3.8) { \textcolor{black}{\small Array contents} };
      \node[anchor=south west] at (0.6, 3.2) { \textcolor{black}{\small $(\alpha, h_z)$} };
      \node[anchor=south west] at (-0.05, 2.1) { \textcolor{orange}{\small $(u, v)$ }};
      \node[anchor=south west] at (2.7, 3.9) { \textcolor{black}{\small $\alpha=0.05$} };
      \node[anchor=south west] at (4.6, 3.9) { \textcolor{black}{\small $\alpha=0.1$} };
      \node[anchor=south west] at (6.7, 3.9) { \textcolor{black}{\small $\alpha=0.5$} };
      \node[anchor=south west, rotate=90] at (2.25, 2.25) { \textcolor{black}{\small Reference} };
      \node[anchor=south west, rotate=90] at (2.25, 0.1) { \textcolor{black}{\small Network O/P} };
     
      \node[anchor=south west, rotate=90] at (0.0, 2.25) { \textcolor{black}{\small Primary} };
      \node[anchor=south west, rotate=90] at (0.0, 0.3) { \textcolor{black}{\small Cascaded} };
\end{tikzpicture}
\caption{\label{fig:isoGgxLookup} Visualization of the learned primary and cascaded arrays for \textit{isotropic GGX} approximation on left and the resulting rendered output (average 40dB PSNR) on right.}

\end{figure}
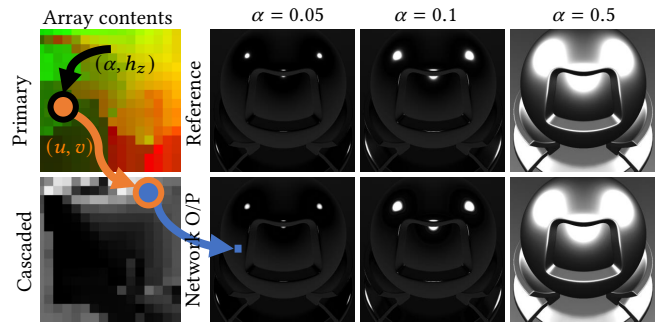

\textit{Isotropic GGX} is a popular \textit{BRDF} used for specular shading, expressed analytically as $D(h_z, \alpha_h) = \alpha_h^4 \cdot (1 + (\alpha_h^4 - 1) \cdot h_z^2)^{-2} / \pi$. This application approximates the analytic expression using \din. We use the input parameters $h_z, \alpha_h$ as the input coordinates to the primary array (figure \ref{fig:isoGgxLookup}) and compare the output of the cascaded array with the corresponding output of the analytic expression. The example serves as a benchmark for comparing various neural primitives in section \ref{sec:implIsoGGx}.

\subsubsection{Principled Disney BRDF approximation\label{sec:disneyApprox}}

\begin{figure*}[t!]
\begin{tikzpicture}
    \node[anchor=south west,inner sep=0] at (0,0){
    \includegraphics[width=17.4cm,trim={0cm 0cm 0 0cm},clip]{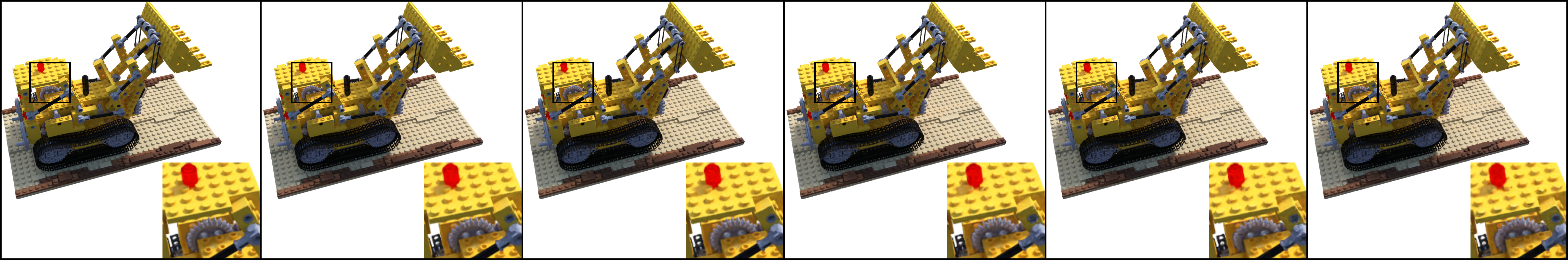}}; ;
    \node[anchor=south west] at (0.4, 2.8) { \textcolor{black}{\small Uncompressed} };
    \node[anchor=south west] at (3.9, 2.8) { \textcolor{black}{\small 5.91$\times$} };
    \node[anchor=south west] at (3.9 + 2.9, 2.8) { \textcolor{black}{\small 9.29$\times$} };
    \node[anchor=south west] at (3.9 + 2*2.9, 2.8) { \textcolor{black}{\small 29.6$\times$} };
    \node[anchor=south west] at (3.9 + 3*2.9, 2.8) { \textcolor{black}{\small 38.3$\times$} };
    \node[anchor=south west] at (3.9 + 4*2.9, 2.8) { \textcolor{black}{\small 59.1$\times$} };
    {
    \newcommand{\wOffset}{0.0}
    \newcommand{\pSkip}{2.9}
    \newcommand{\hUp}{0.3}
    \newcommand{\hDn}{0.00}
    \node[anchor=south west] at (\wOffset + 0*\pSkip, \hUp) {\textcolor{black}{\footnotesize Den: 63.5} };
    \node[anchor=south west] at (\wOffset + 0*\pSkip, \hDn) {\textcolor{black}{\footnotesize RGB: 762} };
    \node[anchor=south west] at (\wOffset + 1*\pSkip, \hUp) {\textcolor{black}{\footnotesize Den: 63.5} };
    \node[anchor=south west] at (\wOffset + 1*\pSkip, \hDn) {\textcolor{black}{\footnotesize RGB: 76.2} };
    \node[anchor=south west] at (\wOffset + 2*\pSkip, \hUp) {\textcolor{black}{\footnotesize Den: 12.7} };
    \node[anchor=south west] at (\wOffset + 2*\pSkip, \hDn) {\textcolor{black}{\footnotesize RGB: 76.2} };
    \node[anchor=south west] at (\wOffset + 3*\pSkip, \hUp) {\textcolor{black}{\footnotesize Den: 12.7} };
    \node[anchor=south west] at (\wOffset + 3*\pSkip, \hDn) {\textcolor{black}{\footnotesize RGB: 15.2} };
    \node[anchor=south west] at (\wOffset + 4*\pSkip, \hUp) {\textcolor{black}{\footnotesize Den: 6.35} };
    \node[anchor=south west] at (\wOffset + 4*\pSkip, \hDn) {\textcolor{black}{\footnotesize RGB: 15.2} };
    \node[anchor=south west] at (\wOffset + 5*\pSkip, \hUp) {\textcolor{black}{\footnotesize Den: 6.35} };
    \node[anchor=south west] at (\wOffset + 5*\pSkip, \hDn) {\textcolor{black}{\footnotesize RGB: 7.62} };
    }
    {
    \newcommand{\wOffset}{0.0}
    \newcommand{\pSkip}{2.9}
    \newcommand{\hUp}{2.5}
    \node[anchor=south west] at (\wOffset + 0*\pSkip, \hUp) {\textcolor{black}{\footnotesize 33.1/96.1} };
    \node[anchor=south west] at (\wOffset + 1*\pSkip, \hUp) {\textcolor{black}{\footnotesize 32.5/95.8} };
    \node[anchor=south west] at (\wOffset + 2*\pSkip, \hUp) {\textcolor{black}{\footnotesize 31.9/95.6} };
    \node[anchor=south west] at (\wOffset + 3*\pSkip, \hUp) {\textcolor{black}{\footnotesize 31.1/95.1} };
    \node[anchor=south west] at (\wOffset + 4*\pSkip, \hUp) {\textcolor{black}{\footnotesize 30.7/94.9} };
    \node[anchor=south west] at (\wOffset + 5*\pSkip, \hUp) {\textcolor{black}{\footnotesize 30.2/94.5} };
    }

    {
    \newcommand{\wOffset}{1.85}
    \newcommand{\pSkip}{2.9}
    \newcommand{\hUp}{0.2}
    \node[anchor=south west, rotate=90] at (\wOffset + 0*\pSkip, \hUp) {\textcolor{black}{\footnotesize 25.4} };
    \node[anchor=south west, rotate=90] at (\wOffset + 1*\pSkip, \hUp) {\textcolor{black}{\footnotesize 24.8} };
    \node[anchor=south west, rotate=90] at (\wOffset + 2*\pSkip, \hUp) {\textcolor{black}{\footnotesize 24.4} };
    \node[anchor=south west, rotate=90] at (\wOffset + 3*\pSkip, \hUp) {\textcolor{black}{\footnotesize 23.7} };
    \node[anchor=south west, rotate=90] at (\wOffset + 4*\pSkip, \hUp) {\textcolor{black}{\footnotesize 23.4} };
    \node[anchor=south west, rotate=90] at (\wOffset + 5*\pSkip, \hUp) {\textcolor{black}{\footnotesize 23.1} };
    }
\end{tikzpicture}
\caption{\label{fig:legoComp} 
Radiance field compression with \textit{DIn.} Top left shows average PSNR($\uparrow$)/FLIP($\uparrow$) w.r.t.\ all images in test set for varying compression ratios. Bottom left shows the size (in MBs) of the Density and RGB fields. PSNR($\uparrow$) for the cutout is provided nearby. Typical dimensions for the density, and RGB grid at 9.3$\times$ compression are $160^3\times3/64^3\times1$, and $280^3\times3/64^3\times12$ respectively.}
\end{figure*}
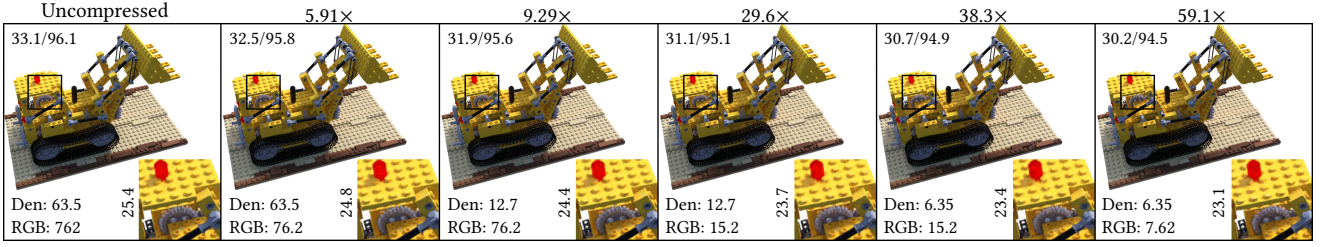

We approximate the \textit{Disney BRDF} using \din, retaining all artist controls with negligible impact on final quality (figure \ref{fig:disneyParamSweep}) while also improving upon the evaluation efficiency (table \ref{tab:disneyRuntimePerf}) compared to the reference analytic implementation. We follow the \textit{Principled Disney BRDF} implementation reference \cite{Li2022DisneyBSDF} to generate our training data. Excluding glass/transmission term, the reference \textit{BRDF} uses 10 artist controllable parameters and seven geometric dot-products as input. It is challenging to handle a high-dimensional input while also being efficient. While hypothetically possible, 17D arrays would be prohibitively expensive in practice. Other primitives using \mlps~such as \mrhe~can scale up with higher dimensions. However, such primitive require large \mlps~to attain desirable results here and do not improve upon the efficiency of reference evaluation.

We use a divide-and-conquer approach to partition the task into several components, leveraging the available domain knowledge in this setting. We start by factoring out the albedo ($\mathbf{\mathcal{A}}$) from final \textit{BRDF} output -- i.e., as referenced in equation 19 in \cite{Li2022DisneyBSDF} and rewrite the reference \textit{BRDF} equation as: 
\begin{equation}
    \text{Disney}({\mathbf{x}, \mathbf{\mathcal{A}}}) = \frac{\mathbf{\mathcal{A}}}{l}p(\mathbf{x}) + q(\mathbf{x}),
\end{equation}
where $\mathbf{x}\in [0, 1)^{17}$ are the control parameters, and $l$ is the luminance computed as a weighted sum of albedo-RGB channels according to $(0.2126, 0.7152, 0.0722)$. The $p$ and $q$ terms are single channel positive scalars obtained directly from our factorization. Note that the learnable quantities $p,q$ do not learn any color information, instead we modulate the albedo with the learned parameters. This is crucial for reducing color bleeding in the final output. We further refactor $p$ and $q$ according to 
\begin{equation}
\begin{split}
p(x) &= c_dD_d + c_{m0}D_m + c_{s0}\\
q(x) &= c_{c}D_c + c_{m1}D_m + c_{s1},
\end{split}
\end{equation}
where $D_d, D_m$, and $D_c$ are the \textit{Disney-diffuse, Disney-metallic, and Disney-clearcoat} distribution (equation 5, 8, and 12 in \cite{Li2022DisneyBSDF}). The remaining $c_*$ terms follow naturally from the factorization. We first approximate each $D_*$ and $c_*$ term independently using \din; the primary aim here is to understand the functional space required to approximate the terms accurately. Once we have the appropriate functional space, we find similar lookups and merge them into fewer indirection pairs. Figure \ref{fig:arch}(a) illustrates the resulting network architecture. Our network also leverages the fact that the artist control parameters can be encoded into a latent vector (separately and completely offline). As such, the primary arrays are also encoders while the cascaded arrays are runtime decoders. We discuss the efficiency advantages of our approach in section \ref{sec:impDisney}.

\subsubsection{Optimized shading pipeline\label{sec:shadingPipelineOpt}}

We improve the quality of our final rendered (figure \ref{fig:teaser}, \ref{fig:horse}) output using an end-to-end optimization of the shading pipeline. Instead of training the texture sampler and the \textit{Disney BRDF} independently, the goal is to make our sampler aware of the learned \textit{BRDF}, thus improving quality. We do so by training the texture sampler with an additional regularization term that compares the final rendered output through the learned \textit{BRDF}. The extra complexity only affects the training pipeline while the networks and inference pipeline essentially stay the same. More details in supplemental section \ref{sec:optPipeline}.

\subsection{Compact radiance fields}

\textit{NeRF} volumetrically represents a scene as a 5D spatio-directional function whose outputs are spatial density and view-dependent emitted radiance. While the original \textit{NeRF}~\cite{NeRF20} uses a deep neural network to represent the density and radiance, subsequent versions obtain better quality, and improved training using coordinate networks based on 3D data-structures. They combine regular grid~\cite{ReluField22} or a tree~\cite{plenoxels21} with fixed function non-linearity such as \textit{SH} or \textit{ReLU}. Recent state of the art \textit{Direct Voxel}~\cite{dvgo22} achieves high quality representation by adaptive scaling of the voxel grid resolution and fine tuning the representation at each update. However, the resulting voxel grid is enormous, requires hundreds of \textit{megabytes} in parameter space. We follow up on their work and improve the compression of their voxel representation by an order of magnitude without losing significant details (figure \ref{fig:legoComp}).

\paragraph{Implementation details} We extract the pre-trained density and view-dependent radiance/RGB grid from the \textit{Direct Voxel} technique and apply \din~to compress the volumes. We use a multi-head network, as shown in figure \ref{fig:arch}(c) for the density and RGB fields. The networks use a spatial 3D coordinate as input to the two primary arrays. The output of the cascaded arrays are trained against tri-linearly interpolated values from the corresponding target grids. We provide more details in supplemental section \ref{sec:suppl:radfield}. 

\section{General implementation details}

\begin{figure*}[t!]
\begin{tikzpicture}
    \node[anchor=south west,inner sep=0] at (0,0){
    \includegraphics[width=17.4cm,trim={0cm 0cm 0 0cm},clip]{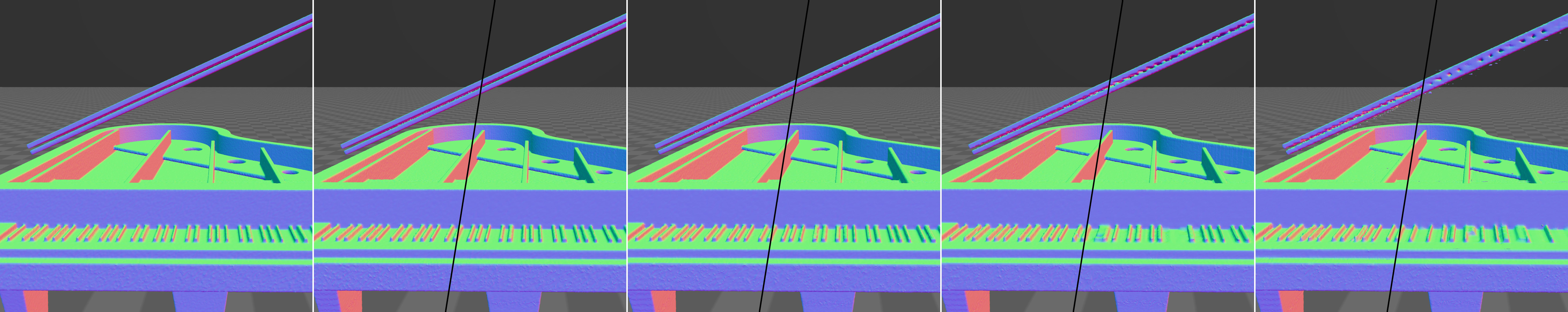}}; ;
    {
    \newcommand{\wOffset}{5}
    \newcommand{\pSkip}{3.5}
    \newcommand{\hUp}{3.4}
    \node[anchor=south west] at (0.0, \hUp) { \textcolor{black}{\small Kd-tree (900M points, 12GB)} };
    \node[anchor=south west] at (\wOffset, \hUp) { \textcolor{black}{\small 96MB} };
    \node[anchor=south west] at (\wOffset + \pSkip, \hUp) { \textcolor{black}{\small 48MB} };
    \node[anchor=south west] at (\wOffset + 2*\pSkip, \hUp) { \textcolor{black}{\small 24MB} };
    \node[anchor=south west] at (\wOffset + 3*\pSkip, \hUp) { \textcolor{black}{\small 12MB} };
    }
    {
    \newcommand{\wOffset}{3.5}
    \newcommand{\wOffsetB}{2.0}
    \newcommand{\pSkip}{3.5}
    \newcommand{\hUp}{3.1}
    \node[anchor=south west] at (0.0, \hUp) { \textcolor{DarkWhite}{\small 0.999/0.000} };
    \node[anchor=south west] at (\wOffset, \hUp) { \textcolor{DarkWhite}{\small 0.988/0.012} };
    \node[anchor=south west] at (\wOffset + \pSkip, \hUp) { \textcolor{DarkWhite}{\small 0.985/0.016} };
    \node[anchor=south west] at (\wOffset + 2*\pSkip, \hUp) { \textcolor{DarkWhite}{\small 0.982/0.018} };
    \node[anchor=south west] at (\wOffset + 3*\pSkip, \hUp) { \textcolor{DarkWhite}{\small 0.977/0.025} };
     \node[anchor=south west] at (\wOffset + \wOffsetB, \hUp) { \textcolor{DarkWhite}{\small 0.985/0.014} };
    \node[anchor=south west] at (\wOffset + \wOffsetB + \pSkip, \hUp) { \textcolor{DarkWhite}{\small 0.982/0.018} };
    \node[anchor=south west] at (\wOffset + \wOffsetB + 2*\pSkip, \hUp) { \textcolor{DarkWhite}{\small 0.977/0.023} };
    \node[anchor=south west] at (\wOffset + \wOffsetB + 3*\pSkip, \hUp) { \textcolor{DarkWhite}{\small 0.972/0.028} };
    }
    \node[anchor=south west] at (3.5, -0.05) { \textcolor{DarkWhite}{\small Ours} };
    \node[anchor=south west] at (5.1, -0.05) { \textcolor{DarkWhite}{\small MRHE} };
    
\end{tikzpicture}
\caption{\label{fig:sdfComp} 
Comparing \din~ (left) with \mrhe~(right) at equal parameter count for \sdf~representation. \mrhe~requires 6 extra memory lookups and $> 512$ extra FLOPs to achieve similar results. Metrics IoU($\uparrow$)/MAE($\downarrow$) are measured on uniform near-surface \sdf~samples. Typical resolution of the primary/cascaded grid at 96MB is $316^3\times3/96^3\times1$. }
\end{figure*}
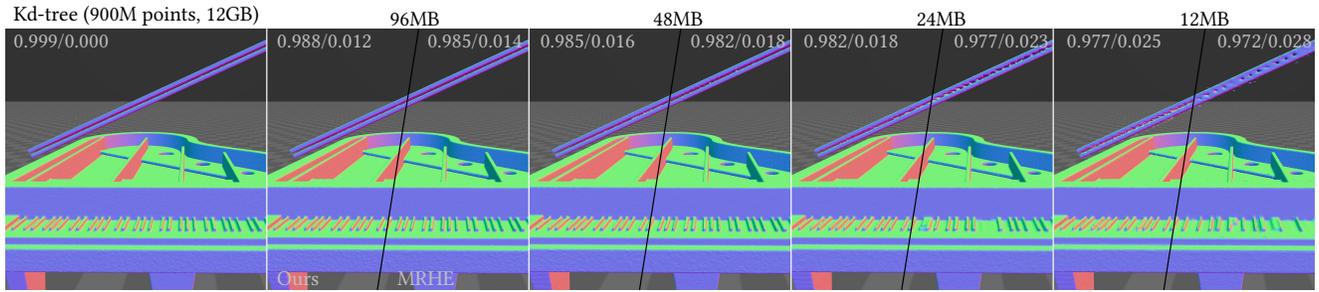

The section provides general implementation details across tasks while more accurate task-specific details are provided in the supplemental. We train our networks using vanilla \textsc{Pytorch} without special optimizations and use 32-bit \textit{full-precision} arithmetic for backpropagation and data generation. We use the \textit{ADAM} optimizer with a learning rate of 0.001 and an \textit{MAE} loss. An important hyperparameter for training is $\rho = N_p/N_c$ -- the ratio of the length of one side of the primary array ($N_p$) to the cascaded array ($N_c$). We set pseudo-optimal values through a hyperparameter search in each setting. In all cases, the optimal $\rho$ is greater than 1. This often results in the cascaded array being much smaller in size (raw bytes) compared to the primary array. In a real-time environment, the access pattern to the primary array is more coherent compared to the cascaded array, but the cascaded array is also much smaller -- a potentially-exploitable property in caching hardware.

Another advantage when $\rho$ is greater than $1$ is in quantization: the cascaded array may contain signed or unbounded values, rendering the effective use of quantization cumbersome and additionally ineffective given the smaller array size; however, for the primary arrays, we quantize values to 8-bits after applying the non-linearity $\mathcal{F}$ to array cells. Quantization either lowers memory size or improves resolution for the primary array. A single level grid, on the other hand, may not benefit from such scheme.

To set the array sizes -- $N_p, N_c$, we specify two inputs: desired total representation size (in bytes) and $\rho$. We also set the length of the sides as the nearest multiple of eight for the primary and a multiple of four for the cascaded. A simple 1-D search suffices to satisfy the constraints whilst also closely matching the desired array sizes. As discussed in section \ref{sec:ovrview}, we initialize a $d$-dim primary array with identity mapping between the input and output.

We implement \textit{Multi Resolution Hash Encoding} in our \textsc{Pytorch}-based framework. When comparing with \mrhe, we use eight grid levels composed of six levels of multi-resolution \textit{dense-grid} arrays and two levels of \textit{hash-grid} arrays. We found quantizing all levels to 8-bit results in better grid resolutions and improved final quality. Each level stores a 2-channel feature vector, totaling 16 latent input channels to the \mlp. The \mlp~is four layers deep and 16 wide. For images, we use bi-linearly interpolated 2D grids, and for \textit{NeRF/SDF} we use tri-linearly interpolated 3D grids. At equal parameter count, \mrhe~always requires $>5$ additional memory accesses and $>1024$ additional \flops~compared to our method. Figure \ref{fig:trainingTime} shows the training characteristics of our technique w.r.t.\ \mrhe.

\section{Results and analysis}
\subsubsection{Isotropic GGX approximation\label{sec:implIsoGGx}} 
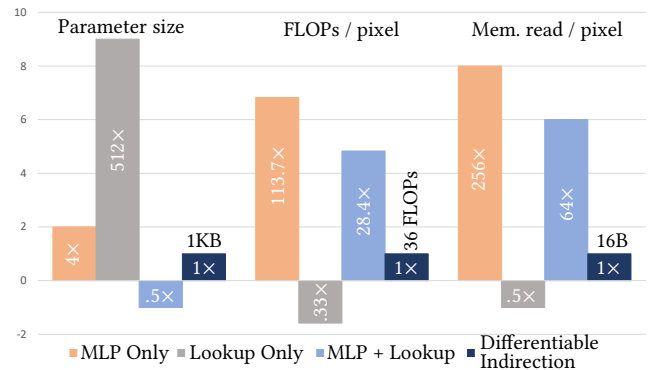
\begin{figure}[b!]
 \begin{tikzpicture}
    \node[anchor=south west,inner sep=0] at (0.0,0.0) {
    \includegraphics[width=11.5cm, trim={0cm 6.8cm 3.8cm 0.1cm},clip]{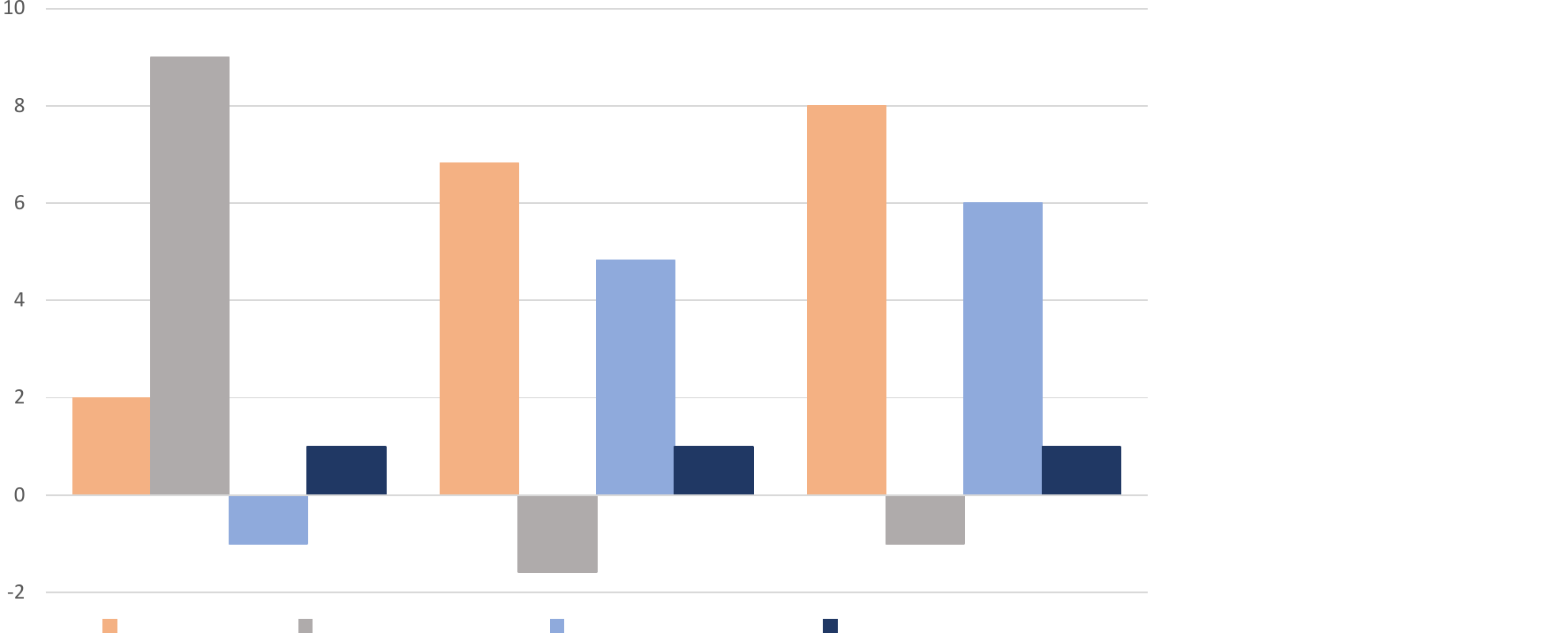}};
    \node[anchor=south west] at (0.5, 4.1+0.1) { \textcolor{black}{\small Parameter size} };
    \node[anchor=south west] at (3.5, 4.0+0.1) { \textcolor{black}{\small FLOPs / pixel} };
    \node[anchor=south west] at (6.0, 4.0+0.1) { \textcolor{black}{\small Mem. read / pixel} };
        
    \node[anchor=south west, rotate=90] at (1.0, 1+0.1) { \textcolor{white}{\small $4\times$} };
    \node[anchor=south west, rotate=90] at (1.60, 2.25+0.1) { \textcolor{white}{\small $512\times$} };
    \node[anchor=south west] at (1.65, 0.525+0.1) { \textcolor{white}{\small $.5\times$} };
    \node[anchor=south west] at (2.275, 0.9+0.1) { \textcolor{white}{\small $1\times$} };
    \node[anchor=south west] at (2.2, 1.25+0.1) { \textcolor{black}{\small 1KB} };

    \node[anchor=south west, rotate=90] at (3.73, 1.8+0.1) { \textcolor{white}{\small $113.7\times$} };
    \node[anchor=south west, rotate=90] at (4.3, 0.25+0.1) { \textcolor{white}{\small $.33\times$} };
    \node[anchor=south west, rotate=90] at (4.87, 1.4+0.1) { \textcolor{white}{\small $28.4\times$} };
    \node[anchor=south west] at (4.96, 0.9+0.1) { \textcolor{white}{\small $1\times$} };
    \node[anchor=south west, rotate=90] at (5.5, 1.2+0.1) { \textcolor{black}{\small 36 FLOPs} };

    \node[anchor=south west, rotate=90] at (6.4, 2.0+0.1) { \textcolor{white}{\small $256\times$} };
    \node[anchor=south west] at (6.45, 0.525+0.1) { \textcolor{white}{\small $.5\times$} };
    \node[anchor=south west, rotate=90] at (4.87+2.7, 1.6+0.1) { \textcolor{white}{\small $64\times$} };
    \node[anchor=south west] at (7.65, 0.9+0.1) { \textcolor{white}{\small $1\times$} };
    \node[anchor=south west, ] at (7.65, 1.25+0.1) { \textcolor{black}{\small 16B} };

    \node[anchor=south west] at (0.8, -0.2) { \textcolor{black}{\small MLP Only} };
    \node[anchor=south west] at (2.25, -0.2) { \textcolor{black}{\small Lookup Only} };
    \node[anchor=south west] at (4.075, -0.2) { \textcolor{black}{\small MLP + Lookup} };
    \node[anchor=south west] at (6.115, -0.15+0.14) { \textcolor{black}{\small Differentiable} };
    \node[anchor=south west] at (6.115, -0.15-0.1) { \textcolor{black}{\small Indirection} };
\end{tikzpicture}
\caption{\label{fig:isoGgxResourcePlot} Comparing relative resource utilization of various primitives on a log scale at equal quality (PSNR~40dB) approximation of \textit{isotropic GGX}.}
\end{figure}

We begin with a discussion of \textit{isotropic GGX} approximation as it serves as a simple testbed to compare various neural primitives and analyze their efficiency. Figure \ref{fig:isoGgxResourcePlot} compares various neural primitives at equal PSNR (> 40dB) measured across a range of roughness as shown in \ref{fig:isoGgxLookup}. More importantly, figure \ref{fig:isoGgxResourcePlot} provides an overview of the neural primitives landscape. Notice the differences in resource utilization of the primitives. An \mlp~with 4 hidden layers and 32 units per layer not only requires thousands of \flops~but also a large memory transfer to fetch the network weights. A single level grid storing the function as texture requires large parameter space but very few \flops~and memory transfer per pixel. A combination of \mlp~and memory-grid strikes a better balance across the three criterion, but our technique improves further. We only require a modest parameter space and few memory transfers, and \flops~per pixel. \flops~required in our technique are for linear interpolation of array cells and often maps directly to hardware texture samplers on GPUs. 

\begin{figure*}[t!]
\begin{tikzpicture}
     \node[anchor=south west,inner sep=0] at (0,0){
    \includegraphics[width=17.4cm,trim={0cm 0cm 0 0cm},clip]{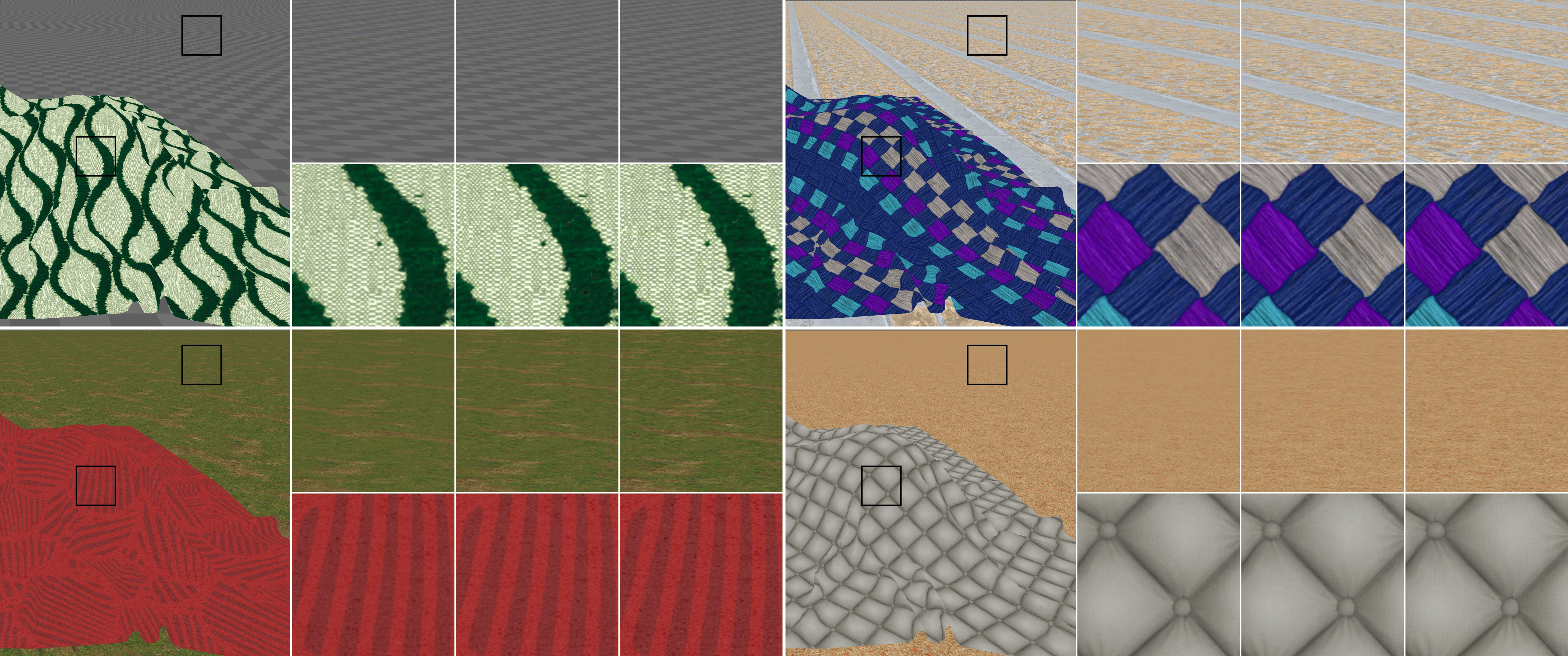}}; ;
     \node[anchor=south west] at (0.4, 7.2) { \textcolor{black}{\small Compression - 6$\times$} };
     \node[anchor=south west] at (3.8, 7.25) { \textcolor{black}{\small 12$\times$} };
     \node[anchor=south west] at (5.7, 7.25) { \textcolor{black}{\small 6$\times$} };
     \node[anchor=south west] at (7.4, 7.25) { \textcolor{black}{\small Ref} };
      \node[anchor=south west] at (1.4 + 8.7, 7.25) { \textcolor{black}{\small 6$\times$} };
     \node[anchor=south west] at (3.8 + 8.7, 7.25) { \textcolor{black}{\small 12$\times$} };
     \node[anchor=south west] at (5.7 + 8.7, 7.25) { \textcolor{black}{\small 6$\times$} };
     \node[anchor=south west] at (7.4 + 8.7, 7.25) { \textcolor{black}{\small Ref} };

     \node[anchor=south west] at (0.0, 6.8) { \textcolor{white}{\footnotesize 32.2/92.9} };
     \node[anchor=south west] at (8.7, 6.8) { \textcolor{black}{\footnotesize 33.8/93.7} };
     \node[anchor=south west] at (0.0, 3.2) { \textcolor{white}{\footnotesize 35.8/94.9} };
     \node[anchor=south west] at (8.7, 3.2) { \textcolor{white}{\footnotesize 34.4/94.1} };

     \node[anchor=south west] at (3.6+1.85, 6.8) { \textcolor{white}{\footnotesize 46.3/97.8} };
     \node[anchor=south west] at (12.3+1.85, 6.8) { \textcolor{black}{\footnotesize 32.8/93.7} };
     \node[anchor=south west] at (3.6+1.85, 3.2) { \textcolor{white}{\footnotesize 32.6/92.2} };
     \node[anchor=south west] at (12.3+1.85, 3.2) { \textcolor{white}{\footnotesize 32.8/93.4} };

     \node[anchor=south west] at (3.2+1.85, 4.15) { \textcolor{black}{\footnotesize 30.4/90.5} };
     \node[anchor=south west] at (12.3+1.85, 5.0) { \textcolor{white}{\footnotesize 32.9/93.9} };
     \node[anchor=south west] at (3.6+1.85, 1.4) { \textcolor{white}{\footnotesize 39.8/96.6} };
     \node[anchor=south west] at (12.3+1.85, 1.4) { \textcolor{white}{\footnotesize 39.4/96.8} };
     
     \node[anchor=south west] at (3.6, 6.8) { \textcolor{white}{\footnotesize 45.5/97.1} };
     \node[anchor=south west] at (12.3, 6.8) { \textcolor{black}{\footnotesize 32.4/93.2} };
     \node[anchor=south west] at (3.6, 3.2) { \textcolor{white}{\footnotesize 32.6/92.1} };
     \node[anchor=south west] at (12.3, 3.2) { \textcolor{white}{\footnotesize 31.2/91.0} };

     \node[anchor=south west] at (3.2, 4.15) { \textcolor{black}{\footnotesize 29.2/89.4} };
     \node[anchor=south west] at (12.3, 5.0) { \textcolor{white}{\footnotesize 31.5/91.6} };
     \node[anchor=south west] at (3.6, 1.4) { \textcolor{white}{\footnotesize 38.0/95.5} };
     \node[anchor=south west] at (12.3, 1.4) { \textcolor{white}{\footnotesize 38.5/95.3} };
     
\end{tikzpicture}
\caption{\label{fig:lodCompress} Compressed texture sampling (and filtering) using \din. Top and bottom cutouts show the sampler output for large and small footprints respectively. PSNR($\uparrow$)/FLIP($\uparrow$) w.r.t.\  uncompressed 16spp anisotropic filtered 1K base textures.}
\end{figure*}
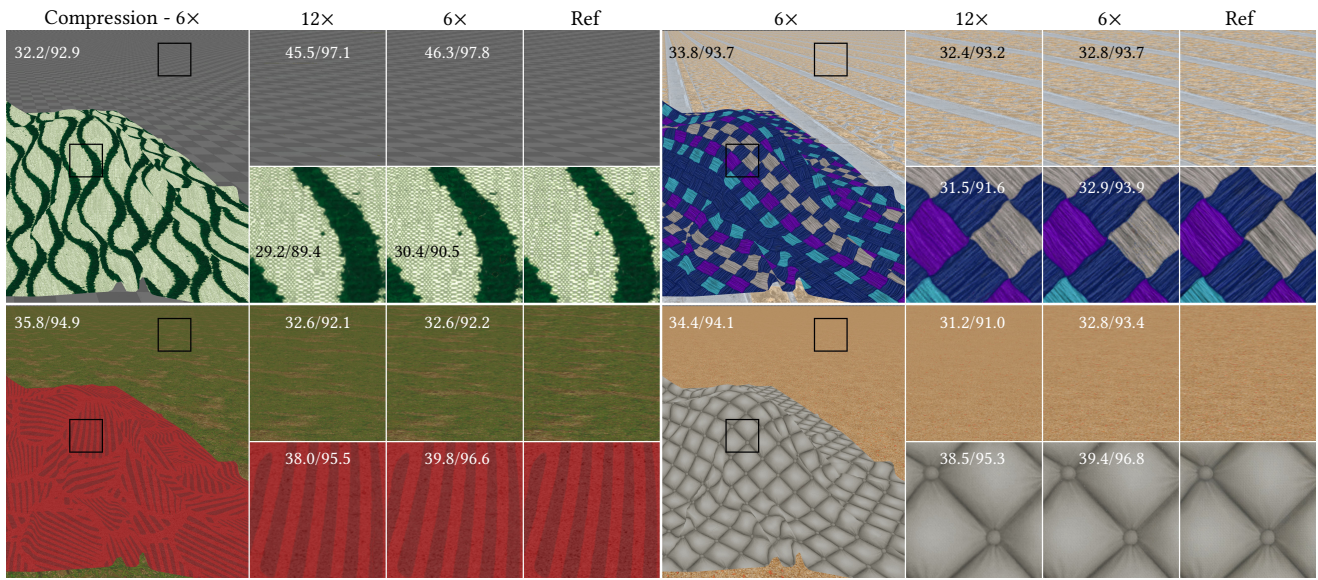

\subsubsection{Signed Distance Fields\label{sec:implSdf}} 
 
Figure \ref{fig:sdfComp} and \ref{fig:psnrPlots}(c) provides a visual and a quantitative analysis of our technique for various parameter sizes. Our technique compares favourably with \mrhe~ at equal parameter count but requires fewer lookups and compute per query.

\subsubsection{Neural Radiance Fields} Figure \ref{fig:legoComp} shows an increasing overall compression due to the underlying compressed density and RGB grids. The RGB grid is more compressible than the density grid and we refer to plot \ref{fig:psnrPlots}(b) to choose a combination of density and RGB compression that retains maximum reconstruction quality. While the density grid is compressible upto 10$\times$, the RGB grid is compressible upto 100$\times$. Both reference grids have the same resolution but the RGB grid has higher number of latent channels. Thus, there are more redundancies to be exploited in the RGB grid which may explain the differences in their compression.

 \subsubsection{Compact Image Representation\label{sec:implTexComp}} Our texture compression technique is resource efficient while also being competitive w.r.t.\ state of the art \textit{block compression} \cite{blockComp79} techniques such as \textit{ASTC} (figure \ref{fig:psnrPlots}). We require few bytes ($\leq12$) of memory read per pixel and a few linear interpolations ($\leq20$) to decode a texture. Other neural techniques~\cite{ntc2023} may use per material (or a group of materials) \mlp~decoder. This necessitates thousands of additional \flops~and \textit{kilobytes} of memory transfer to fetch the \mlp~weights per pixel. This is potentially problematic when several materials are present on screen, especially on low end hardware. \din~has a constant resource utilization independent of the number of materials present on screen.
 
 We compare our technique with \textit{ASTC}~\cite{citeAstc}, \textit{ETC2}~\cite{etc2ref} and \textit{MRHE} as shown in plot \ref{fig:psnrPlots}(a). \textit{ETC2} has a fixed compression of 6$\times$ while \textit{ASTC} has a variable compression up to 24$\times$. Ours and \textit{MRHE}, being learned techniques, achieve unbounded variable compression. In figure \ref{fig:psnrPlots}(a), we downsample the image for compression beyond their respective maximum for \textit{ETC2} and \textit{ASTC}. Our technique is generic, yet comparable in quality with specialized \textit{ASTC}.

  \subsubsection{Texture sampling} Figure 10 shows the output of our texture sampler at 6$\times$, and 12$\times$ compression. Figure \ref{fig:astcCompLod}, compares a single evaluation of our network (1-spp) with nearest neighbor and anisotropic sampled \textit{ASTC} for Albedo, Normal, and AO textures. Note the cost of evaluation of our network is comparable to nearest neighbor \textit{ASTC} while retaining quality much superior to 1-spp \textit{ASTC}. Thus by amortizing texture compression and filtering in one network, we can extract higher quality per sample than we could with isolated compression and sampling.

\begin{table}[b!]
\caption{\label{tab:disneyRuntimePerf} HLSL runtime performance of approximate and reference Disney BRDF at 4K resolution on Mobile 3070Ti.}
\begin{tabular}{|c|cccc|}
\hline
\multirow{2}{*}{Technique}                                                   & \multicolumn{4}{c|}{\begin{tabular}[c]{@{}c@{}}Point emitter count\\ (time in ms)\end{tabular}} \\ \cline{2-5} 
                                                                             & \multicolumn{1}{c|}{1}      & \multicolumn{1}{c|}{2}      & \multicolumn{1}{c|}{3}     & 4     \\ \hline
Differentiable Indirection                                                           & \multicolumn{1}{c|}{0.654}  & \multicolumn{1}{c|}{0.710}  & \multicolumn{1}{c|}{0.798} & 0.923 \\ \hline
Reference Disney                                                             & \multicolumn{1}{c|}{0.690}  & \multicolumn{1}{c|}{0.838}  & \multicolumn{1}{c|}{1.11}  & 1.41  \\ \hline
\begin{tabular}[c]{@{}c@{}}Baseline Diffuse\\ (shader overhead)\end{tabular} & \multicolumn{1}{c|}{0.642}  & \multicolumn{1}{c|}{0.648}  & \multicolumn{1}{c|}{0.656} & 0.668 \\ \hline
\end{tabular}
\end{table}

\subsubsection{Disney BRDF approximation\label{sec:impDisney}} Here, the primary and cascaded arrays are split into an offline-encoder and a runtime-decoder (figure \ref{fig:arch}(a)). The encoder transforms the artist control parameters or \textit{PBR} textures into latent encoded textures of the same resolution. Similar to \textit{PBR} textures, these encoded textures are uv-sampled at runtime and used as an input to the decoder along with other geometric dot products. This lends the opportunity to make the encoder arrays much larger - 2k$\times$2k in resolution, as they are completely offline. The cascaded arrays (decoder) are decidedly small - 16$\times$16 in resolution, so they fit in lowest tier caches/\textit{SRAM} and accessed with minimal latency at runtime. 

\begin{figure*}[t!]
\begin{tikzpicture}
    \node[anchor=south west,inner sep=0] at (0,0){
    \includegraphics[width=17.1cm,trim={0cm 0cm 0 0cm},clip]{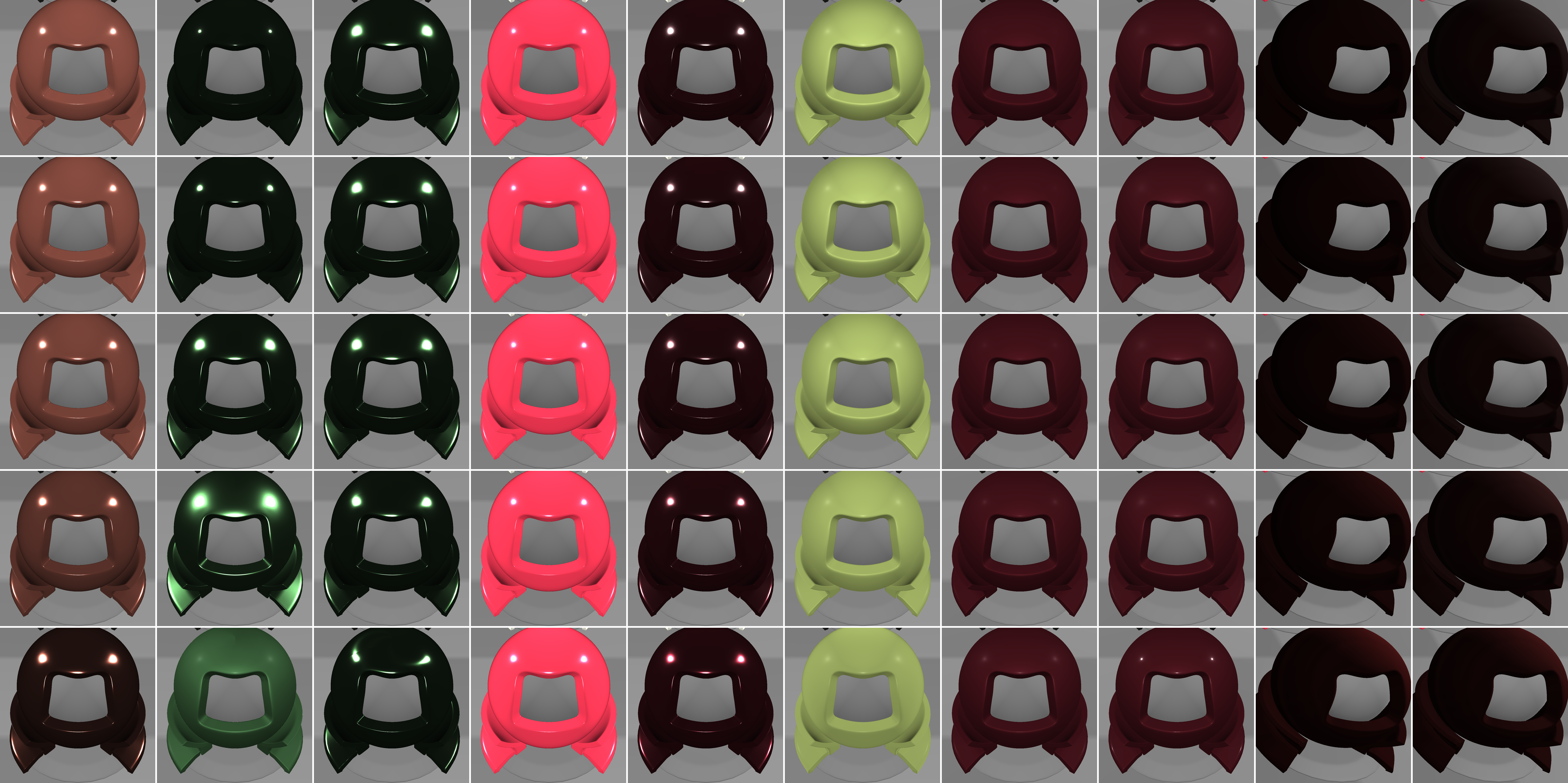}};
    \node[anchor=south west, rotate=90] at (0.0, 7.4) { \textcolor{black}{\small 0.05} };
    \node[anchor=south west, rotate=90] at (0.0, 5.7) { \textcolor{black}{\small 0.1} };
    \node[anchor=south west, rotate=90] at (0.0, 4.0) { \textcolor{black}{\small 0.2} };
    \node[anchor=south west, rotate=90] at (0.0, 2.33) { \textcolor{black}{\small 0.4} };
    \node[anchor=south west, rotate=90] at (0.0, 0.63) { \textcolor{black}{\small 0.8} };
    \node[anchor=south west] at (0.3, 8.65) { \textcolor{black}{\small Metallic} };
    \node[anchor=south west] at (1.8, 8.58) { \textcolor{black}{\small Roughness} };
    \node[anchor=south west] at (3.45, 8.58) { \textcolor{black}{\small Anisotropic} };
    \node[anchor=south west] at (5.4, 8.58) { \textcolor{black}{\small Specular} };
    \node[anchor=south west] at (7.1, 8.7) { \textcolor{black}{\small Specular-} };
    \node[anchor=south west] at (7.4, 8.5) { \textcolor{black}{\small tint} };
    \node[anchor=south west] at (8.65, 8.6) { \textcolor{black}{\small Subsurface} };
    \node[anchor=south west] at (10.5, 8.7) { \textcolor{black}{\small Clearcoat-} };
    \node[anchor=south west] at (10.6, 8.43) { \textcolor{black}{\small strength} };
    \node[anchor=south west] at (12.1, 8.7) { \textcolor{black}{\small Clearcoat-} };
    \node[anchor=south west] at (12.4, 8.43) { \textcolor{black}{\small gloss} };
    \node[anchor=south west] at (14.1, 8.6) { \textcolor{black}{\small Sheen} };
    \node[anchor=south west] at (15.8, 8.7) { \textcolor{black}{\small Sheen-} };
    \node[anchor=south west] at (15.9, 8.5) { \textcolor{black}{\small tint} };
    \node[anchor=south west] at (0.4, 3.37) { \textcolor{black}{\small 54.70} }; 
    \node[anchor=south west] at (0.4, -0.05) { \textcolor{black}{\small 52.63} }; 
    \node[anchor=south west] at (2.2, 6.8) { \textcolor{black}{\small 59.45} }; 
    \node[anchor=south west] at (2.2, -0.05) { \textcolor{black}{\small 38.27} }; 
    \node[anchor=south west] at (3.9, 1.67) { \textcolor{black}{\small 51.71} }; 
    \node[anchor=south west] at (3.9, -0.05) { \textcolor{black}{\small 44.30} }; 
    \node[anchor=south west] at (5.6, 6.8) { \textcolor{black}{\small 54.88} }; 
    \node[anchor=south west] at (5.6, -0.05) { \textcolor{black}{\small 49.23} }; 
    \node[anchor=south west] at (7.3, 6.8) { \textcolor{black}{\small 48.05} }; 
    \node[anchor=south west] at (7.3, -0.05) { \textcolor{black}{\small 47.95} }; 
    \node[anchor=south west] at (9, 3.37) { \textcolor{black}{\small 42.49} }; 
    \node[anchor=south west] at (9, -0.05) { \textcolor{black}{\small 37.78} }; 
    \node[anchor=south west] at (10.75, 6.8) { \textcolor{black}{\small 44.38} }; 
    \node[anchor=south west] at (10.75, -0.05) { \textcolor{black}{\small 44.27} }; 
    \node[anchor=south west] at (12.45, 1.67) { \textcolor{black}{\small 44.23} }; 
    \node[anchor=south west] at (12.45, -0.05) { \textcolor{black}{\small 40.12} }; 
    \node[anchor=south west] at (14.2, 5.1) { \textcolor{black}{\small 48.87} }; 
    \node[anchor=south west] at (14.2, -0.05) { \textcolor{black}{\small 44.81} }; 
    \node[anchor=south west] at (15.9, 6.8) { \textcolor{black}{\small 46.69} }; 
    \node[anchor=south west] at (15.9, -0.05) { \textcolor{black}{\small 45.21} }; 
\end{tikzpicture}
\caption{\label{fig:disneyParamSweep} Figure demonstrating the retention of all artist-control parameters using our efficient \textit{Disney BRDF} approximation using \din. The minimum and maximum PSNR when compared with analytic evaluation is provided for each row of varying parameter value along the vertical axis.}
\end{figure*}

\begin{figure*}[b!]
\begin{tikzpicture}
     \node[anchor=south west,inner sep=0] at (0,0){
    \includegraphics[width=17.4cm,trim={0cm 0cm 0 0cm},clip]{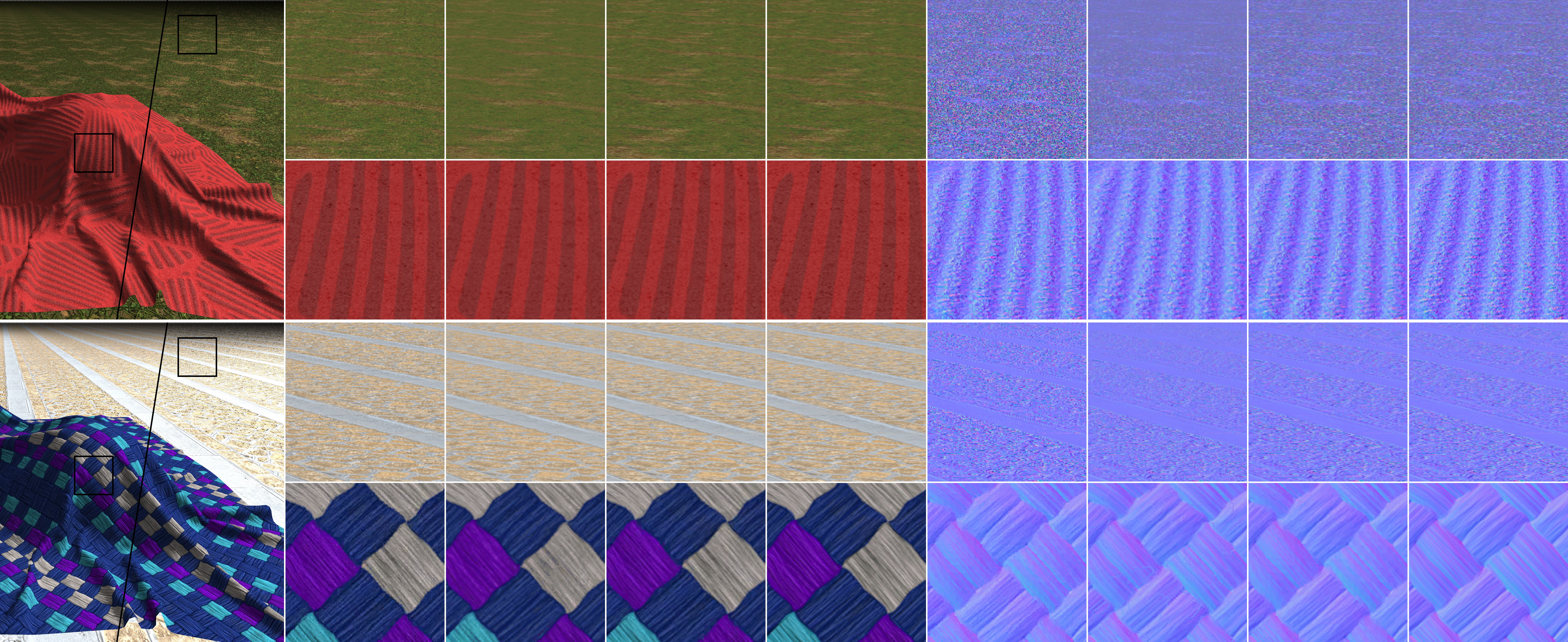}}; ;
     \node[anchor=south west] at (0.4, 7.35) { \textcolor{black}{\small Ours-} };
     \node[anchor=south west] at (0.45, 7.05) { \textcolor{black}{\small 1spp} };
     \node[anchor=south west] at (1.9, 7.35) { \textcolor{black}{\small ASTC-} };
     \node[anchor=south west] at (1.95, 7.05) { \textcolor{black}{\small 1spp} };
     \node[anchor=south west] at (3.6, 7.35) { \textcolor{black}{\small ASTC-} };
     \node[anchor=south west] at (3.65, 7.05) { \textcolor{black}{\small 1spp} };
     \node[anchor=south west] at (5.4, 7.35) { \textcolor{black}{\small Ours-} };
     \node[anchor=south west] at (5.45, 7.05) { \textcolor{black}{\small 1spp} };
     \node[anchor=south west] at (7.1, 7.35) { \textcolor{black}{\small ASTC-} };
     \node[anchor=south west] at (7.1, 7.05) { \textcolor{black}{\small 16spp} };
     \node[anchor=south west] at (8.7, 7.35) { \textcolor{black}{\small Uncomp-} };
     \node[anchor=south west] at (8.85, 7.05) { \textcolor{black}{\small 16spp} };
     \node[anchor=south west] at (3.6+7.2, 7.35) { \textcolor{black}{\small ASTC-} };
     \node[anchor=south west] at (3.65+7.2, 7.05) { \textcolor{black}{\small 1spp} };
     \node[anchor=south west] at (5.4+7.2, 7.35) { \textcolor{black}{\small Ours-} };
     \node[anchor=south west] at (5.45+7.2, 7.05) { \textcolor{black}{\small 1spp} };
     \node[anchor=south west] at (7.1+7.2, 7.35) { \textcolor{black}{\small ASTC-} };
     \node[anchor=south west] at (7.1+7.2, 7.05) { \textcolor{black}{\small 16spp} };
     \node[anchor=south west] at (8.7+7.2, 7.35) { \textcolor{black}{\small Uncomp-} };
     \node[anchor=south west] at (8.85+7.2, 7.05) { \textcolor{black}{\small 16spp} };
     {
     \newcommand{\wOffset}{0.0}
    \newcommand{\pSkip}{2.10}
    \newcommand{\hUp}{6.1}
    
     \node[anchor=south west] at (\wOffset + 0*\pSkip, \hUp) {\textcolor{white}{\footnotesize 26.8/87.3} };
    \node[anchor=south west] at (\wOffset + 1*\pSkip, \hUp) {\textcolor{white}{\footnotesize 24.5/85.9} };
     }
     {
     \newcommand{\wOffset}{0.0}
    \newcommand{\pSkip}{2.1}
    \newcommand{\hUp}{2.5}
    
     \node[anchor=south west] at (\wOffset + 0*\pSkip, \hUp) {\textcolor{black}{\footnotesize 27.8/89.8} };
    \node[anchor=south west] at (\wOffset + 1*\pSkip, \hUp) {\textcolor{black}{\footnotesize 24.4/88.3} };
     }
     
    {
    \newcommand{\wOffset}{3.5}
    \newcommand{\pSkip}{1.8}
    \newcommand{\hUp}{6.7}
    \newcommand{\hDn}{4.95}
    \node[anchor=south west] at (\wOffset + 0*\pSkip, \hUp) {\textcolor{white}{\footnotesize 29.1/91.9} };
    \node[anchor=south west] at (\wOffset + 0*\pSkip, \hDn) {\textcolor{white}{\footnotesize 33.9/94.2} };
    \node[anchor=south west] at (\wOffset + 1*\pSkip, \hUp) {\textcolor{white}{\footnotesize 32.6/92.1} };
    \node[anchor=south west] at (\wOffset + 1*\pSkip, \hDn) {\textcolor{white}{\footnotesize 38.0/95.5} };
    \node[anchor=south west] at (\wOffset + 2*\pSkip, \hUp) {\textcolor{white}{\footnotesize 36.9/95.4} };
    \node[anchor=south west] at (\wOffset + 2*\pSkip, \hDn) {\textcolor{white}{\footnotesize 37.4/95.3} };
    \node[anchor=south west] at (\wOffset + 4*\pSkip, \hUp) {\textcolor{white}{\footnotesize 18.9/79.7} };
    \node[anchor=south west] at (\wOffset + 4*\pSkip, \hDn) {\textcolor{white}{\footnotesize 24.1/83.7} };
    \node[anchor=south west] at (\wOffset + 5*\pSkip, \hUp) {\textcolor{white}{\footnotesize 23.9/85.4} };
    \node[anchor=south west] at (\wOffset + 5*\pSkip, \hDn) {\textcolor{white}{\footnotesize 25.8/85.6} };
    \node[anchor=south west] at (\wOffset + 6*\pSkip, \hUp) {\textcolor{white}{\footnotesize 27.0/90.0} };
    \node[anchor=south west] at (\wOffset + 6*\pSkip, \hDn) {\textcolor{white}{\footnotesize 26.7/86.0} };
    }
    {
    \newcommand{\wOffset}{3.5}
    \newcommand{\pSkip}{1.8}
    \newcommand{\hUp}{3.15}
    \newcommand{\hDn}{1.4}
    \node[anchor=south west] at (\wOffset + 0*\pSkip, \hUp) {\textcolor{black}{\footnotesize 29.3/92.9} };
    \node[anchor=south west] at (\wOffset + 0*\pSkip, \hDn) {\textcolor{white}{\footnotesize 29.2/90.9} };
    \node[anchor=south west] at (\wOffset + 1*\pSkip, \hUp) {\textcolor{black}{\footnotesize 32.4/93.2} };
    \node[anchor=south west] at (\wOffset + 1*\pSkip, \hDn) {\textcolor{white}{\footnotesize 31.5/91.6} };
    \node[anchor=south west] at (\wOffset + 2*\pSkip, \hUp) {\textcolor{black}{\footnotesize 36.7/96.1} };
    \node[anchor=south west] at (\wOffset + 2*\pSkip, \hDn) {\textcolor{white}{\footnotesize 34.0/93.5} };
    \node[anchor=south west] at (\wOffset + 4*\pSkip, \hUp) {\textcolor{white}{\footnotesize 25.5/89.3} };
    \node[anchor=south west] at (\wOffset + 4*\pSkip, \hDn) {\textcolor{white}{\footnotesize 33.2/90.5} };
    \node[anchor=south west] at (\wOffset + 5*\pSkip, \hUp) {\textcolor{white}{\footnotesize 29.0/90.3} };
    \node[anchor=south west] at (\wOffset + 5*\pSkip, \hDn) {\textcolor{white}{\footnotesize 34.8/92.7} };
    \node[anchor=south west] at (\wOffset + 6*\pSkip, \hUp) {\textcolor{white}{\footnotesize 33.1/94.3} };
    \node[anchor=south west] at (\wOffset + 6*\pSkip, \hDn) {\textcolor{white}{\footnotesize 35.9/91.6} };
    }
\end{tikzpicture}
\caption{\label{fig:astcCompLod} Figure comparing combined effect of texture compression (12x) and filtering using \textit{differentiable indirection} w.r.t.\ \textit{ASTC} at varying pixel footprints and filtering sample count. Our technique is computationally comparable to \textit{ASTC}-1spp, yet yields better filtered results compared to the same. PSNR($\uparrow$)/FLIP($\uparrow$) w.r.t.\ uncompressed 16spp reference is provided. Uncompressed textures are have a base resolution of 1K.}
\end{figure*}
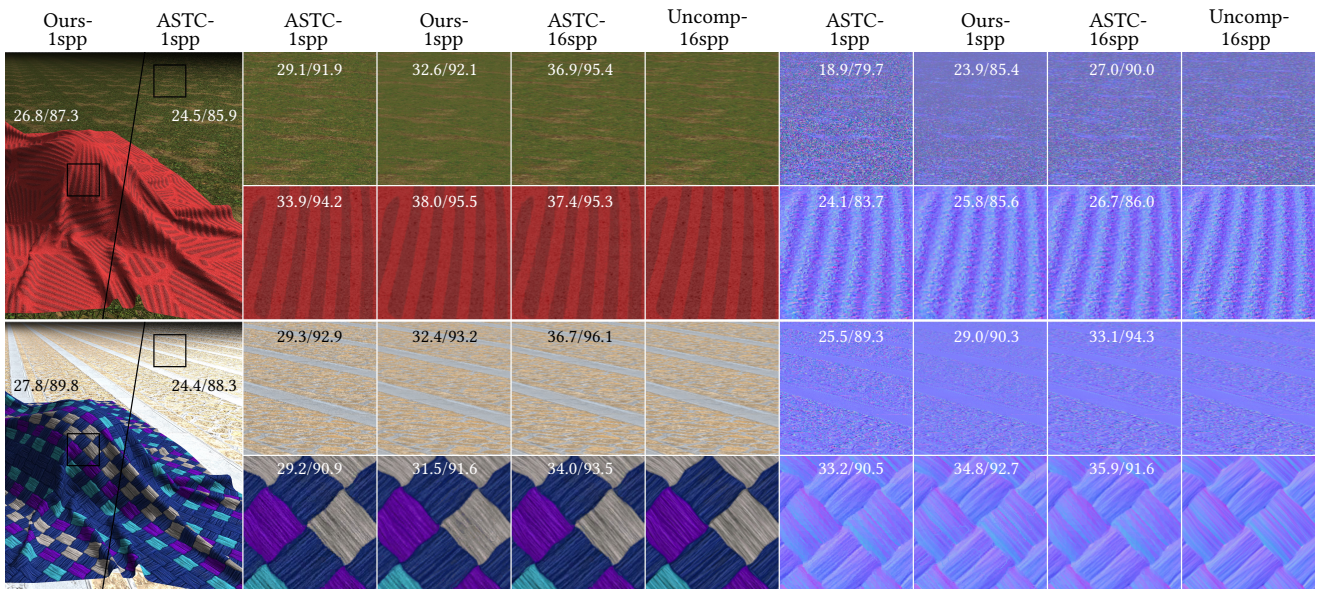

A sweep across all artist parameters is shown in figure \ref{fig:disneyParamSweep}. The runtime part of the network requires four 2D-bilinear lookups and a single 1D lookup. We require an additional 41 \flops~to combine the output of the lookups into final result. The reference analytic implementation requires 240+ \flops~in total. Our runtime arrays are 16x16 resolution and $\leq4$ channels deep, making it easier to test on a commodity GPU. Table \ref{tab:disneyRuntimePerf} shows the runtime performance of our technique running at 4K resolution for up to 4 point light sources. A performance advantage is obtained when using the hardware texture sampler for bilinear interpolations.

\subsubsection{Shading pipeline optimization} 
\begin{figure*}[t!]
\begin{tikzpicture}
    \node[anchor=south west,inner sep=0] at (0,0){\includegraphics[width=\textwidth, trim={0cm 0cm 0cm 0.0cm},clip]{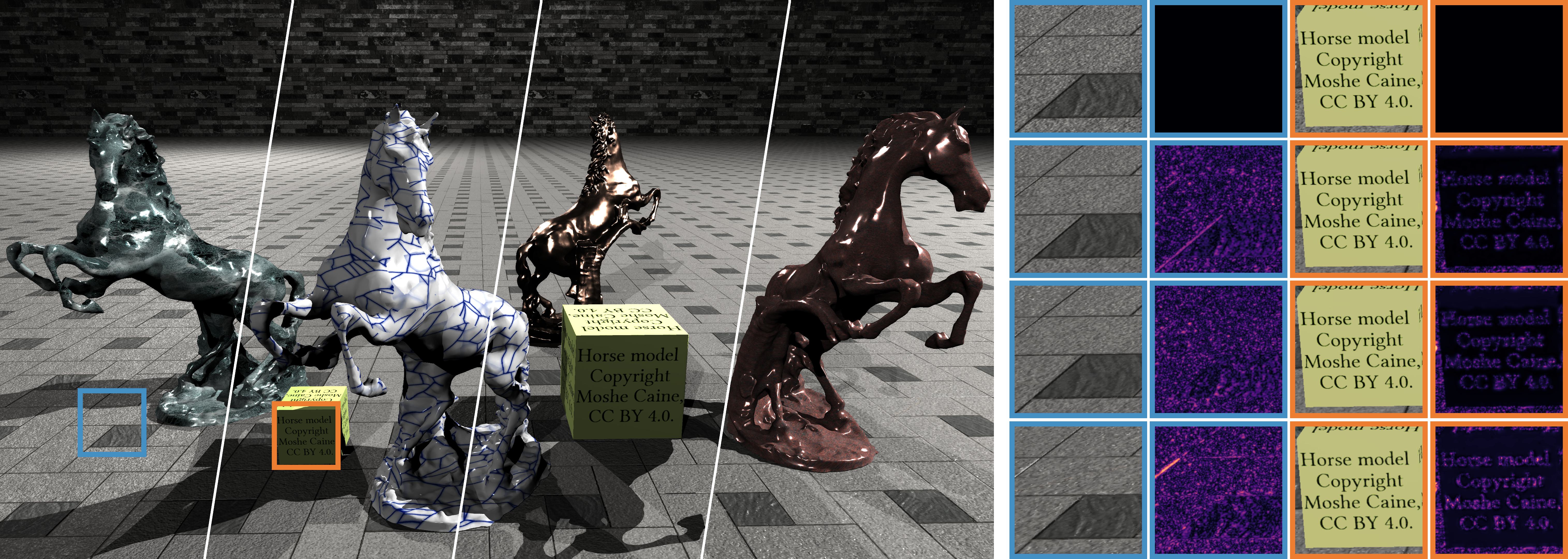}}; 
    \node[anchor=south west] at (0.0,5.9) {\textcolor{white}{\footnotesize Texture: Uncompressed}};
    \node[anchor=south west] at (0.0,5.6) {\textcolor{white}{\footnotesize Spp: 16}};
    \node[anchor=south west] at (0.0,5.3) {\textcolor{white}{\footnotesize Shading: Reference Disney}};
    \node[anchor=south west] at (3.3,5.9) {\textcolor{white}{\footnotesize Texture: Ours 6$\times$}};
    \node[anchor=south west] at (3.3,5.6) {\textcolor{white}{\footnotesize Spp: 1}};
    \node[anchor=south west] at (3.3,5.3) {\textcolor{white}{\footnotesize Shading: Ours Disney}};
    \node[anchor=south west] at (6.2,5.9) {\textcolor{white}{\footnotesize Texture: Ours 12$\times$}};
    \node[anchor=south west] at (9.1,5.9) {\textcolor{white}{\footnotesize Texture: Ours 24$\times$}};
    \node[anchor=south west] at (2.4,0.0) {\textcolor{white}{\footnotesize PSNR($\uparrow$)/Flip($\uparrow$):29.3/92.5}};
    \node[anchor=south west] at (5.8,0.0) {\textcolor{white}{\footnotesize 28.8/92.1}};
    \node[anchor=south west] at (9.2,0.0) {\textcolor{white}{\footnotesize 27.4/91.2}};
     \node[anchor=south west, rotate=90] at (11.55,4.9) {\textcolor{black}{\footnotesize Reference}};
     \node[anchor=south west, rotate=90] at (11.55,3.5) {\textcolor{black}{\footnotesize 6$\times$}};
     \node[anchor=south west, rotate=90] at (11.55,2.0) {\textcolor{black}{\footnotesize 12$\times$}};
     \node[anchor=south west, rotate=90] at (11.55,0.6) {\textcolor{black}{\footnotesize 24$\times$}};
     \node[anchor=south west] at (13.2,4.8) {\textcolor{white}{\footnotesize Flip-Error}};
     \node[anchor=south west] at (16.4,4.8) {\textcolor{white}{\footnotesize Flip-Error}};
     {
    \newcommand{\wLeft}{11.75}
    \newcommand{\wRight}{14.9}
    \newcommand{\hOffset}{3.15}
    \newcommand{\pSkip}{1.6}
    \node[anchor=south west] at (\wLeft, \hOffset - 0*\pSkip) {\textcolor{white}{\footnotesize 28.0/89.5} };
    \node[anchor=south west] at (\wRight, \hOffset - 0*\pSkip) {\textcolor{white}{\footnotesize 31.8/95.4} };
    \node[anchor=south west] at (\wLeft, \hOffset - 1*\pSkip) {\textcolor{white}{\footnotesize 24.4/86.6} };
    \node[anchor=south west] at (\wRight, \hOffset - 1*\pSkip) {\textcolor{white}{\footnotesize 28.7/93.5} };
    \node[anchor=south west] at (\wLeft, \hOffset - 2*\pSkip) {\textcolor{white}{\footnotesize 22.2/83.6} };
    \node[anchor=south west] at (\wRight, \hOffset - 2*\pSkip) {\textcolor{white}{\footnotesize 26.2/90.2} };
    }
\end{tikzpicture}
\caption{\label{fig:horse} Figure shows the use of our technique on end-to-end \textit{PBR} shading. We jointly optimize our compressed latent texture representation with our differentiable \textit{Disney BRDF} for a 5\% improvement in PSNR over independent texuring and shading. PSNR($\uparrow$)/FLIP($\uparrow$) w.r.t.\ uncompressed 16spp reference is provided. Uncompressed textures are have a base resolution of 1K. Horse model \copyright~Moshe Caine~\cite{caine23}, CC-BY-4.0.}
\end{figure*}
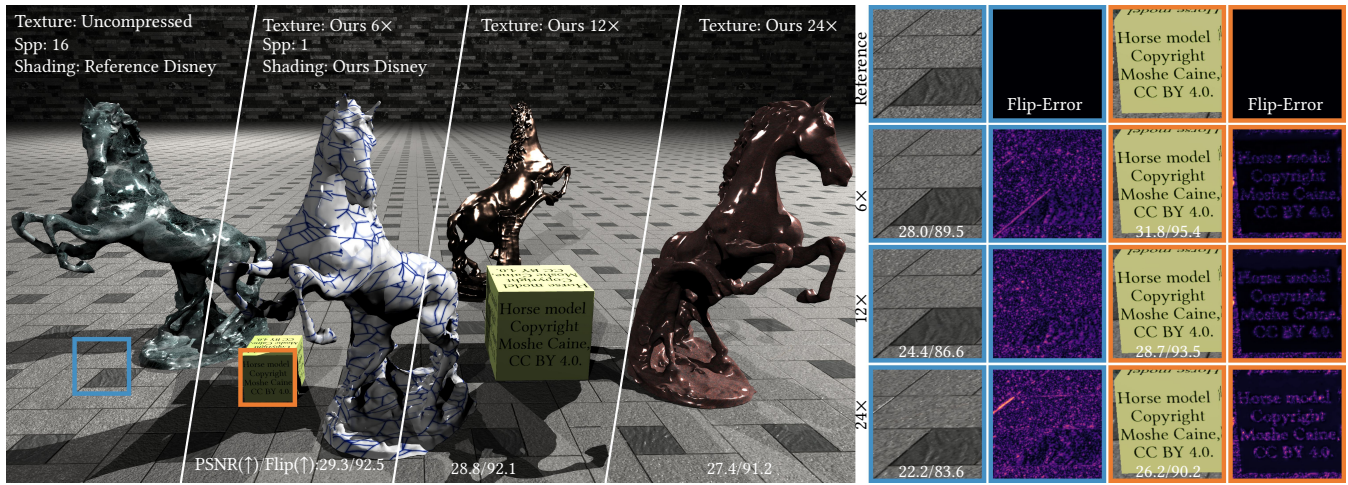

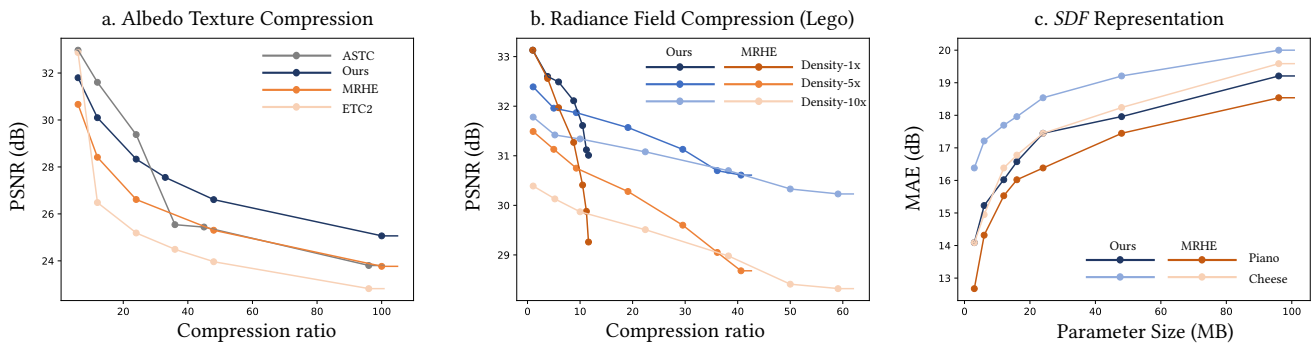
\begin{figure*}[ht!]
\begin{tikzpicture}
    \node[anchor=south west,inner sep=0] at (0,0){
    \includegraphics[width=17.4cm,trim={0cm 11.05cm 1.52cm 0.95cm},clip]{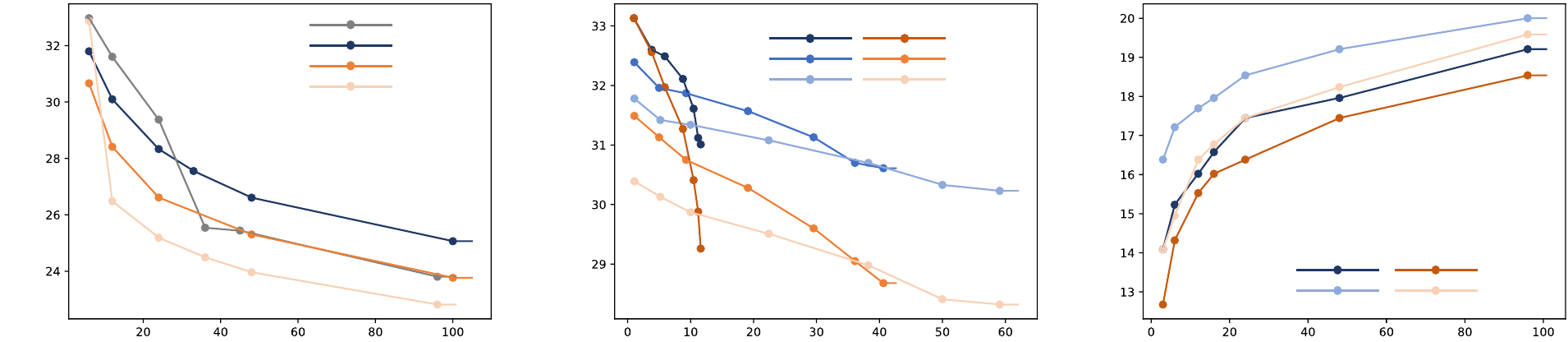}}; ;
    \node[anchor=south west] at (1.2, 3.75) { \textcolor{black}{\small a. Albedo Texture Compression} };
    \node[anchor=south west] at (6.9, 3.75) { \textcolor{black}{\small b. Radiance Field Compression (Lego)} };
    \node[anchor=south west] at (13.6, 3.75) { \textcolor{black}{\small c. \sdf~Representation} };

    \node[anchor=south west] at (1.0 + 1.2, -0.4) { \textcolor{black}{\small Compression ratio} };
    \node[anchor=south west] at (1.0 + 6.9, -0.4) { \textcolor{black}{\small Compression ratio} };
    \node[anchor=south west] at (0.5 + 13.4, -0.4) { \textcolor{black}{\small Parameter Size (MB)} };

    \node[anchor=south west] at (4.4, 3.35) { \textcolor{black}{\tiny ASTC} };
    \node[anchor=south west] at (4.4, 3.15) { \textcolor{black}{\tiny Ours} };
    \node[anchor=south west] at (4.4, 2.92) { \textcolor{black}{\tiny MRHE} };
    \node[anchor=south west] at (4.4, 2.65) { \textcolor{black}{\tiny ETC2} };
    
    \node[anchor=south west] at (8.7, 3.41) { \textcolor{black}{\tiny Ours} };
    \node[anchor=south west] at (9.7, 3.4) { \textcolor{black}{\tiny MRHE} };
    \node[anchor=south west] at (10.5, 3.2) { \textcolor{black}{\tiny Density-1x} };
    \node[anchor=south west] at (10.5, 2.95) { \textcolor{black}{\tiny Density-5x} };
    \node[anchor=south west] at (10.5, 2.7) { \textcolor{black}{\tiny Density-10x} };
    \node[anchor=south west] at (14.6, 0.83) { \textcolor{black}{\tiny Ours} };
    \node[anchor=south west] at (15.55, 0.82) { \textcolor{black}{\tiny MRHE} };
    \node[anchor=south west] at (16.45, 0.66) { \textcolor{black}{\tiny Piano} };
    \node[anchor=south west] at (16.45, 0.38) { \textcolor{black}{\tiny Cheese} };
    
    \node[anchor=south west, rotate=90] at (0.45, 1.35) { \textcolor{black}{\small PSNR (dB)} };
    \node[anchor=south west, rotate=90] at (6.5, 1.35) { \textcolor{black}{\small PSNR (dB)} };
    \node[anchor=south west, rotate=90] at (12.35, 1.35) { \textcolor{black}{\small MAE (dB)} };
\end{tikzpicture}
\caption{\label{fig:psnrPlots} a. Figure compares texture compression behavior for various techniques for 1k textures. \textit{ASTC}, and \textit{ETC2} textures are downsampled beyond their maximum compression ratio of 24$\times$, and 6$\times$ respectively. b. Figure shows various iso-lines corresponding to a fixed density-grid compression with varying RGB-grid compression. Total uncompressed grid size is 826MB. c. Figure shows variation in \textit{MAE} (dB) with parameter size for \sdf~representation. \textit{MAE} is computed only on perturbed near-surface samples.}
\end{figure*}

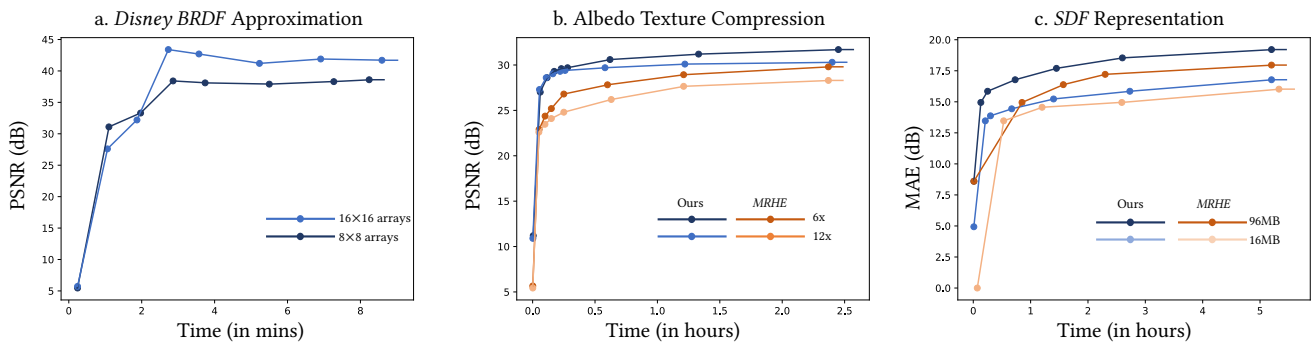
\begin{figure*}[b!]
\begin{tikzpicture}
    \node[anchor=south west,inner sep=0] at (0,0){
    \includegraphics[width=17.4cm,trim={0cm 11.05cm 1.52cm 0.95cm},clip]{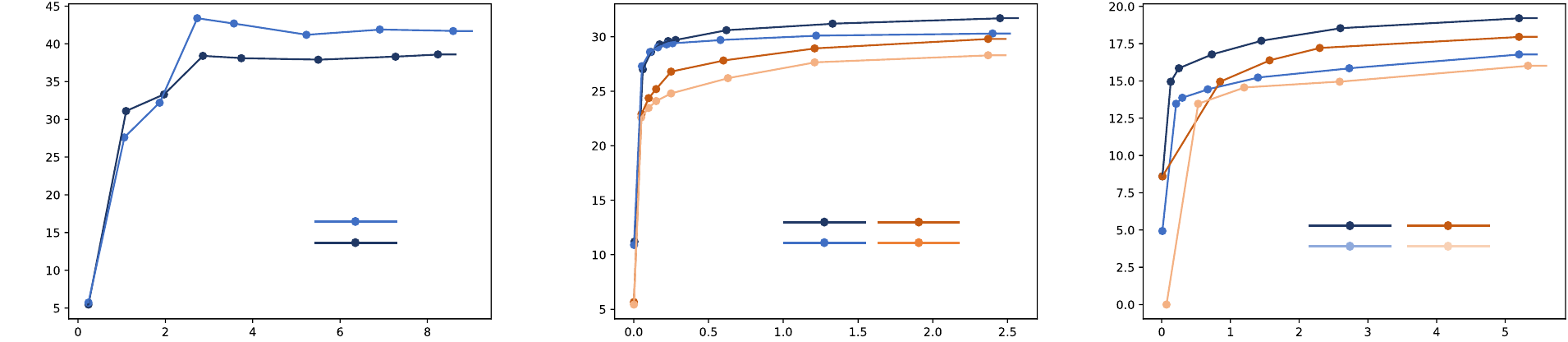}}; ;
    \node[anchor=south west] at (1.1, 3.75) { \textcolor{black}{\small a. \textit{Disney BRDF} Approximation} };
    \node[anchor=south west] at (7.2, 3.75) { \textcolor{black}{\small b. Albedo Texture Compression} };
    \node[anchor=south west] at (13.6, 3.75) { \textcolor{black}{\small c. \sdf~Representation} };

    \node[anchor=south west] at (1.0 + 1.2, -0.4) { \textcolor{black}{\small Time (in mins)} };
    \node[anchor=south west] at (1.0 + 6.9, -0.4) { \textcolor{black}{\small Time (in hours)} };
    \node[anchor=south west] at (0.5 + 13.4, -0.4) { \textcolor{black}{\small Time (in hours)} };

    \node[anchor=south west] at (4.35, 1.15) { \textcolor{black}{\tiny 16$\times$16 arrays} };
    \node[anchor=south west] at (4.35, 0.9) { \textcolor{black}{\tiny 8$\times$8 arrays} };
    
    \node[anchor=south west] at (8.85, 1.39) { \textcolor{black}{\tiny Ours} };
    \node[anchor=south west] at (9.85, 1.39) { \textcolor{black}{\tiny \mrhe} };
    \node[anchor=south west] at (10.65, 1.2) { \textcolor{black}{\tiny 6x} };
    \node[anchor=south west] at (10.65, 0.95) { \textcolor{black}{\tiny 12x} };
    
    \node[anchor=south west] at (14.7, 1.38) { \textcolor{black}{\tiny Ours} };
    \node[anchor=south west] at (15.75, 1.38) { \textcolor{black}{\tiny \mrhe} };
    \node[anchor=south west] at (16.45, 1.15) { \textcolor{black}{\tiny 96MB} };
    \node[anchor=south west] at (16.45, 0.9) { \textcolor{black}{\tiny 16MB} };
    
    \node[anchor=south west, rotate=90] at (0.45, 1.35) { \textcolor{black}{\small PSNR (dB)} };
    \node[anchor=south west, rotate=90] at (6.5, 1.35) { \textcolor{black}{\small PSNR (dB)} };
    \node[anchor=south west, rotate=90] at (12.35, 1.35) { \textcolor{black}{\small MAE (dB)} };
\end{tikzpicture}
\caption{\label{fig:trainingTime} Figure shows the convergence characteristics of \din~and \textit{MRHE} for various tasks in the same framework with similar training overheads.}
\end{figure*}

Figure \ref{fig:teaser} and \ref{fig:horse} shows the final results of our end-to-end optimized shading pipeline. We obtain an additional 8\%, 5\% better PSNR in the first, and second figure respectively using this approach. As shown figure \ref{fig:teaser}, the runtime performance at 4K resolution is < 1.5ms for our learned texture sampling and shading on a Nvidia 3090 GPU. In absence of \textit{ASTC} hardware, we report runtime with \textit{BC7} in figure \ref{fig:teaser}.

\section{Conclusion and future work}
 We show \textit{differentiable indirection} as a powerful primitive that efficiently represents \textit{data} and \textit{compute} across neural graphics pipeline with applications potentially beyond graphics. While our technique is \textit{bandwidth}, \textit{compute}, and \textit{space} efficient, the technique is challenging to apply at higher dimensions. Section \ref{sec:disneyApprox} shows a recipe to overcome this using parameter space factorization and exploitation of the problem structure. More generally, we suspect factorization techniques such as spectral or tensor decomposition may prove useful. The effectiveness our technique is also improved by scaling the grid resolution, feature count, or by augmenting the latent representation with fixed function logic such as \textit{SH, PE} or other parameter-free embeddings. While our technique is aimed at improving runtime efficiency, the effectiveness of our technique in the context of direct reconstruction or inverse-rendering tasks is yet to be explored. Finally, we look forward to interesting arrangements of differentiable arrays as regular layers in neural/array networks. 

 \begin{acks}
We thank Cheng Chang, Sushant Kondguli, Anton Michels, Warren Hunt, and Abhinav Golas for their valuable input and the reviewers for their constructive feedback. We also thank Moshe Caine for the horse-model ~\cite{caine23} with \textit{CC-BY-4.0} license, and \textsc{Adobe Substance-3D} ~\cite{adobe23} for the \textit{PBR} textures. This work was done when Sayantan was an intern at Meta Reality Labs Research. While at McGill University, he was also supported by a Ph.D.\ scholarship from the \textit{Fonds de recherche du Québec -- nature et technologies}.
\end{acks}

\bibliographystyle{ACM-Reference-Format}
\bibliography{main}


\setcounter{section}{0}
\setcounter{equation}{0}
\setcounter{figure}{0}
\setcounter{table}{0}
\clearpage
\noindent \textbf{\Large Efficient Graphics Representation with Differentiable Indirection - Supplemental}

\section{Additional details}
This section provides additional implementation details, results, and visualizations for several task. 

\subsection{Isotropic GGX approximation \label{sec:isoGgx}}
\begin{figure*}[t!]
\begin{tikzpicture}
    \node[anchor=south west,inner sep=0] at (0,0){
    \includegraphics[width=17.4cm,trim={0cm 6.5cm 0cm 0.0cm},clip]{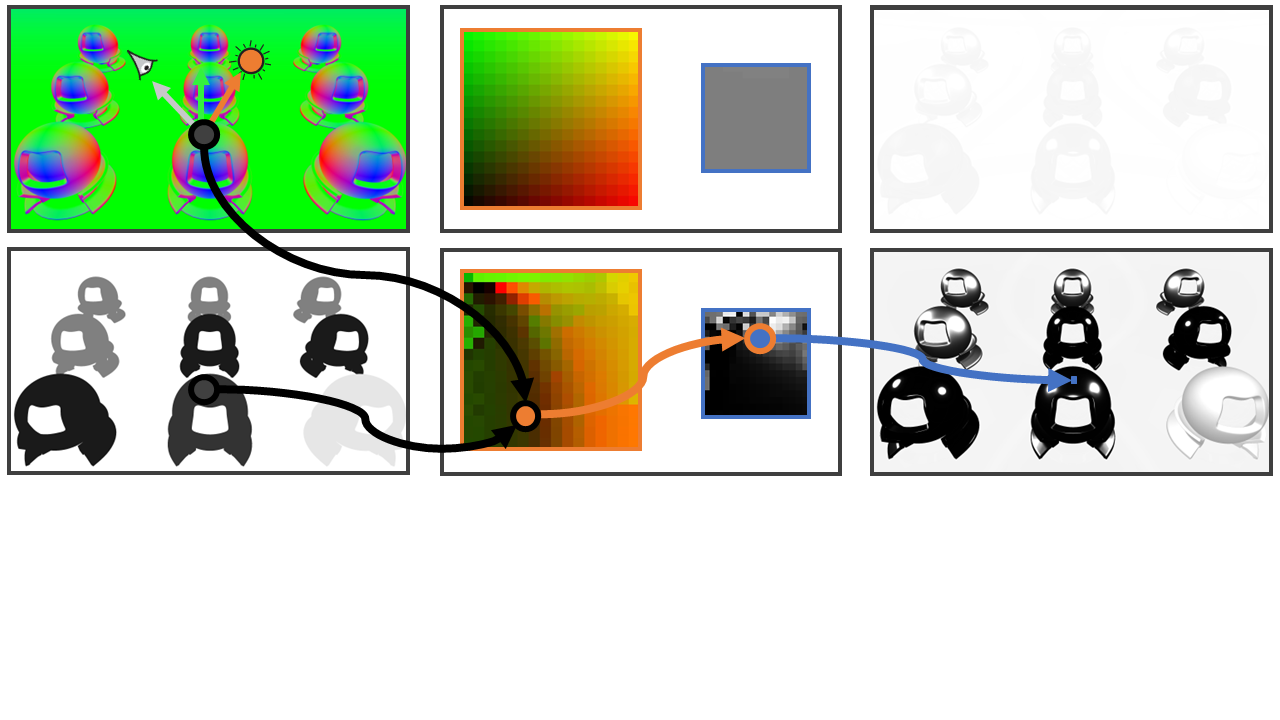}};;
    \node[anchor=south west] at (1.25, 6.3) {\textcolor{black}{\small Input G-Buffer visualization}};
    \node[anchor=south west] at (7.25, 6.3) {\textcolor{black}{\small Isotropic-GGX network}};
    \node[anchor=south west] at (13.5, 6.3) {\textcolor{black}{\small Output pixel color}};
    \node[anchor=south west] at (0.2, 5.95) {\textcolor{black}{\small Surface normals}};
    \node[anchor=south west] at (0.2, 2.6) {\textcolor{black}{\small Roughness ($\alpha_h$)}};
    \node[anchor=south west] at (8, 6) {\textcolor{black}{\small Initial content}};
    \node[anchor=south west] at (8, 2.65) {\textcolor{black}{\small After training}};
    \node[anchor=south west] at (11.95, 5.9) {\textcolor{black}{\small Initial output (PSNR: 9dB)}};
    \node[anchor=south west] at (11.95, 2.65) {\textcolor{black}{\small After training (PSNR: 41dB)}};
     \node[anchor=south west] at (6.9, 4.7) {\textcolor{black}{\small Primary}};
     \node[anchor=south west] at (6.85, 4.4) {\textcolor{black}{\small ($16^2\times2$)}};
     \node[anchor=south west] at (9.6, 4.7) {\textcolor{black}{\small Cascaded}};
     \node[anchor=south west] at (9.7, 4.4) {\textcolor{black}{\small ($8^2\times1$)}};
     \node[anchor=south west] at (5.5, 2.65) {\textcolor{black}{\small $h_z$}};
     \node[anchor=south west] at (5.5, 0.45) {\textcolor{black}{\small $\alpha_h$}};
\end{tikzpicture}
\caption{\label{fig:isoGggxInference} Figure visualizes the \textit{isotropic GGX} inference pipeline. The primary array is queried with the parameters $(\alpha_h, h_z)$ computed from various G-Buffer components as visualized on the left.  Content of primary and cascaded array is visualized in the centre while the network output (pixel color) on a variety glossiness ($\alpha_h$) is shown on the right, before and after training.}
\end{figure*}

\textit{Isotropic GGX} is a popular \textit{BRDF} (\textit{Bidirectional Reflectance Distribution Function}) for modelling glossy reflections in variety of real-time applications. The function itself is very simple, given by:
\begin{equation}
    D(h_z, \alpha_h) = \frac{\alpha_h^4}{\pi \cdot (1 + (\alpha_h^4 - 1) \cdot h_z^2)^2 },
    \label{eq:isoGGX}
\end{equation}
where $\alpha_h \in [0,1)$ controls the glossiness of a surface and $h_z \in [0, 1)$ is the dot-product of the half-vector bisecting the camera and emitter direction with the surface normal. An artist decides the glossiness value $\alpha_h$; the second parameter $h_z$ is obtained from the surface normal, emitter and camera direction at the location of shading, as shown in figure \ref{fig:isoGggxInference}. When rendering, equation \ref{eq:isoGGX} is evaluated independently for each pixel on screen. We refer the readers to online tutorials~\cite{learnOpenGL} for a lightweight introduction to \textit{shading} and other resources~\cite{waltMicro07, pbrtReflection} for a more rigorous description. A crucial aspect to note here is the difference in how $\alpha_h$ and $h_z$ is obtained in a real-time environment. $\alpha_h$ being an artist control parameter, is associated with the material properties of a 3D mesh, as authored by an  artist. Such artist control parameters are often stored as textures -- uv-mapped to the surface of a mesh. On the other hand, $h_z$ is associated with the geometry, and the location of light and camera w.r.t.\ the shade point. As such, $h_z$ is computed on the fly while $\alpha_h$ is fetched from memory. We exploit this distinction in the next section when we approximate the \textit{Disney BRDF}.

\paragraph{Training and inference} For \textit{isotropic GGX} training, we use a cascade of two 2D arrays. The primary array is initialized with an undistorted uv-map while the cascaded array with a constant 0.5. Since the cascaded array stores values that are beyond $[0, 1)$ range, we do not use any non-linearity for the cascaded array. For \textit{isotropic GGX}, we set $\rho = 2$. As described in the main document, $\rho$ is the ratio of the length of one side of the primary array to the cascaded array. Thus, the primary array is 16x16 and the cascaded array is 8x8 resolution. We obtain training samples from appropriate distributions for the two inputs; $h_z$ is sampled from a cosine-hemisphere~\cite{pbrtCosineSampling} distribution and $\alpha_h$ from an exponential-like distribution with more samples biased towards lower roughness values. Output of the network is compared with the reference (equation \ref{eq:isoGGX}) and the loss is backpropagated to train the arrays. The training is performed in a local coordinate frame with normal vector pointing at (0,0,1).
\paragraph{Visualizations and conclusion} Figure \ref{fig:isoGggxInference} visualizes the inference pipeline for \textit{isotropic GGX} case. The pipeline replaces many compute \flops~required to evaluate equation \ref{eq:isoGGX} with two memory lookups. Our approach may produces inherent advantage with increasing resolution as the cost of memory lookups may scale sub-linearly with resolution due to caching, while compute \flops~may scale linearly with resolution. Compared to super-resolution which amortizes cost by exploiting spatial (and/or temporal) locality in screen-space, the case of folding compute \flops~as lookups may be interpreted as amortization through memory.

\subsection{Disney BRDF approximation}

Section \ref{sec:isoGgx} discussed a simplified \textit{BRDF}. Practical \textit{BRDFs}, such as \textit{Disney}, is composed of multiple components which enable artists to manipulate and develop plausible materials by controlling a set of parameters. Our implementation, based on the reference~\cite{Li2022DisneyBSDF} document, employs four components: \textbf{diffuse}, \textbf{metallic}, \textbf{clearcoat}, and \textbf{sheen}. Controlling these components are the 10 artist control parameters, often stored as uv-mapped textures. These parameters are categorically similar to $\alpha_h$ in section \ref{sec:isoGgx}, except we have many more them. \textit{Disney} also requires 7 geometric dot-products, similar to $h_z$ in section \ref{sec:isoGgx}. Thus, \textit{Disney} has total 17 input parameters as opposed to just 2 for \textit{isotropic GGX} in section \ref{sec:isoGgx}.

Our goal is to fold as many compute operations (for evaluating the reference \textit{BRDF}) into memory lookups while retaining quality and also improving runtime performance. This is challenging as we must minimize the number of memory lookups. Each lookup is potentially 7-10$\times$ more expensive compared to a single \textit{MAC/FMA} operation, even assuming the data resides in closest tier cache or \textit{SRAM}. Here we show one of the many possible instantiations of \textit{differentiable indirection} to approximate \textit{Disney BRDF} while adhering to the aforementioned constraints. While the exact recipe is not crucial, it is useful to show some of the practical constraints that we consider in our implementation.

Following from the main paper, we refactor $p$ and $q$ according to 
\begin{equation}
\begin{split}
p(x) &= c_dD_d + c_{m0}D_m + c_{s0}\\
q(x) &= c_{c}D_c + c_{m1}D_m + c_{s1},
\end{split}
\label{eq:disneySplit}
\end{equation}
where $D_d, D_m$, and $D_c$ are the \textit{Disney-diffuse}, \textit{Disney-metallic}, and \textit{Disney-clearcoat} distributions (equation 5, 8, and 12 in reference document~\cite{Li2022DisneyBSDF}). Such factorization minimizes color bleeding by separating out the albedo from final \textit{BRDF} expression. We first detail the $D_d$, $D_m$, and $D_c$ term followed by a discussion of rest of the $C_*$ terms. The $D_*$ terms constitutes the distinctive makeup of the \textit{BRDF}, hence, we approximate them with best possible quality.
\paragraph{Disney-metallic ($D_m$)} The \textit{Disney-metallic} term is modelled using the \textit{anisotropic GGX} function as described by equation 8 in the reference~\cite{Li2022DisneyBSDF} document. 
\textit{anisotropic GGX} requires four input -- the roughness $\alpha_x, \alpha_y$ along the surface tangents, and the dot-products $h_x, h_y$ of the half-vector (similar to \textit{isotropic GGX} case) with the surface tangents. Hypothetically it is possible to model \textit{anisotropic GGX} using \textit{differentiable indirection} consisting of a 4D primary -- queried using the coordinates $(\alpha_x, \alpha_y, h_z, h_y)$, and a corresponding 4D cascaded -- queried using the pointers stored in the primary. However, doing so is inefficient due to several reasons. First, the raw size of 4D arrays are too big to accommodate in lowest tier caches or \textit{SRAM}. Second, two lookups adds to latency, and 4D interpolations requires many \flops~ -- which our technique is intended to replace. Also, such process would not exploit the fact that some computations can be baked into the textures as an offline process. Instead, we rewrite equation 8 as follows:
\begin{equation*}
    D_m(\alpha_x, \alpha_y, h_x, h_y) = \frac{1}{d_0(\alpha_x, \alpha_y)\cdot h_x^2 + d_1(\alpha_x, \alpha_y)\cdot h_y^2 + d_2(\alpha_x, \alpha_y)},
\end{equation*}
and learn the coefficients $d_*(\alpha_x, \alpha_y)$ using \textit{differentiable indirection}. Thus the primary is a 2D array queried using the coordinates $(\alpha_x, \alpha_y)$. The cascaded is also a 2D-array (with 3 channels). The channels in the cascaded corresponds to the triplet -- $d_0, d_1, d_2$ in the above equation. Notice we only use \textit{DIn} to process the artist control parameters $\alpha_x, \alpha_y$. The geometric dot products $h_x, h_y$ are mixed in with the output of \textit{DIn} to produce the final output. Such factorization has the advantage that the primary array can be large (2k$\times$2k) and works completely offline. The primary array acts as an encoder that takes in $\alpha_x, \alpha_y$ and outputs an encoded information or pointer. 
As discussed earlier, artist control parameter $\alpha_x, \alpha_y$ are stored as uv-mapped textures. In our case, we use the primary array to process each texel in the texture and store the encoded texture instead. The encoded texture is accessed similar to a standard texture fetch and processed using the cascaded array. The cascaded array is kept small (16$\times$16) so that it fits in lower tier cache or \textit{SRAM}. Notice only the cascaded array is required at runtime, minimizing the number of lookups and \flops.
\paragraph{Disney-diffuse ($D_d$)} We express the \textit{Disney-diffuse} term as a weighted sum of two functions -- $f_i$ and $g_i$. The first function is parameterized using a subset of the 10 artist control parameters, we call $\mathbf{\alpha_d} \in [0, 1)^2$ and the second function using some subset of geometry dot-product, we call $\mathbf{h_d} \in [0, 1)^3$, as shown below: 
\begin{equation*}
    D_d(\mathbf{\alpha_d}, \mathbf{h_d}) \approx \sum\limits_{i=0}^{K}f_i(\mathbf{\alpha_d})g_i(\mathbf{h_d}).
\end{equation*}
The accuracy is improved with more terms but empirical observations show $K=3$ is sufficient to attain accuracy over 35dB. Due to reasons similar to \textit{Disney-metallic}, the artist parameter functions $f_i$ are modelled using \textit{differentiable indirection} while the $g_i$ terms are directly computed.
\paragraph{Disney-clearcoat ($D_c$)} The \textit{Disney-clearcoat} term is modelled similar to the \textit{metallic} term except it is simpler. We again rewrite equation 12 in the reference document as:
\begin{equation*}
    D_c(\alpha_c, h_z) = \frac{1}{c_0(\alpha_c)\cdot h_z^2 + c_1(\alpha_c)},
\end{equation*}
where we model the functions $c_i$ using \textit{differentiable indirection}.
\paragraph{Coefficient terms} The coefficient terms $c_*$ in equation \ref{eq:disneySplit} are trained together with the $D_*$ terms. That is, we have independent loss functions for $D_*$ terms but we train $c_*D_*$ together. Hence, the individual $c_*$ terms maybe less accurate but acceptable as long as they complement the $D_*$ terms in their product. Many of the $c_*$ and $D_*$ terms share similar lookups, further amortizing cost. Figure \ref{fig:disneyInterimResults} visualizes several $c_*$ terms.

\begin{figure*}[t!]
\begin{tikzpicture}
    \node[anchor=south west,inner sep=0] at (0,0){
    \includegraphics[width=17.4cm,trim={0cm 0cm 0cm 0.0cm},clip]{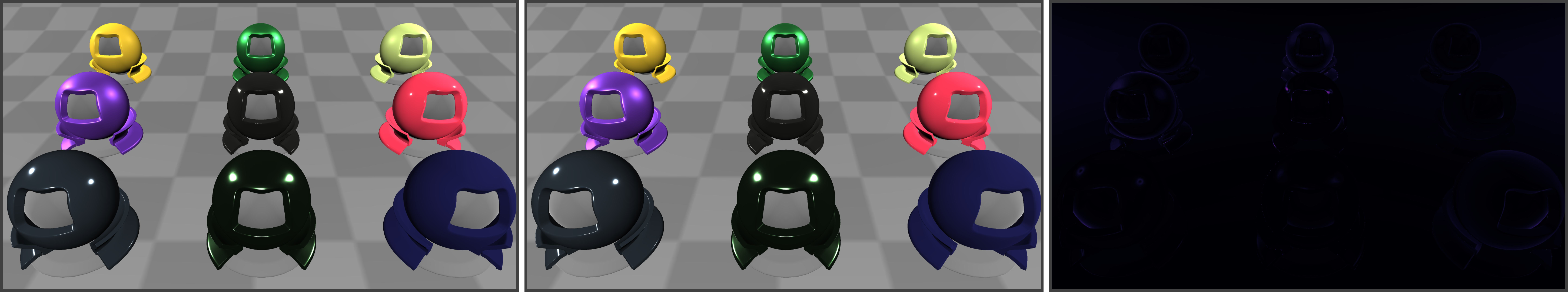}};;
    \node[anchor=south west] at (0.0, 2.8) {\textcolor{black}{\small Network output}};
     \node[anchor=south west] at (5.8, 2.8) {\textcolor{black}{\small Reference}};
     \node[anchor=south west] at (11.6, 2.8) {\textcolor{white}{\small Flip Error}};
\end{tikzpicture}
\caption{\label{fig:disneyResults} Figure visualizes the final output of \textit{Disney BRDF} approximation for a variety of material configuration.}
\end{figure*}

\begin{figure*}[h!]
\begin{tikzpicture}
    \node[anchor=south west,inner sep=0] at (0,0){
    \includegraphics[width=17.4cm,trim={0cm 0cm 0cm 0.0cm},clip]{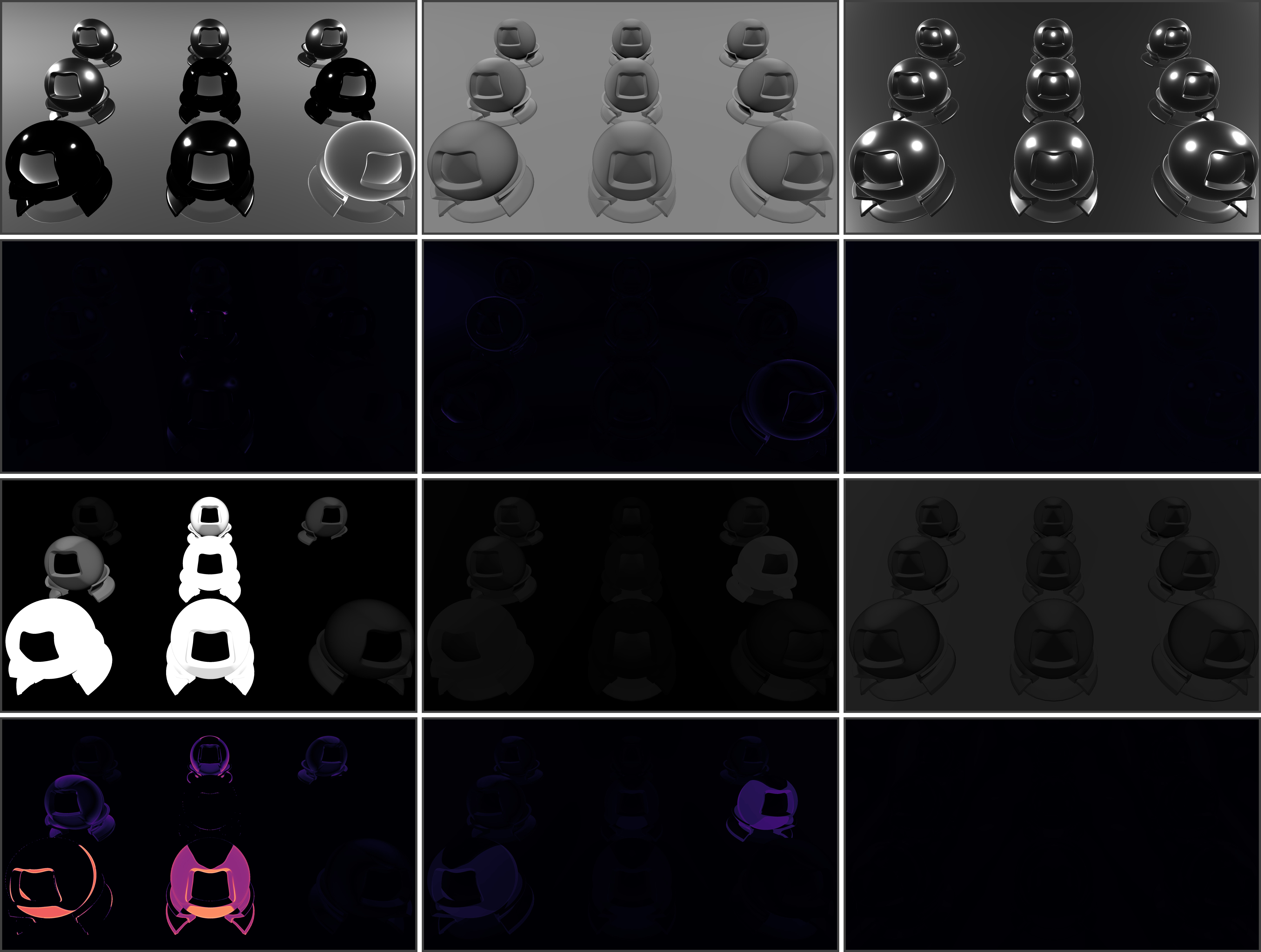}};;
     \node[anchor=south west] at (0.0, 12.65) {\textcolor{white}{\small \textit{Disney-metallic}--$D_m$}};
     \node[anchor=south west] at (5.8, 12.65) {\textcolor{white}{\small \textit{Disney-diffuse}--$D_d$}};
     \node[anchor=south west] at (11.6, 12.65) {\textcolor{white}{\small \textit{Clearcoat}--$D_c$}};

     \node[anchor=south west] at (0.0, 9.35) {\textcolor{white}{\small Flip Error--$D_m$}};
     \node[anchor=south west] at (5.8, 9.35) {\textcolor{white}{\small Flip Error--$D_d$}};
     \node[anchor=south west] at (11.6, 9.35) {\textcolor{white}{\small Flip Error--$D_c$}};

     \node[anchor=south west] at (0.0, 6.05) {\textcolor{white}{\small $c_{m0}$}};
     \node[anchor=south west] at (5.8, 6.05) {\textcolor{white}{\small $c_{m1}$}};
     \node[anchor=south west] at (11.6, 6.05) {\textcolor{white}{\small $c_d$}};

     \node[anchor=south west] at (0.0, 6.05-3.3) {\textcolor{white}{\small Flip Error--$c_{m0}$}};
     \node[anchor=south west] at (5.8, 6.05-3.3) {\textcolor{white}{\small Flip Error--$c_{m1}$}};
     \node[anchor=south west] at (11.6, 6.05-3.3) {\textcolor{white}{\small Flip Error--$c_d$}};
\end{tikzpicture}
\caption{\label{fig:disneyInterimResults} Figure visualizes the intermediate outputs of \textit{Disney BRDF} approximation under a variety of material configuration. Notice errors in $c_{m0}$ is masked by the $D_m$ term as the product $c_{m0}D_m$ is jointly optimized while $D_m$ is also optimized separately.}
\end{figure*}

\begin{table}[b!]
\caption{\label{tab:lookups} Brief description of various lookups associated with \textit{Disney BRDF} approximation.}
\begin{tabular}{|c|c|c|}
\hline
\begin{tabular}[c]{@{}c@{}}Dimension/\\ Interpolation\end{tabular} & O/P Channels & Associated Output                                                             \\ \hline
16x16 / Bilinear                                                   & 3            & Anisotropic GGX                                                               \\ \hline
16x16 / Bilinear                                                   & 1            & Smith Masking                                                                 \\ \hline
16x16 / Bilinear                                                   & 4            & \begin{tabular}[c]{@{}c@{}}Disney Diffuse, \\ Clear coat masking\end{tabular} \\ \hline
16x16 / Bilinear                                                   & 4            & Metallic, Clear coat, Sheen                                                   \\ \hline
16 / Linear                                                        & 2            & Clear coat gloss                                                              \\ \hline
\end{tabular}
\end{table}
\paragraph{Training and inference} Training pipeline is similar to section \ref{sec:isoGgx}, except we have many more parameters that we sample from the appropriate distributions. We sample the roughness parameters for \textit{metallic} and \textit{clearcoat} from an exponential like distribution while rest of the artist parameters and albedo from a uniform distribution. We sample the emitter and camera direction from a hemispherical distribution with normal vector pointing at (0,0,1). One important note is that we directly learn the denominator in the expressions for $D_m, D_c$, thus partially avoiding the unbounded non-linear behavior of the two functions. During inference, only the cascaded arrays, as outlined in table \ref{tab:lookups}, are required. The output of the primary arrays are encoded as standard textures.  

\begin{figure*}[h!]
\begin{tikzpicture}
    \node[anchor=south west,inner sep=0] at (0,0){
    \includegraphics[width=17.4cm,trim={0cm 0.0cm 0cm 0.0cm},clip]{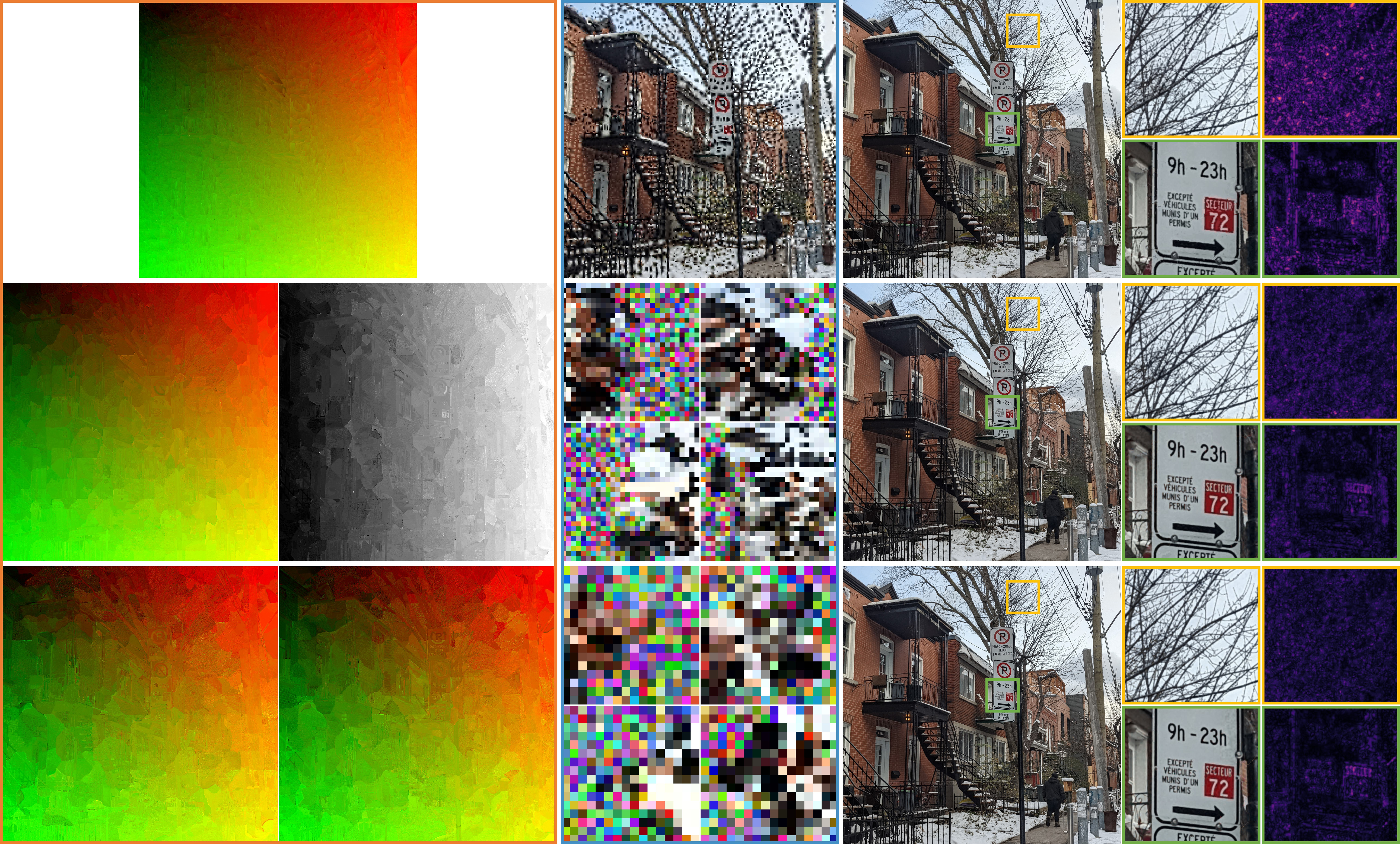}};;
     \node[anchor=south west] at (3.0, 10.5) {\textcolor{black}{\small Primary}};
     \node[anchor=south west] at (8.0, 10.5) {\textcolor{black}{\small Cascaded}};
     \node[anchor=south west] at (13.5, 10.5) {\textcolor{black}{\small Output}};
    \node[anchor=south west, rotate=90] at (0, 8.0) {\textcolor{black}{\small Cascaded-2D}};
     \node[anchor=south west, rotate=90] at (0, 4.5) {\textcolor{black}{\small Cascaded-3D}};
     \node[anchor=south west, rotate=90] at (0, 1.0) {\textcolor{black}{\small Cascaded-4D}};
     \node[anchor=south west] at (10.5, 7.1) {\textcolor{white}{PSNR: 27.2}};
     \node[anchor=south west] at (10.5, 3.55) {\textcolor{white}{PSNR: 30.7}};
     \node[anchor=south west] at (10.5, 0.0) {\textcolor{white}{PSNR: 32.3}};

     \node[anchor=south west] at (15.9, 8.8) {\textcolor{white}{Flip-Err}};

     \node[anchor=south west] at (2.3, 7.0) {\textcolor{black}{$976^2$, 2 channels}};
     \node[anchor=south west] at (7.5, 7.0) {\textcolor{white}{$244^2$, 3 channels}};

    \node[anchor=south west] at (2.3, 3.5) {\textcolor{gray}{$880^2$, 3 channels}};
    \node[anchor=south west] at (7.5, 3.5) {\textcolor{white}{$28^3$, 3 channels}};

    \node[anchor=south west] at (2.3, 0) {\textcolor{black}{$736^2$, 4 channels}};
    \node[anchor=south west] at (7.5, 0) {\textcolor{white}{$16^4$, 3 channels}};
\end{tikzpicture}
\caption{\label{fig:primaryCascadedContent} Figure visualizes the contents of the primary and cascaded arrays for varying (vertically) network configurations. All network compresses a 2k image by 6$\times$ using \textit{differentiable indirection}. In case of 3D and 4D cascaded arrays, 2D slices of the multi-dimensional volume is visualized in the $2^{nd}$ and $3^{rd}$ row. Result improve with the dimension of the cascaded array.}
\end{figure*}
\paragraph{FLOPs calculation and performance} The reference \textit{Disney BRDF} uses 92 additions, 150 multiplications, 17 divisions, 5 square-roots, and 1 logarithm. Our implementation, in addition to the 5 lookups as detailed below in table \ref{tab:lookups}, requires 11 additions, 26 multiplications, and 2 divisions. The extra arithmetic is required to combine the output of the lookups into the final \textit{BRDF} output. To compare compute instructions as a single number, we assume additions and multiplications require 1 \flop, divisions 2 \flops, square root and logarithms 4 \flops~each. These are likely conservative estimates on modern hardware and low power hardware may require more \flops~for complex instructions. Plugging these values, we estimate the required number of \flops~for reference implementation is about 300 \flops. We reduce the number of \flops~by a factor of 0.8 to account for optimization inefficiencies in our implementation and arrive at a ~240 \flops~estimate for reference implementation. Our approximate implementation requires 5 lookups and 41 \flops.

We note that our implementation is not unique but one of several possible implementations using \textit{differentiable indirection}. However, our factorization yields good results close to reference with performance advantage even on commodity GPU hardware. Note the total array size is 6KB (half-precision) and 12KB (full-precision) and easily stored in on-chip memory such as \textit{SRAM} or \textit{L1}-cache. To obtain a performance advantage over reference implementation on a commdity GPU hardware such as Nvidia 3000 series, it is recommended to use hardware texture sampler with hardware bi-linear interpolations to fetch the results from the cascaded arrays.
Figure \ref{fig:disneyResults} shows the final output of our technique for a variety of materials. Figure \ref{fig:disneyInterimResults} shows the various intermediate output of our technique. Notice how errors in $c_{m0}$ complements $D_m$.

\subsection{Compact Image Representation \label{sec:compactImg}}

\begin{figure*}[h!]
\begin{tikzpicture}
    \node[anchor=south west,inner sep=0] at (0,0){
    \includegraphics[width=17.4cm,trim={0cm 0.0cm 0cm 0.0cm},clip]{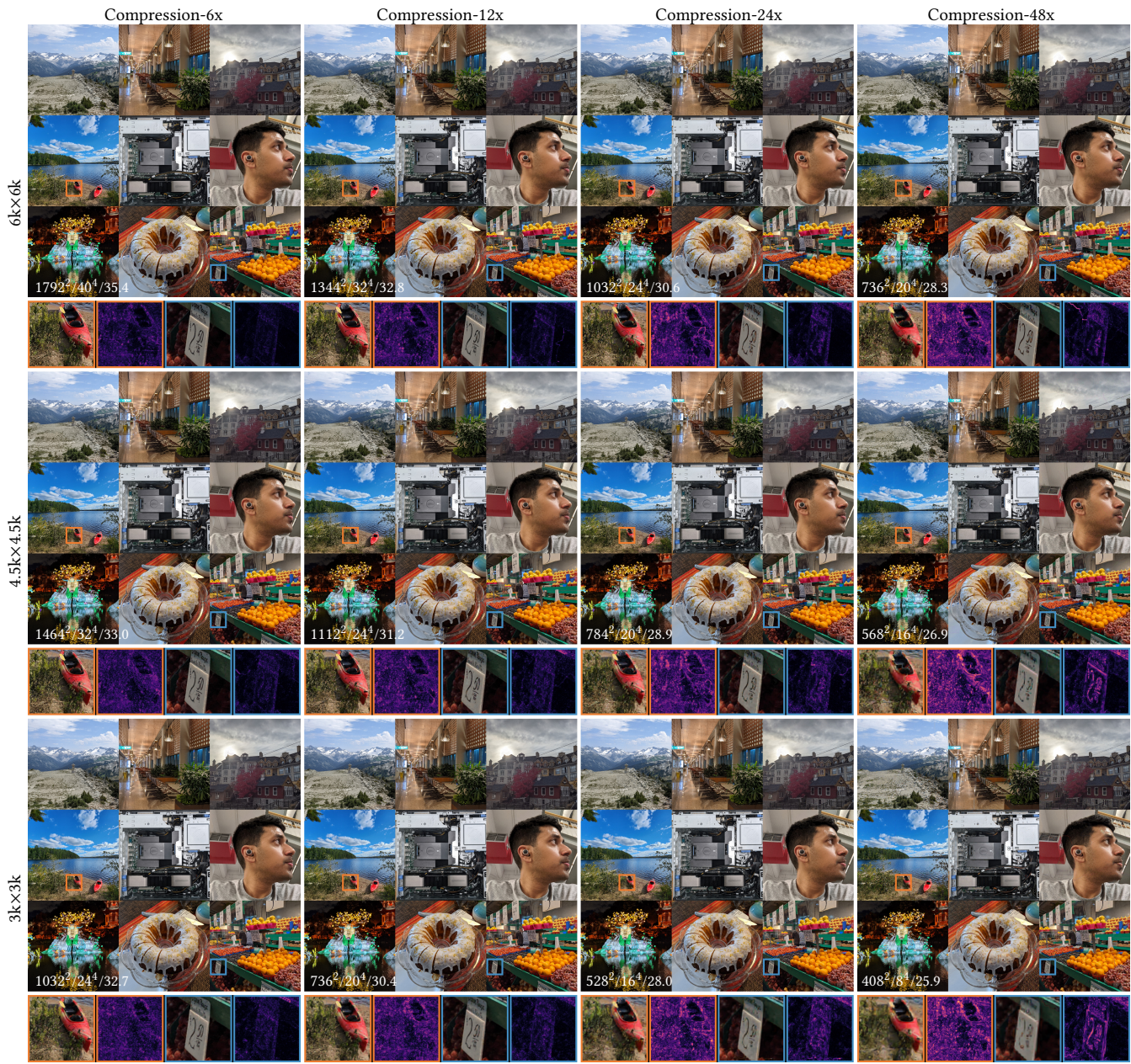}};;
    \node[anchor=south west] at (0.8+0.3, 16.3) {\textcolor{black}{\small Compression-6x}};
     \node[anchor=south west] at (5.1+0.3, 16.3) {\textcolor{black}{\small Compression-12x}};
     \node[anchor=south west] at (9.5+0.3,16.3) {\textcolor{black}{\small Compression-24x}};
      \node[anchor=south west] at (13.8+0.3, 16.3) {\textcolor{black}{\small Compression-48x}};
      \node[anchor=south west, rotate=90] at (0, 13.1) {\textcolor{black}{\small 6k$\times$6k}};
    \node[anchor=south west, rotate=90] at (0, 7.4) {\textcolor{black}{\small 4.5k$\times$4.5k}};
    \node[anchor=south west, rotate=90] at (0, 2.1) {\textcolor{black}{\small 3k$\times$3k}};
    {
    \newcommand{\wOffset}{0.0}
    \newcommand{\wSkip}{4.35}
    \newcommand{\vOffset}{12.05}
    \newcommand{\vSkip}{-5.48}
     \newcommand{\texcol}{white}
    \node[anchor=south west] at (\wOffset + 0*\wSkip,  \vOffset + 0*\vSkip) {\textcolor{\texcol}{\footnotesize $1792^2$/$40^4$/35.4} };
    \node[anchor=south west] at (\wOffset + 1*\wSkip,  \vOffset + 0*\vSkip) {\textcolor{\texcol}{\footnotesize $1344^2$/$32^4$/32.8} };
    \node[anchor=south west] at (\wOffset + 2*\wSkip,  \vOffset + 0*\vSkip) {\textcolor{\texcol}{\footnotesize $1032^2$/$24^4$/30.6} };
    \node[anchor=south west] at (\wOffset + 3*\wSkip,  \vOffset + 0*\vSkip) {\textcolor{\texcol}{\footnotesize $736^2$/$20^4$/28.3} };
    \node[anchor=south west] at (\wOffset + 0*\wSkip,  \vOffset + 1*\vSkip) {\textcolor{\texcol}{\footnotesize $1464^2$/$32^4$/33.0} };
    \node[anchor=south west] at (\wOffset + 1*\wSkip,  \vOffset + 1*\vSkip) {\textcolor{\texcol}{\footnotesize $1112^2$/$24^4$/31.2} };
    \node[anchor=south west] at (\wOffset + 2*\wSkip,  \vOffset + 1*\vSkip) {\textcolor{\texcol}{\footnotesize $784^2$/$20^4$/28.9} };
    \node[anchor=south west] at (\wOffset + 3*\wSkip,  \vOffset + 1*\vSkip) {\textcolor{\texcol}{\footnotesize $568^2$/$16^4$/26.9} };
    \node[anchor=south west] at (\wOffset + 0*\wSkip,  \vOffset + 2*\vSkip) {\textcolor{\texcol}{\footnotesize $1032^2$/$24^4$/32.7} };
    \node[anchor=south west] at (\wOffset + 1*\wSkip,  \vOffset + 2*\vSkip) {\textcolor{\texcol}{\footnotesize $736^2$/$20^4$/30.4} };
    \node[anchor=south west] at (\wOffset + 2*\wSkip,  \vOffset + 2*\vSkip) {\textcolor{\texcol}{\footnotesize $528^2$/$16^4$/28.0} };
    \node[anchor=south west] at (\wOffset + 3*\wSkip,  \vOffset + 2*\vSkip) {\textcolor{\texcol}{\footnotesize $408^2$/$8^4$/25.9} };
    }
\end{tikzpicture}
\caption{\label{fig:imgParamSweep} Figure shows the effect of compression (horizontally) on varying image resolution (vertically). While the $1^{st}$ row compresses a native resolution image, the $2^{nd}$ and $3^{rd}$ row compresses a pre-downsampled version of the native image. We use a 2D-primary/4D-cascaded network configuration for all images in this figure. The resolution of the primary, cascaded arrays, and the associated output PSNRs are provided for all images. We note, it is more difficult to compress an already downsampled image, as indicated by a reducing PSNR along each column. Cutouts underneath each image highlight two interesting regions as zoomed-in views along with associated \textit{FLIP} error map.}
\end{figure*}

Images are multi-channel 2D arrays storing a variety of presumed spatially redundant data. Images are natural -- such as photographs taken by a smartphone camera or synthetic -- such as video-game textures. Images are also used as containers to store other forms of data such as \textit{normal}~\cite{normalMappingLearnOGL, glintSurvey} maps used to alter shading computations or \textit{displacement}~\cite{dispalcementMapGpuGem, displacementMapSurvey} maps to alter the underlying geometry. Usually photographs or textures have 3 channels for storing red, green, and blue colors for each pixel. With \textit{PBR} shading, textures may have more than 3 channels. Extra channels store information like \textit{roughness} maps, \textit{normal} maps, \textit{ambient-occlusion} maps or other spatially varying parameters required to reproduce the behavior of a material. A compact representation of images is thus of utmost importance. In a real-time context, lowering the number of compute \flops, memory bandwidth required to decode a pixel and the overall representation size is crucial.

\textit{Block compression}~\cite{blockComp79, citeAstc} techniques exploit spatial locality of images by regressing a line through a block of pixels and only storing an index identifying a point on the line for each pixel. The technique also stores the end-points of a line per block. Our technique may be considered a generalized extension of block compression where we learn the indices (in the primary array) and the regressed line (in the cascaded array) using gradient decent and \textit{differentiable indirection}. There are several possible arrangements for primary and cascaded array in our case. Figure \ref{fig:primaryCascadedContent} shows a hyper-parameter sweep across various primary/cascaded arrangements such as 2D/2D, 2D/3D, and 2D/4D.

\begin{table}[t!]
\caption{\label{tab:rgbRatios} Empirically obtained optimal resolution ratios ($\rho=N_{p}/N_c$) for varying compression of 4K RGB-textures.}
\begin{tabular}{l|llll|}
\cline{2-5}
\multirow{2}{*}{}                 & \multicolumn{4}{l|}{Compression Ratio}                                                                   \\ \cline{2-5} 
                                  & \multicolumn{1}{c|}{3x}  & \multicolumn{1}{c|}{6x}  & \multicolumn{1}{c|}{12x} & \multicolumn{1}{c|}{24x} \\ \hline
\multicolumn{1}{|l|}{RGB} & \multicolumn{1}{l|}{128} & \multicolumn{1}{l|}{128} & \multicolumn{1}{l|}{80}  & 72                       \\ \hline
\end{tabular}
\end{table}

\paragraph{Setup} We use a 2D array with 4 channels as the primary and a 4D array with $k$-channels as the cascaded array, where $k$ is the number channels in a texture or image. We initialize the primary array with uv-ramp repeated twice for 4 channels. We initialize the cascaded array with grey, white, light-blue (0.5,0.5,1), and grey for albedo, ao, normal, and roughness channels respectively. Other channels are set zero. Let us assume the length of one side of the primary array is $N_p$ and the cascaded array is $N_c$. The ratio of the two sides is indicated by $\rho = N_p/N_c$. Some pseudo-optimal values of $\rho$ are provided in table \ref{tab:rgbRatios}. The values are obtained by varying $\rho$ for a given resolution and compression and selecting the one with best output quality. Our use cases have $\rho \in [40, 128]$. A safe default value for 4K textures is $\rho=96$ and for 1K or 2K textures is $\rho=64$ at 12x compression ratios.

 Generally, the optimal $\rho$ is proportional to the redundancies in a texture. Usually, natural images and high resolution images have more redundant pixels and prefers a higher $\rho$. The optimal $\rho$ also varies inversely with the required compression ratios. Using the defaults values under various tested circumstances produce results within 5\% from optimal.

To set up the resolution of the primary and cascaded array, we use three information - uncompressed size of the texture in bytes denoted by B, expected compression denoted by $e$, and $\rho$. The size of our representation in bytes is given by:
\begin{equation}
\begin{split}
    B_{comp} &= 4 N_p^2 + kN_c^4 \\
             &= 4 \rho^2N_c^2 + kN_c^4, \,\, \text{using} \,\, \rho = \frac{N_p}{N_c}.
\end{split}
\label{eq:compactImgSz}
\end{equation}
Note that $e = B/B_{comp}$; thus using equation \ref{eq:compactImgSz}, we can solve for $N_c$. However, we add some additional constraints. We require $N_p$, and $N_c$ are integer multiples of 8, and 4 respectively to avoid memory alignment issues. Thus instead of directly solving the variable, we do a linear search in $N_c$ that minimizes the difference $e - B/B_{comp}$ while also satisfying all constraints.
\paragraph{Training and inference} We train the network using {uv, color} pairs and use stratified random sampling to generate the training uv coordinates. Each strata corresponds to a texel in the base texture. We obtain the target color values using a bi-linear sampler. We tested with \textit{VGG-19}~\cite{Vgg19} and \textit{SSIM} losses which require 2D patches of texels to perform convolutions. However, the convolutional losses did not significantly impact quality but slows the training. As such, we only use \textit{MAE} as loss. We use \textit{ADAM} optimizer with 0.001 learning rate. For inference, we quantize both primary and the cascaded array to 8-bit. An interesting effect is that the cascaded array is often $\leq1$MB for 4K textures, which may fit in a lower tier cache.
\paragraph{Results and visualizations} Figure \ref{fig:primaryCascadedContent} shows the content of primary and cascaded arrays for 2D primary and varying (2-4D) cascaded network configurations. Figure \ref{fig:imgParamSweep} shows the effect compression (6-48$\times$) on a range of resolutions. 

\subsection{Neural texture sampling \label{sec:texSamp}}

Using uv-mapped textures in 3D scenes requires appropriate \textit{texture filtering} to avoid aliasing. Aliasing occurs when the pixels on screen do not align one-to-one with the texels on a texture. The mapping is either one-to-many (minification-filtering) or many-to-one (magnification-filtering), depending on the size of the projected size of pixel footprint in the texture-space. In modern GPUs, filtering is performed inside hardware using a chain of mip-maps obtained from the base texture. At runtime, the appropriate mip-levels are selected based on the pixel-footprint and \textit{tri-linearly} interpolated between adjacent mip-levels. More advanced filtering involves \textit{anisotropic} filtering which not only takes into account the size of the pixel footprint but also its orientation in texture-space. Our goal is to approximate a \textit{tri-linear} texture sampler using \textit{differentiable indirection} without storing an explicit mip-chain.     

\paragraph{Setup} Our texture sampler takes two input - a uv coordinate and an estimate of the pixel footprint. In real-time systems~\cite{realTimeSystem}, the latter is computed using shader derivatives. Shader derivatives are \textit{numerical finite difference} derivatives of a quantity w.r.t.\ x-axis (horizontal, called \textit{ddx}) and y-axis (vertical, called \textit{ddy} \textit{Direct 3D}) in screen space. These derivatives are generally hardware accelerated and computed in \textit{pixel/fragment} shaders. For inference purposes, we collect the shader derivatives of the uv-coordinates as part of the \textit{GBuffer} generation among many other parameters required for shading. We compute the pixel footprint as the magnitude of the cross product between the of the two derivatives of the uv coordinates w.r.t.\ horizontal and vertical axis in screen-space.

Our network consist of two primary arrays: one corresponding to the uv-input, identified by its resolution $N_{p0}$. Second primary corresponding to the pixel footprint input, called $N_{p1}$. The shapes of the arrays are provided in the table \ref{tab:configSampler}. 

\begin{table}[!t]
\caption{\label{tab:configSampler} Network configuration details for texure sampler.}
\begin{tabular}{|c|c|c|c|}
\hline
Table Name & Shape/Resolution       & Input                                                         & \begin{tabular}[c]{@{}c@{}}o/p \\ channels\end{tabular} \\ \hline
Primary-0  & 2D, $N_{p0} \times N_{p0}$ & uv-coords                                                     & 3                                                       \\ \hline
Primary-1  & 1D, $N_{p1}$               & Pixel-footprint                                               & 1                                                       \\ \hline
Cascaded   & 4D, $N_{lod} \times N_c^3$ & \begin{tabular}[c]{@{}c@{}}Primary-1\\ Primary-0\end{tabular} & k                                                       \\ \hline
\end{tabular}
\end{table}

We use a single cascaded array with 4-dimesnions. The resolution of the cascaded is not uniform along all dimensions. $N_c$ is the resolution of the sides attached to $N_{p0}$ (primary array for uv) and $N_{lod}$ is the resolution of the side attached to $N_{p1}$ (primary array for footprint). For instance, we would compute the volume of the cascaded array as $N_{lod}\times N_c^3$. 

We set $N_{lod}\approx log_2(N_{base})$, where $N_{base}$ is the base texture resolution. We further fine tune $N_{lod}$ with hyper parameter search. The pseudo-optimal values are $N_{lod} =$ 8, 12, 12 for 1K, 2k, 4k textures respectively. We set $N_{p1} = 4N_{lod}$. The total bytes required for our representation is thus 

\begin{table}[b!]
\caption{\label{tab:samplerRatios} Empirically obtained optimal resolution ratios ($N_{p0}/N_c$) for varying compression of 4K RGB and material texture for combined compression and sampling.}
\begin{tabular}{c|cccc|}
\cline{2-5}
\multicolumn{1}{l|}{}                                                                        & \multicolumn{4}{l|}{Compression ratio}                                              \\ \cline{2-5} 
                                                                                             & \multicolumn{1}{c|}{3x}  & \multicolumn{1}{c|}{6x} & \multicolumn{1}{c|}{12x} & 24x \\ \hline
\multicolumn{1}{|c|}{\begin{tabular}[c]{@{}c@{}}RGB\\ (3-channels)\end{tabular}}             & \multicolumn{1}{c|}{80}  & \multicolumn{1}{c|}{64} & \multicolumn{1}{c|}{64}  & 48  \\ \hline
\multicolumn{1}{|c|}{\begin{tabular}[c]{@{}c@{}}RGB, Normal, AO\\ (7-channels)\end{tabular}} & \multicolumn{1}{c|}{128} & \multicolumn{1}{c|}{96} & \multicolumn{1}{c|}{96}  & 72  \\ \hline
\end{tabular}
\end{table}

\begin{equation}
\begin{split}
    B_{comp} &= 3N_{p0}^2 + N_{p1} + kN_c^3N_{lod} \\
             &= 3 \rho^2N_c^2 + N_{p1} + kN_c^3N_{lod}, \,\, \text{using} \,\, \rho = \frac{N_{p0}}{N_c}.
\end{split}
\label{eq:texSampImgSz}
\end{equation}
Rest of the values - $N_{p0}$, and $N_c$ are estimated similar to section \ref{sec:compactImg} with pseudo-optimal values values of $\rho$ provided in table \ref{tab:samplerRatios}. 

\paragraph{Training} To generate the target data for training, we emulate a proxy \textit{tri-linear} texture sampler that mimics a \textit{tri-linear} texture sampler in real-time 3D APIs such as \textit{OpenGL} or \textit{D3D}. The proxy sampler, only used for target data generation, works by generating a mip-chain of the base texture and \textit{tri-linearly} interpolate between the mip-levels according to uv-coordinates and pixel footprint. Thus the proxy sampler is a soft-emulation of the actual hardware. For training, we randomly sample the uv-coordinates using stratified-random sampling, similar to section \ref{sec:compactImg}. We also randomly sample the pixel footprint values $\in [0, 1)$, where 0 corresponds to most detailed and 1 to least detailed level-of-detail (LOD) in the mip-chain. We pass the same inputs (uv and pixel footprint) to our network and the proxy texture sampler and compare their output to train our network.

We ensure half of the random samples belong to LOD-0 by generating uniform random samples and raising it to the power $n$, where $n = -log(N_{base}) / log(p)$. In our case, $p=0.5$ - corresponding to 50\% samples in LOD-0. We provide a proof in the next section. 
\paragraph{Sampling LODs for training data generation} Our goal is to generate more training samples from the detailed and less samples from the coarser mip-maps. One way achieve this is to sample the pixel footprint values from an exponential distribution, however, exponential distribution tends to put too few samples for less detailed mip-maps. The samples however, are better distributed by simply raising the samples collected from a uniform distribution to the power of $n$ as described in previous paragraph. This section calculates the optimal value of $n$. We generate the pixel footprint samples ($x_{f}$) as    

\begin{figure}[t!]
 \begin{tikzpicture}
    \node[anchor=south west,inner sep=0] at (0.0,0.0) {
    \includegraphics[width=11.5cm, trim={0.0cm 13.2cm 19.50cm 0.0cm},clip]{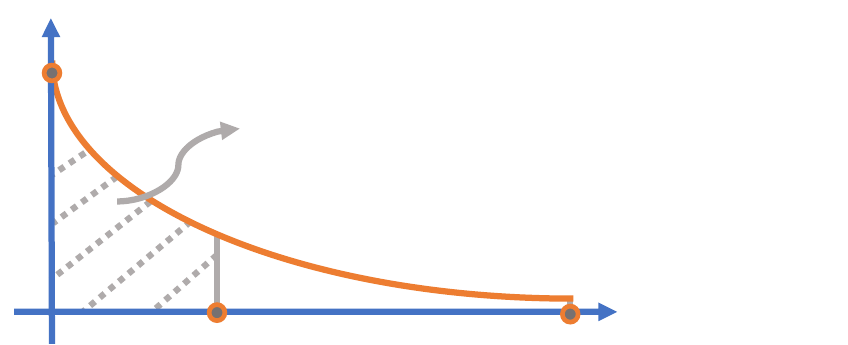}};
    \node[anchor=south west] at (0.2, 3.5) { \textcolor{black}{\small 1} };
    \node[anchor=south west, rotate=90] at (0.6, 1.2) { \textcolor{black}{Density} };
    \node[anchor=south west] at (0.3, 0.0) { \textcolor{black}{0} };
    \node[anchor=south west] at (2.8, -0.05) { \textcolor{black}{$t$} };
    \node[anchor=south west] at (5, 0.0) { \textcolor{black}{$x$} };
    \node[anchor=south west] at (5, 0.0) { \textcolor{black}{$x$} };
    \node[anchor=south west] at (7.6, -0.05) { \textcolor{black}{1} };
    \node[anchor=south west] at (7.6, -0.05) { \textcolor{black}{1} };
    \node[anchor=south west] at (3.2, 2.6) {\textcolor{black}{$\int_0^t \text{pdf}(x) dx = p$} };
\end{tikzpicture}
\caption{\label{fig:sampleFrac} Figure illustrating the fraction of samples $p$ below a threshold $t$ as being equivalent to evaluating the \textit{CDF} at $t$.}
\end{figure}

\begin{equation}
    x_f = u^n,\,\, u \sim U(0, 1),
    \label{eq:uniformPow}
\end{equation}
where $U$ indicates a uniform distribution. Let us assume we generate fraction $p (\neq \rho)$ of the total samples from pixel footprint value $t$ or below. The \textit{cumulative distribution function} of $U$ is given as

\begin{equation}
    \text{CDF}_U(u) = P(U \leq u) = u.
    \label{eq:uniformCDF}
\end{equation}
We calculate the \textit{cumulative distribution function} of $X_f$ as

\begin{equation}
\begin{split}
    \text{CDF}_{X_f}(x_f) &= P(X_f \leq x_f), \text{ using eq. \ref{eq:uniformPow}} \\
                  &= P(U^n \leq x_f)  \\
                  &= P(U \leq x_f^{\frac{1}{n}}), \text{ using eq. \ref{eq:uniformCDF}} \\
                  &= x_f^{\frac{1}{n}}.
\end{split}
\end{equation}
From figure \ref{fig:sampleFrac}, note that fraction of samples $p$ below a threshold $t$ as being equivalent to evaluating the \textit{CDF} at $t$. Therefore,

\begin{equation}
\begin{split}
    CDF_{X_f}(t) = t^{\frac{1}{n}} = p
    \text{ or }n = log_p(t).
\end{split}
\end{equation}

In our application, $t = 1/N_{base}$, and $p=0.5$. 
\\

\noindent \textbf{Optional -- Exponential sampling:} We perform a similar analysis as last section for selecting the correct parameter ($\lambda_e$) for sampling the footprints from an exponential distribution given by $exp(-\lambda_ex)$.
\begin{equation}
    x_f = -\frac{1}{\lambda_e}ln(u)
    \label{eq:uniformExp}
\end{equation}
\begin{equation}
    \text{CDF}_{X_f}(x_f) = 1 - e^{-\lambda \cdot x_f} 
    \label{eq:expCdf}
\end{equation}
\begin{equation}
    \lambda_e = -\frac{ln(1 - p)} {t}
\end{equation}
Even with appropriate parameters, the low resolution LODs receive too few samples, hence not used for our application.

\subsection{Optimized shading pipeline \label{sec:optPipeline}}
The section aims at improving the quality of the final render by making the neural texture sampler aware of the approximate \textit{BRDF}. We do so by enforcing two loss functions. First loss compares the output of our neural texture sampler with the output of the reference proxy texture sampler similar to section \ref{sec:texSamp}. 

The second loss compares the output of neural texture sampler through our approximate \textit{Disney BRDF} with the corresponding target reference. To generate the reference, we collect 16 uniformly distributed samples across the pixel footprint, and aggregate the samples post evaluation through the reference \textit{Disney BRDF}; process referred as \textit{appearance filtering} in the literature. However, to minimize aliasing at large pixel footprints without increasing sample count, we also pre-filter each individual sample with tri-linear filtering. We assume axis-aligned pixel footprint. For backpropagation, we use our pre-trained learned \textit{Disney BRDF} as a differentiable fixed function layer - i.e. we freeze the contents of the pre-trained decoder arrays in our approximate \textit{BRDF} while allowing gradients to back-propagate through the decoder arrays.

 We blend in the two losses using a hyper-parameter $\lambda \in [0,1]$, where $\lambda=0$ indicates purely the first loss while 1 indicates pure second loss. We notice some interesting details.
\begin{figure}[t!]
 \begin{tikzpicture}
    \node[anchor=south west,inner sep=0] at (0.0,0.0) {
    \includegraphics[width=7.5cm, trim={0.0cm 7.7cm 19.1cm 1.0cm},clip]{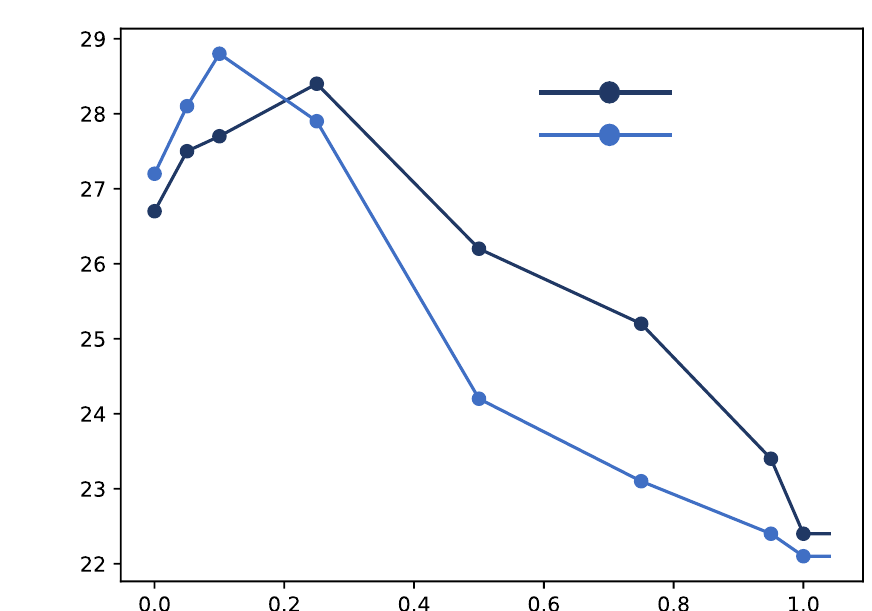}};
    \node[anchor=south west, rotate=90] at (0.6, 1.9) { \textcolor{black}{PSNR (dB)} };
    \node[anchor=south west] at (4, -0.4) { \textcolor{black}{$\lambda$} };
    \node[anchor=south west] at (5.8, 4.3) { \textcolor{black}{\small Fabric scene} };
    \node[anchor=south west] at (5.8, 3.9) { \textcolor{black}{\small Horse scene} };

\end{tikzpicture}
\caption{\label{plot:psnrLambda} Plot illustrating variation in quality with increasing $\lambda$ for two scenes. $\lambda$ indicates the fraction of shading loss used for training as discussed in \ref{sec:optPipeline}}
\end{figure}

\begin{table}[b!]
\caption{\label{tab:sdfRatios} Empirically obtained optimal resolution ratios ($N_p/N_c$) for varying parameter size in the SDF task.}
\begin{tabular}{l|cccccc|}
\cline{2-7}
                            & \multicolumn{6}{c|}{Parameter size}                                                                                                             \\ \cline{2-7} 
                            & \multicolumn{1}{c|}{96MB} & \multicolumn{1}{c|}{48MB} & \multicolumn{1}{c|}{24MB} & \multicolumn{1}{c|}{16MB} & \multicolumn{1}{c|}{12MB} & 6MB \\ \hline
\multicolumn{1}{|c|}{Piano} & \multicolumn{1}{c|}{2.56} & \multicolumn{1}{c|}{2.71} & \multicolumn{1}{c|}{2.08} & \multicolumn{1}{c|}{2.81} & \multicolumn{1}{c|}{2.56} & 2   \\ \hline
\end{tabular}
\end{table}

Optimizing our learned sampler directly using the shaded samples (i.e. $\lambda=1$) results in training instability. Our test with a simple diffuse \textit{BRDF} also indicates similar issues. We believe there are two main issues. First, the optimization is underconstrained i.e. different combinations of albedo, normal, AO may yield the same shaded result. Second, the unbounded non-linearities in the \textit{Disney BRDF} may also cause the training to diverge. We solve the first issue by adding a regularization term, in our case we do so by setting $\lambda< 1$. Setting $\lambda$ strictly less than 1 essentially uses the first loss as regularization. We fix the second issue by clipping the gradients backpropagating through the non-linear metallic component of the \textit{BRDF}. A learning rate scheduler may also improve convergence in this case. A variation in quality due to increasing $\lambda$ is shown in plot \ref{plot:psnrLambda}. For the \textit{Horse scene}, and the \textit{Fabric scene} in figure 1, 13 of the main paper shows an improvement in quality with at $\lambda=0.1, 0.25$ for the two scenes respectively.

\subsection{Signed Distance Fields \label{sec:sdf}}
In the \sdf~case, we use an \textit{ADAM} optimizer paired with a learning rate scheduler. Also, as described in the main paper, our training samples mostly consist of near surface samples and uses the \textit{MAE} loss. We quantize the distances to $\pm1$ and learn a truncated \sdf~representation. For faster convergence, we also initialize the cascaded 3D-grid with an approximate \sdf~representation. At each voxel of the cascaded array, we use a Kd-Tree to lookup the nearest sample in the training set and set the voxel's content as the corresponding truncated distance. We notice, pre-initializing the cascaded array minimizes artifacts. The primary 3D-array is initialized as uniform ramp as usual. We set up the array resolutions similar to section \ref{sec:compactImg} and use $\rho \in [1.5,2.5]$ as shown in table \ref{tab:sdfRatios}. During inference we quantize both the primary and cascaded arrays to 8-bit. 

\begin{table}[b!]
\caption{\label{tab:nerfRatios} Empirically obtained optimal resolution ratios ($N_p/N_c$) for varying compression ratios in NeRF task. Uncompressed density and RGB grids have a resolution of $256\times256\times256$ with 1 and 12 channels respectively.}
\begin{tabular}{l|cccccc|}
\cline{2-7}
\multirow{2}{*}{}                                                                        & \multicolumn{6}{c|}{Compression Ratio}                                                                                                                                \\ \cline{2-7} 
                                                                                         & \multicolumn{1}{c|}{5x}   & \multicolumn{1}{c|}{10x}  & \multicolumn{1}{c|}{25x}  & \multicolumn{1}{c|}{50x}  & \multicolumn{1}{c|}{75x}  & 100x                      \\ \hline
\multicolumn{1}{|c|}{\begin{tabular}[c]{@{}c@{}}Density Grid\\ (1 channel)\end{tabular}} & \multicolumn{1}{c|}{3.57} & \multicolumn{1}{c|}{3.8}  & \multicolumn{1}{c|}{2.69} & \multicolumn{1}{c|}{1.97} & \multicolumn{1}{c|}{2.82} & 2.51                      \\ \hline
\multicolumn{1}{|c|}{\begin{tabular}[c]{@{}c@{}}RGB Grid\\ (12-channels)\end{tabular}}   & \multicolumn{1}{l|}{5.05} & \multicolumn{1}{l|}{3.42} & \multicolumn{1}{l|}{4.14} & \multicolumn{1}{l|}{4.53} & \multicolumn{1}{l|}{3.58} & \multicolumn{1}{l|}{5.74} \\ \hline
\end{tabular}
\end{table}

\subsection{Radiance field compression \label{sec:suppl:radfield}}
Similar to section \ref{sec:sdf}, we initialize the cascaded 3D-array with the a scaled version of the target grid extracted from the \textit{Direct Voxel} technique. Such initialization is optional but improves training convergence. However, unlike section \ref{sec:sdf}, we only quantize the primary array to 8-bit during inference. $\rho$ values are provided in table \ref{tab:nerfRatios}. We use standard \textit{MAE} loss and \textit{ADAM} optimizer. We use the \textit{Direct Voxel} code base to train the grid-representation, albeit with some minor modification -- we remove the \mlp~used at the output of the RGB field and replace it with \textit{positional encoding} (\textit{PE}). The view directions are mixed in with the latent vectors from the grid representation using \textit{PE}. The output of the \textit{PE} is summed and passed through a \textit{sigmoid} function to generate the final output. We note the \textit{Direct Voxel} code base already uses \textit{PE}, we thus only replace the \mlp~with a summation.

Extracting the grid representation and replacing with our primary/cascaded network is challenging as the gird representation is deeply embedded and requires several modifications in the inference code. One needs to carefully match the output of the grid representation for a given input; as such, it is important to use the same sampling recipe used in the \textit{Direct Voxel} technique to generate the target data with. We also carefully align the data to initialize the cascaded array such that the output is as close to the reference as possible. A misaligned initialization may cause slower training.      

\section{Known issues and fixes}

\begin{figure}[t!]
 \begin{tikzpicture}
    \node[anchor=south west,inner sep=0] at (0.0,0.0) {
    \includegraphics[width=8.5cm, trim={0.0cm 0.0cm 0.0cm 0.0cm},clip]{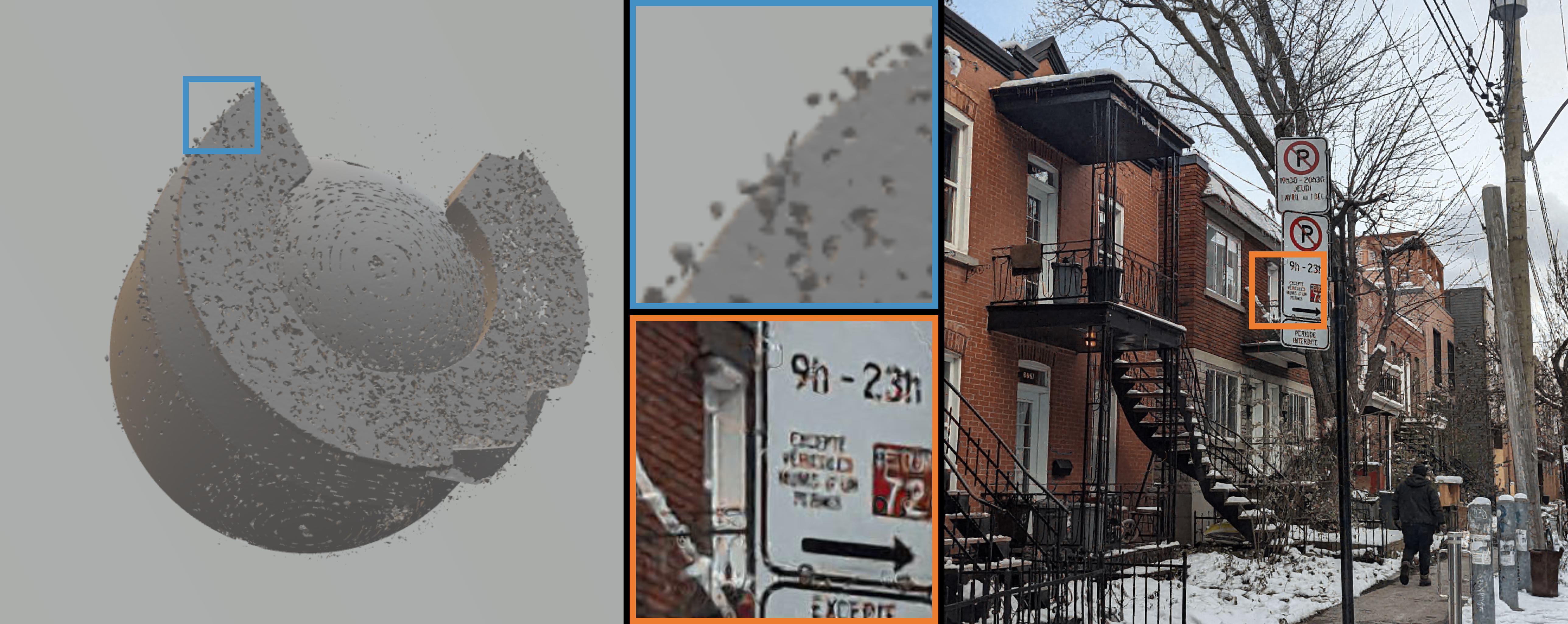}};
     \node[anchor=south west] at (0.3, 3.35) { \textcolor{black}{\sdf~representation} };
     \node[anchor=south west] at (5.5, 3.35) { \textcolor{black}{Image compression} };
\end{tikzpicture}
\caption{\label{fig:dinArtefacts} Figure illustrating common artefacts with \sdf~representation and image compression. The \textit{floater} artefacts seen on the surface of geometry is mitigated using truncated \sdf. For image compression, artefacts like nonlinear-seams may appear at high compression rate -- 48$\times$ in this case.  }
\end{figure}

\textit{Differentiable indirection} generally works out of the box; however, we document some known issues, their fixes, and potential improvements. Figure \ref{fig:dinArtefacts} shows some of the issues we encountered when using \textit{DIn}. When compressing 2D images, we noticed voronoi-seam like artefacts in the compressed representation, especially at higher compression ratios. Similar issue manifest itself in 3D as \textit{floater} artefacts in \sdf~representation as shown in the figure \ref{fig:dinArtefacts}.

\paragraph{Analysis} Remember that we initialize the primary array with \textbf{monotonically} increasing identity uv-map and let gradient descent locally distort the uv-map. The key to reducing artefact is controlling this distortion. Ideally, for an artefact free output, we should preserve monotonicity in the primary array. However, forcing monotonicity may also reduce the quality of compression, as the primary array cannot adjust freely. These opposing forces must be balanced depending on the circumstances.
\paragraph{Fixes} We start with two simple strategies. These are the first line of defense against artefacts and often extremely effective. First strategy involves controlling the magnitude of the gradient. For a given compression, we aim to limit the amount of unnecessary distortion (non-monotonic) to the identity uv-map stored in the primary. Large gradients may put unnecessary kinks in the fabric of primary which may not iron itself out during training. So we explicitly or implicitly control the gradient going from the cascaded to primary. Some example of implicit gradient control are as follows:
\begin{itemize}
    \item Pre-initialized cascaded array: The magnitude of the gradient going from cascaded to primary depends on correctness of the values stored in the cascaded. Hypothetically, if the cascaded already contains values that perfectly match the output, there is no gradient distorting the identity uv-map in the primary. As such, initialize the cascaded with good values as often as possible. In case of \sdf~and \textit{NeRF} cascaded is initialized with down-sampled version of the uncompressed representations. For shading and textures, we use a constant DC-value, generally the mean of the output.
    \item Better loss and bounded output: Avoid losses that produce unbounded output while operating on bounded values. An example of bad loss function in our context is \textit{MAPE}, generally used in for training \sdfs. The loss is infinite when the reference input is close to zero. This leads to spurious zero crossing in the compressed representation due to extremely high gradients close to a geometry surface, manifesting as \textit{floater} artifacts. We replace \textit{MAPE} with \textit{MAE} that uniformly penalize all values. However, we also quantize the output of the \sdf~represntation to $\pm1$, a.k.a \textit{truncated sdf} to emphasis the surface boundary without using unbounded loss values. In case of shading, which often produces unbounded output, we learn the reciprocal of the output which has bounded range and avoids bad gradients.      
    \item Smaller cascaded array: We highlight that the backpropagating gradients are amplified proportional to the resolution of the cascaded array. When situation permits, it is better to use smaller cascaded arrays, minimizing the risk of large gradients. We think this information may be vital for developing multi-level or deep indirection where such amplification may play crucial role in training stability.  
\end{itemize}\leavevmode\newline

Next we provide some explicit forms of gradient control:
\begin{itemize}
\item Learning rate scheduler: A learning rate scheduler, often built-in the standard libraries such as  \textit{PyTorch}, is a very useful form of gradient control. The purpose of a scheduler is to gradually lower the learning rate to reduce overshooting the optima when large (modestly) gradients are presents.
\item Gradient clipping: If it is necessary to manually restrict gradients, clipping might be the easiest solution. It is straightforward to access the gradient tensor of the primary array and clip the values before the next gradient-descent step. We determine the clipping bounds in the following paragraph. Usually gradient clipping is \textbf{not} required when other gradient control measures are set appropriately.     
\end{itemize}\leavevmode\newline

When set correctly, gradient clipping should enforce monotonicity in the \textbf{primary} array, as discussed in the next paragraph. However, strict monotonicity may be too restrictive for most cases, as it imposes harsh constraint on the optimization process and reduces reconstruction PSNR. In this paragraph, we discuss gradient clipping to minimize large kinks (not necessarily enforce monotonicity) in the primary array. An effective clipping bound is thus the average length of the diagonal of a hyper-cube cell. For a 1-dimensional primary array, the bounds are $\pm(\text{Array resolution} \times \text{learning rate})^{-1}/2$. One may also use these bounds for monitoring purposes to detect spurious behavior. We recommend tuning other parameters discussed previously before any gradient clipping. 

This paragraph discusses gradient clipping for enforcing strict monotonicity in the primary. Although from our experience, such is neither required nor recommended. To maintain strict monotonicity, one must assign a clipping range for each cell individually; in other words, a gradient descent step should not increase the value of the current cell beyond the value of the next cell and should not reduce its value below the value of the previous cell. We argue that the non-linearity $\mathcal{F}$ is not required when strict monotonicity is maintained as the content of the array is always positive and always restricted to $[0, 1)$ range. In a 1-dimensional context, we pin the first array cell to 0 and last array cell to 1; gradients do not affect the values in the boundary cells. For the $i^{th}$ cell of a 1-D array, the lower ($\epsilon_l$) and upper ($\epsilon_u$) clipping bounds are:

\begin{equation*}
    \epsilon_u[i] < \frac{c[i + 1] - c[i]}{2\alpha},
\end{equation*}
\begin{equation*}
\epsilon_l[i] > \frac{c[i - 1] - c[i]}{2\alpha},
\end{equation*}
 where $c[i]$ is the content of the array and $\alpha$ is the learning rate. We clip the gradients in the aforementioned range for the $i^{th}$ cell in the array to enforce strict monotonicity.

 We have now covered our first strategy involving controlling the gradient, either implicitly or explicitly; we are ready to discuss our second strategy. Our second strategy involves increasing the number of features or channels in the primary array and accordingly adjusting the dimensionality of the cascaded array. The additional dimensions allows for robustness against sharp kinks/distortions in the primary array. The idea is simple; at any given input location in the primary array, the probability that a large fraction of channels have sharp kinks/distortions at the same location is reduced. For example, it would be bad if all 4 channels in the primary array in figure \ref{fig:primaryCascadedContent} (last row) had a kink at the exact same location. However, the probability of such event is lowered with increasing channels. Generally, having one or more extra channels/dimensions than required is great for improving robustness against artefacts.
 
\paragraph{Advanced fixes} We have now discussed the main mechanisms to build robustness against artefacts. We next discuss some more advanced techniques, however, we found they are rarely required in practice. We introduce the concept of soft monotonicity. Unlike strict monotonicity, our goal is to encourage monotonicity in the \textbf{primary} array without forcing it.  Assuming 1-D array, we define monotonicity as:
\begin{equation*}
    \mathcal{F}(c[i + 1]) - \mathcal{F}(c[i]) > \epsilon,
\end{equation*}
where $c[i]$ indicates the content of the array at $i^{th}$ location, $\mathcal{F}$ is the output non-linearity and $\epsilon \leq 0$ is a small negative value.  Note that the left side of the equation is computing a numerical finite difference derivative using \textit{forward difference} method. One may use more advanced numerical derivative such as \textit{central difference} to compute a more accurate finite difference derivative on the left side. On the right side, setting $\epsilon=0$, would indicate strict monotonicity, while a small negative value indicates some tolerance for non-monotonic behavior. We set $\epsilon \propto -N^{-1}$, where $N$ is array resolution. We put the above equation as regularization as follows:
\begin{equation*}
     \frac{\kappa}{N} \sum\limits_{i=0}^{N-1} ReLU(\epsilon + \mathcal{F}(c[i]) - \mathcal{F}(c[i + 1])),
\end{equation*}
where $\kappa$ is a positive regularization constant. The regularization term is added with other losses and optimized using gradient descent. One small implementation details is that we first compute $\mathcal{F}(c[i])$, copy the result to another array without gradient tensor, pad the copied array with border values, and offset the indices to compute appropriate finite difference.  

\paragraph{Future work} We have outlined all important techniques that we have tested for improving robustness and reducing artefacts. We discuss another potential idea that may help reduce artifacts. Recall our arrays are linearly interpolated. A higher order interpolant such as cubic may improve smoothness and reduce artefacts but at the cost of increased computation and reduced compression capacity due cubic constraints. 

\end{document}